\documentclass[11pt,a4paper]{article}
\usepackage{amsfonts}
\usepackage{amsmath}
\usepackage{amssymb}
\usepackage{amsthm}
\usepackage{bbm}
\usepackage{color}
\usepackage{comment}
\usepackage{enumitem}
\usepackage[margin=2.41cm]{geometry}
\usepackage{graphicx}
\usepackage{hyperref}
\usepackage[utf8]{inputenc}
\usepackage{mathrsfs}
\usepackage{mathtools}
\usepackage{rotating}
\usepackage{tikz}
\usepackage{upgreek}
\usepackage{varwidth}
\usepackage[all,cmtip]{xy}
\usetikzlibrary{shapes.geometric}

\usepackage[backend=biber, sorting=nyt, style=alphabetic, uniquename=init, giveninits=true, date=year, url=false, doi=false, isbn=false, eprint=false]{biblatex}

\AtEveryBibitem{\clearfield{number}}            
\AtEveryBibitem{\clearlist{language}}
\AtEveryBibitem{\clearfield{howpublished}}

\DeclareBibliographyCategory{cited}
\AtEveryCitekey{\addtocategory{cited}{\thefield{entrykey}}}

\renewbibmacro{in:}{}                           
\DeclareFieldFormat[article,periodical]{volume}{\mkbibbold{#1}}     
\DeclareFieldFormat{journaltitle}{#1\isdot}     

\renewbibmacro*{publisher+location+date}{%
  \printlist{publisher}%
  \iflistundef{location}
    {\setunit*{\addcomma\space}}
    {\setunit*{\addcolon\space}}%
  \printlist{location}%
  \setunit*{\addcomma\space}%
  \usebibmacro{date}%
  \newunit}

\definecolor{darkred}{rgb}{0.8,0.1,0.1}
\hypersetup{
    colorlinks=false,         
    linkcolor=darkred,
    citecolor=blue,
}

\theoremstyle{plain}
\newtheorem{theo}{Theorem}[section]

\newtheorem{propo}[theo]{Proposition}
\newtheorem{cor}[theo]{Corollary}

\theoremstyle{definition}
\newtheorem{defi}[theo]{Definition}
\newtheorem{constr}[theo]{Construction}
\newtheorem{setup}[theo]{Set-up}

\newtheorem{exx}[theo]{Example}
\newenvironment{ex}
  {\pushQED{\qed}\exx}
  {\popQED\endexx}

\newtheorem{remm}[theo]{Remark}
\newenvironment{rem}
  {\pushQED{\qed}\remm}
  {\popQED\endremm}

\numberwithin{equation}{section}

\def\dR{\mathrm{dR}}
\def\cc{\mathrm{c}}
\def\colim{\mathrm{colim}}

\def\dd{\mathrm{d}}

\def\ev{\mathrm{ev}}

\def\id{\mathrm{id}}
\def\oone{\mathbbm{1}}
\def\bfone{\mathbf{1}}
\def\op{\mathrm{op}}

\def\pr{\mathrm{pr}}
\def\sc{\mathrm{sc}}
\def\supp{\mathrm{supp}}

\def\vol{\mathrm{vol}}

\def\CC{\mathbf{C}}
\def\DD{\mathbf{D}}
\def\EE{\mathbf{E}}

\def\MM{\mathbf{M}}

\def\calH{\mathcal{H}}

\def\AMod{{_A\Mod}}
\def\BMod{{_B\Mod}}

\def\CCO{\mathbf{CCO}}
\def\Ch{\mathbf{Ch}}

\def\DGA{\mathbf{DGA}}

\def\Loc{\mathbf{Loc}}

\def\Fun{\mathbf{Fun}}

\def\Mod{\mathbf{Mod}}
\def\Mon{\mathbf{Mon}}

\def\Net{\mathbf{Net}}
\def\Rep{\mathbf{Rep}}
\def\RepA{{\Rep(\AAA)}}
\def\RepB{{\Rep(\BBB)}}

\def\Vec{\mathbf{Vec}}

\def\data{\mathbf{data}}

\def\bbC{\mathbb{C}}
\def\bbK{\mathbb{K}}

\def\bbR{\mathbb{R}}

\def\bbZ{\mathbb{Z}}

\def\AAA{\mathfrak{A}}
\def\BBB{\mathfrak{B}}

\def\VVV{\mathfrak{V}}

\def\Sol{\mathfrak{Sol}}
\def\Data{\mathfrak{D}}
\def\Res{\mathfrak{Res}}
\def\Ext{\mathfrak{Ext}}

\def\FFF{\mathcal{F}}
\def\GGG{\mathcal{G}}
\def\LLL{\mathcal{L}}
\def\MMM{\mathcal{M}}

\def\V{\mathcal{V}}
\def\W{\mathcal{W}}

\def\1{I}

\newcommand\und[1]{\underline{#1}}

\def\sk{\vspace{1mm}}

\makeatletter
\let\@fnsymbol\@alph
\makeatother

\title{%
Homotopy theory of net representations
}

\author{%
Angelos Anastopoulos$^{1,a}$\ and\
Marco Benini$^{1,2,b}$\vspace{4mm}\\
{\small ${}^1$ Dipartimento di Matematica, Universit\`a di Genova,}\\
{\small Via Dodecaneso 35, 16146 Genova, Italy.}\vspace{2mm}\\
{\small ${}^2$ INFN, Sezione di Genova,}\\
{\small Via Dodecaneso 33, 16146 Genova, Italy.}\vspace{4mm}\\
{\small \begin{tabular}{ll}
Email: & ${}^a$~\texttt{anastopoulos@dima.unige.it}\\
& ${}^b$~\texttt{benini@dima.unige.it}\vspace{2mm}
\end{tabular}
}
}

\date{July 2022}

\begin{document}

\maketitle


\begin{abstract}
\noindent The homotopy theory of representations of nets of algebras over a (small) category with values in a closed symmetric monoidal model category is developed. We illustrate how each morphism of nets of algebras determines a change-of-net Quillen adjunction between the model categories of net representations, which is furthermore a Quillen equivalence when the morphism is a weak equivalence. These techniques are applied in the context of homotopy algebraic quantum field theory with values in cochain complexes. In particular, an explicit construction is presented that produces constant net representations for Maxwell $p$-forms on a fixed oriented and time-oriented globally hyperbolic Lorentzian manifold.
\end{abstract}




\renewcommand{\baselinestretch}{0.8}\normalsize
\tableofcontents
\renewcommand{\baselinestretch}{1.0}\normalsize

\newpage


\section{\label{sec:intro}Introduction and summary}
Historically quantum field theory on Minkowski spacetime 
was invented in the form of operators acting on a Hilbert 
space. This structure combines two related, but different, pieces of data, namely 
(1)~the algebra of canonical commutation relations (CCR),
witnessing the quantum nature of the theory, and 
(2)~a distinguished choice of representation 
for the (abstract) CCR algebra, whose role is to draw 
a bridge between the CCR algebra 
and its familiar implementation 
by operators acting on a Hilbert space. 
One of the fundamental achievements of the algebraic 
approach to quantum field theory 
\cite{HaagKastler_1964_AlgebraicApproach} 
was to recognize and isolate the two different and 
complementary roles played by the data (1) and (2). 
On the one hand, the general axioms of quantum field theory 
can be postulated abstractly, at the algebraic level; 
on the other hand, the (typically many) inequivalent 
representations provide incarnations of a CCR algebra 
corresponding to different states of the physical system. 
\sk

It is remarkable that the viewpoint of the algebraic 
approach to quantum field theory paved the way to vast 
generalizations, enlarging both the domain and the target categories 
of a quantum field theory. For instance, algebraic quantum 
field theories in the sense of Haag and Kastler 
\cite{HaagKastler_1964_AlgebraicApproach} were originally defined 
on the directed set of causally convex open subsets 
of Minkowski spacetime, later generalised to any fixed oriented and time-oriented globally hyperbolic 
Lorentzian manifold \cite{Dimock_1980_AlgebrasLocal}. 
More generally, locally covariant quantum field theories 
\cite{BrunettiFredenhagenVerch_2003_GenerallyCovariant} 
are defined on the category of all oriented and time-oriented 
globally hyperbolic Lorentzian manifolds, while locally covariant conformal field theories 
\cite{Pinamonti_2009_ConformalGenerally, CrawfordRejznerVicedo_2021_Lorentzian2d, BeniniGiorgettiSchenkel_2021_SkeletalModel} 
are defined on a similar category, however with orientation 
and time-orientation preserving conformal, instead of 
isometric, open embeddings whose image is causally convex.
\sk

As far as the target category is concerned, while the prime example 
is certainly the category of complex vector spaces, 
perturbative algebraic quantum field theory 
\cite{HollandsWald_2001_LocalWick, BrunettiFredenhagen_2000_MicrolocalAnalysis, BrunettiDutschFredenhagen_2009_PerturbativeAlgebraic} 
replaces the field of complex numbers with a ring 
of formal power series. Furthermore, the construction of 
quantum gauge theories through 
the Batalin-Vilkovisky formalism, which is in the center of our interest, forces one 
to work in the category of cochain complexes 
\cite{Hollands_2008_RenormalizedQuantum, FredenhagenRejzner_2012_BatalinVilkoviskyFormalism, FredenhagenRejzner_2013_BatalinVilkoviskyFormalism}.
\sk

From a conceptual point of view 
the Batalin-Vilkovisky formalism crucially relies 
on the notion of quasi-isomorphisms 
(namely the weak equivalences in the standard model 
structure on the category of cochain complexes). 
For instance, it is by means of quasi-isomorphisms 
that the Batalin-Vilkovisky formalism introduces its 
auxiliary fields, which are eventually responsible 
of the efficacy of this approach. 
Informally, this means that not only isomorphic, but more generally, weakly equivalent 
cochain complexes should be regarded as ``being the same'' for physical purposes. 
This, however, brings along the complication 
that any physically relevant construction one performs 
has to respect weak equivalences too. 
This is the starting point, and one of the main 
motivations, of the homotopy algebraic quantum field 
theory programme 
\cite{BeniniSchenkelWoike_2019_HomotopyTheory, BeniniSchenkel_2019_HigherStructures, BeniniBruinsmaSchenkel_2020_LinearYang, BruinsmaFewsterSchenkel_2021_RelativeCauchy}.
\sk

So far the above mentioned homotopy algebraic quantum field 
theory programme focused mainly on datum~(1) 
of a quantum field theory, i.e.\ the CCR algebras, 
as mentioned above. The main purpose of the 
present paper is to investigate also datum~(2),
the representations of the CCR algebras
and, in particular, their homotopy theory. 
\vspace{3.5ex}

To define representations of a quantum field theory, 
we adopt the framework of nets of algebras, 
see e.g.\ \cite{RuzziVasselli_2012_NewLight, RuzziVasselli_2012_RepresentationsNets}. 
In view of the variety of domain $\CC$ and target $\MM$ categories considered in the literature, 
some of which are listed in the previous paragraph, 
we take $\CC$ to be any (small) category and 
$\MM$ to be any complete and cocomplete closed 
symmetric monoidal category. From this general perspective 
the category of nets of algebras is simply the category of 
functors $\AAA$ from $\CC$ to the category of monoids in $\MM$ 
with morphisms given by natural transformations. 
Furthermore, a net representation $\LLL$ of $\AAA$ consists 
of an assignment of left $\AAA(c)$-modules $\LLL_c$ for all 
objects $c \in \CC$, which is coherent (in a suitable, 
but straightforward sense) with respect to all morphisms 
in $\CC$.
\sk

Our first task is to investigate the assignment 
to a net of algebras $\AAA$ of the corresponding category 
of net representations $\RepA$. In particular, we are 
interested in the relation between the categories of net 
representations $\RepA$ and $\RepB$ that is determined 
by a morphism of nets $\Phi: \AAA \to \BBB$. 
This goal is achieved via a suitable change-of-net adjunction 
$\Ext_\Phi \dashv \Res_\Phi: \RepB \to \RepA$, inspired 
by the classical change-of-monoid adjunction for left modules 
over monoids. As one expects, isomorphisms between nets 
of algebras are associated with (adjoint) equivalences 
between the associated categories of net representations. Thus, implementing the principle that, since isomorphic nets 
of algebras ``are the same'', the corresponding 
categories of net representations must also ``be the same''. 
Note, however, that ``being the same'' for categories 
of net representations means being equivalent 
(but not necessarily isomorphic) as categories. 
\sk

The above question about the categories of net 
representations ``being the same'' when the nets of algebras 
they are associated with ``are the same'' becomes 
more intricate when the target category $\MM$ 
comes endowed with a notion of weak equivalences 
that is weaker than that of isomorphisms. 
(Recall from above that this situation is encountered 
in the contexts of the Batalin-Vilkovisky formalism 
and of homotopy algebraic quantum field theory, 
where the target category $\MM = \Ch_\bbC$ of cochain 
complexes has weak equivalences given by quasi-isomorphisms.) 
The latter induces a notion of weak equivalences 
between nets of algebras, simply given by natural 
transformations whose components are weak equivalences 
in $\MM$. As a consequence, one would like to ensure 
that nets of algebras that ``are the same'', 
i.e.\ weakly equivalent, correspond to categories of net 
representations that ``are the same'' too. 
It turns out, however, that weakly equivalent nets of algebras 
may fail to be associated with categories of net representations 
that are equivalent in the ordinary categorical sense.
\sk

Indeed, in Example \ref{ex:KG} we show that two weakly equivalent nets of algebras implementing 
the simple Klein-Gordon field (the standard Klein-Gordon net 
and the one coming from the Batalin-Vilkovisky formalism) 
correspond to categories of net representations 
that are manifestly inequivalent in the ordinary categorical sense. 
Informally speaking, two equivalent 
mathematical models for the same quantum field theory 
end up to two inequivalent physical descriptions.
\sk

This evident shortcoming is the problem that we address
in the first part of this paper, Section \ref{sec:Rep}, 
namely to find the appropriate replacement of the concept of ordinary 
categorical equivalence such that the assignment of categories 
of net representations to nets of algebras becomes ``invariant'' 
with respect to weak equivalences of nets of algebras. 
In other words, the goal is to understand the appropriate notion 
of weak equivalence between net representations. 
Achieving this goal solves the concrete issue raised 
above with the Klein-Gordon field. 
\sk

It turns out that achieving this goal consists of three steps: 
(1)~Endow the categories of net representations 
with a suitable model structure (including 
a notion of weak equivalences between net representations), 
(2)~promote the change-of-net adjunction to 
a Quillen adjunction between the model categories 
of net representations from step~(1), 
(3)~show that the change-of-net Quillen adjunction 
from step~(2) is also a Quillen equivalence when it is 
associated with a weak equivalence of nets of algebras. 
By doing so, the principle that, since weakly equivalent nets 
of algebras ``are the same'', the associated 
categories of net representations must ``be the same'' too, 
is restored by formalizing the
notion of ``being the same'' with the concept of Quillen equivalence, 
which is in general weaker than the concept of ordinary categorical equivalence. 
\sk

However, it is noteworthy that the concept of Quillen equivalence
between two model categories descends to the concept of ordinary categorical equivalence
once we pass to their associated homotopy categories \cite[Ch.\ 1]{Hovey_1999_ModelCategories}
(i.e.\  the categories obtained by inverting all weak equivalences).
Therefore, the three-step approach described above establishes
that the homotopy categories of net representations 
are actually equivalent in the ordinary categorical sense when the
associated nets of algebras are weakly equivalent. 
\vspace{5mm}

With the homotopy theory of net representations at hand 
and having applications in the context of homotopy algebraic 
quantum field theory in mind, we turn to the second part of this paper, 
namely Section \ref{sec:Maxwell}, 
that addresses the problem of constructing 
explicit net representations for a given 
net of algebras valued in the target category $\MM = \Ch_\bbC$ 
of cochain complexes.
Concretely, we consider 
the net of algebras associated with Maxwell $p$-forms 
\cite{HenneauxTeitelboim_1986_FormElectrodynamics, HenneauxTeitelboim_1992_QuantizationGauge}. 
Note that Maxwell $p$-forms 
correspond to the massless Klein-Gordon field for $p=0$ 
and to linear Yang-Mills theory for $p=1$.
For this example we consider a fixed 
oriented and time-oriented globally hyperbolic 
Lorentzian manifold $M$, which for simplicity 
we assume ultra-static 
and admitting a compact spacelike Cauchy surface.
As explained in more detail below, we proceed constructing a net representation by first constructing
a representation of the algebra of observables that corresponds
to the whole spacetime $M$ and then use it to produce
a representation for the net of algebras.
\sk

The construction of a representation of the global algebra 
of observables involves two steps.
In the first step we construct a two-point function 
$\omega_2$ as a cochain map defined on the complex of linear 
observables on $M$. The fact that $\omega_2$ is a cochain map 
simultaneously encodes its compatibility both with 
the action of gauge transformations 
and with the equation of motion. 
(Incidentally, for $p=1$ we observe that $\omega_2$ recovers 
in degree $0$ cohomology the Hadamard two-point function 
constructed in \cite{FewsterPfenning_2003_QuantumWeak} 
for the electromagnetic vector potential, thus drawing 
a bridge between the approach we propose 
and well-established ones.) 
The second step mimics the first stage 
of the Gelfand-Naimark-Segal construction to define 
from the two-point function $\omega_2$ a left module 
on the global algebra of observables.
\sk

Having now a representation of the global algebra of observables
we derive a constant (in the sense of Construction \ref{constr:eval-const-adj}) net representation, 
which essentially amounts to restricting
the global algebra representation to the local algebras. 
(Constant net representations are similar in spirit to \cite[Sec.\ 4.2]{RuzziVasselli_2012_NewLight}.) 
\sk

Finally, we exploit the homotopy theory of net 
representations developed in the first part 
to present a very explicit description of the data 
of all constant (in the sense of Construction \ref{constr:eval-const-adj}) net representations up to weak equivalence 
in the simplest scenario, i.e.\ $p=1$ 
and $M$ the two-dimensional flat Lorentz cylinder. 
\vspace{3.5ex}

Let us now summarize the contents of this paper. 
Section \ref{sec:Rep} recalls the categories of nets 
of algebras and of net representations for any domain category 
$\CC$ and target symmetric monoidal category $\MM$, 
generalizing the standard concepts in a straightforward way. 
In passing, we present an adjunction relating representations 
of a net of algebras and left modules over a single monoid 
from the net. Constant net representations arise through 
a specific instance of this adjunction, which we shall use 
later on to construct 
a concrete net representation. Next, we move on to illustrate 
the change-of-net adjunction associated with a morphism 
of nets, which is manifestly an adjoint 
equivalence in the case of an isomorphism. 
As those will be useful tools when we shall later endow 
the categories of net representations with a model structure, 
we devote some time to discuss the $M$-tensoring, powering 
and enriched hom on the category of net representations. 
Motivated by the fact that the assignment of representation categories 
to nets of algebras is {\it not} invariant in the ordinary categorical 
sense with respect to weak equivalences of nets of algebras 
(Example \ref{ex:KG}), we conclude Section \ref{sec:Rep} endowing 
the category of net representations with a model structure (Corollary 
\ref{cor:Rep-model-str}) obtained through a standard (right) 
transfer construction (Theorem \ref{th:transfer}). 
For this purpose we assume $\MM$ to be a suitable 
(in the sense of Set-up \ref{setup}) closed symmetric monoidal 
model category. This allows us to promote the change-of-net 
adjunction to a Quillen adjunction, which is furthermore 
a Quillen equivalence when it is associated with 
a weak equivalence of nets of algebras 
(Proposition \ref{propo:change-net-vs-model-str}). 
Section \ref{sec:Maxwell} focuses on constructing concrete net 
representations when $\CC$ is the category of causally convex 
open subsets of a fixed oriented and time-oriented globally 
hyperbolic Lorentzian manifold $M$, $\MM = \Ch_\bbC$ 
is the closed symmetric monoidal model category 
of cochain complexes over $\bbC$ and the net of algebras is the one 
of Maxwell $p$-forms. First, we construct the 
net of algebras of Maxwell $p$-forms via CCR quantization \eqref{eq:CCR} 
of the complex of linear observables for Maxwell $p$-forms
from \eqref{eq:L} endowed with the Poisson structure 
\eqref{eq:PoissonStr}.
Then we show that the resulting net of algebras
is actually a homotopy algebraic 
quantum field theory in the sense of 
\cite{BeniniSchenkelWoike_2019_HomotopyTheory, BeniniBruinsmaSchenkel_2020_LinearYang}, 
i.e.\ it is actually defined on the category of all $m$-dimensional 
oriented and time-oriented globally hyperbolic Lorentzian 
manifolds and it fulfils both the causality axiom 
and the homotopy time-slice axiom. Then, restricting this net of algebras to 
all causally convex open subsets of a fixed oriented and time-oriented 
globally hyperbolic Lorentzian manifold $M$, 
which for simplicity we assume ultra-static 
and admitting a compact spacelike Cauchy surface, 
we present a simple construction of a two-point function 
$\omega_2$, that is the cochain map we use 
to define a concrete constant net representation. 
We conclude with a simple and explicit description 
of all constant net representations up to weak 
equivalence for Maxwell $1$-forms on the flat Lorentz 
cylinder. Finally, Appendix \ref{app:monoids-modules} 
collects some useful facts and constructions about 
left modules over a monoid and their homotopy theory.


\section{\label{sec:Rep}Representation theory for nets of algebras}
In this section we first recall the basic concepts of nets of algebras 
over a (small) category $\CC$ 
internal to a symmetric monoidal category $\MM$ and the associated 
categories of net representations. These concepts are straightforward 
generalizations of their more familiar analogs in the $C^\ast$-algebraic setting, 
see e.g.\ \cite{RuzziVasselli_2012_NewLight, RuzziVasselli_2012_RepresentationsNets}. 
Then we move on to illustrate how 
morphisms of nets of algebras induce 
change-of-net adjunctions between 
the associated categories of net representations, 
essentially unfolding the concept of change-of-monoid adjunction
from Section \ref{app:monoids-modules}. 
Finally, assuming that $\MM$ is a closed symmetric monoidal model 
category, we endow the category of nets of algebras 
and the category of net representations 
with canonical model structures combining the analogous 
results for monoids and modules recalled in Appendix 
\ref{subsec:model-str} and a well-known transfer theorem 
for cofibrantly generated model structures. 
The main goal of the model structures we propose 
is to obtain a Quillen equivalence 
between model categories of net representations
from a weak equivalence of nets of algebra. 
This goal is motivated by the simple Example \ref{ex:KG}, 
which shows that the ordinary categorical adjunction 
underlying the above mentioned Quillen equivalence fails to be an ordinary 
categorical equivalence for many examples of weak equivalences 
of nets of algebras that feature in concrete applications. 
Therefore, the outcome of this section is that, 
when the concept of ``being the same'' for nets of algebras is encoded 
by weak equivalence (for instance in the Batalin-Vilkovisky formalism, 
see Section \ref{sec:intro}), the correct formalization 
of ``being the same'' for the corresponding categories of net 
representations is provided by the concept of Quillen equivalence 
and not by the concept of ordinary categorical equivalence. 
Speaking more loosely, the proposed framework explains that 
suitable model structures on the categories 
of net representations are crucial for the assignment 
of categories of net representations to be ``invariant'' 
with respect to weak equivalences of nets of algebras.

\subsection{\label{subsec:AQFT}Nets of algebras and their representations}
Let $\MM = (\MM,\otimes,\oone)$ be a complete and cocomplete 
closed symmetric monoidal category. 
Let us briefly recall the concepts of monoids and their left modules 
in $\MM$. (The constructions and structures involving monoids 
and their left modules that are relevant for the present paper 
are briefly collected in Appendix \ref{app:monoids-modules}.) 
A {\it monoid} $A = (A,\mu,\bfone)$ in $\MM$ 
consists of an object $A \in \MM$ endowed with two morphisms 
$\mu: A \otimes A \to A$ and $\bfone: \oone \to A$ in $M$, 
called {\it multiplication} and {\it unit} respectively, 
subject to the usual associativity 
($\mu (\mu \otimes \id_A) = \mu (\id_A \otimes \mu)$) 
and unitality ($\mu (\bfone \otimes \id_A) = \id_A = \mu (\id_A \otimes \bfone)$) axioms. 
Furthermore, one defines a {\it morphism of monoids} 
$\varphi: A \to B$ as a morphism $\varphi: A \to B$ in $\MM$ 
that preserves multiplications 
($\mu_B\, (\varphi \otimes \varphi) = \varphi\, \mu_A$) 
and units ($\bfone_B = \varphi\, \bfone_A$). 
$\Mon(\MM)$ denotes the category of monoids in $\MM$. 
Taking a monoid $A \in \Mon(\MM)$, we recall that a 
{\it left $A$-module} $L = (L,\lambda)$ in $\MM$ consists of 
an object $L \in \MM$ endowed with a morphism 
$\lambda: A \otimes L \to L$ in $\MM$, 
called {\it left $A$-action}, subject to the usual axioms 
($\lambda (\id_A \otimes \lambda) = \lambda (\mu \otimes \id_L)$ 
and $\lambda (\bfone \otimes \id_L) = \id_L$). Furthermore, 
one defines a {\it morphism of left $A$-modules} 
$F: L \to L^\prime$ as a morphism $F: L \to L^\prime$ in $\MM$ 
that preserves the left $A$-actions 
($\lambda^\prime (\id_A \otimes F) = F \lambda$). 
This defines the category $\AMod$ of left $A$-modules in $\MM$. 
(Along the same lines one defines the category $\Mod_A$ 
of right $A$-modules in $\MM$.) 
\sk

Given a (small) category $\CC$, 
consider the category $\Fun(\CC,\MM)$ of functors 
$\CC \to \MM$ with natural transformations as morphisms. 
Forming tensor products of functors object-wise and considering 
the constant functor $\oone \in \Fun(\CC,\MM)$, 
which assigns $\oone \in \MM$ to all objects in $\CC$, 
endows $\Fun(\CC,\MM)$ with a symmetric monoidal structure. 
Because $\MM$ is a complete and cocomplete closed symmetric monoidal 
category, $\Fun(\CC,\MM) = (\Fun(\CC,\MM),\otimes,\oone)$ 
is a complete and cocomplete closed symmetric monoidal category too. 
(Recall that limits and colimits in functor categories 
are computed object-wise and that the internal hom of $\Fun(\CC,\MM)$ 
is computed combining limits and the internal hom of $\MM$.) 
The symmetric monoidal functor category $\Fun(\CC,\MM)$ 
leads to a particularly concise definition of nets of algebras. 

\begin{defi}\label{def:net}
Consider a (small) category $\CC$ and a complete and cocomplete 
closed symmetric monoidal category $\MM$. 
The category $\Net_{\CC}^{\MM}$ of $\MM$-valued {\it nets of algebras} 
over $\CC$ is the category $\Mon(\Fun(\CC,\MM))$ of monoids 
in the symmetric monoidal functor category $\Fun(\CC,\MM)$, 
see \ref{defi:monoid}. 
\end{defi}

\begin{rem}\label{rem:net}
Unpacking this compact definition recovers (a generalization of) 
the usual concepts of a net of algebras 
as a functor from $\CC$ to $\Mon(\MM)$
and of a morphism between nets of algebras as a natural transformation between them.
In other words, the categories 
$\Mon(\Fun(\CC,\MM))$ and $\Fun(\CC, \Mon(\MM))$ coincide manifestly, 
as explained below.
\sk

Indeed, $\AAA \in \Net_{\CC}^{\MM}$ consists of an underlying functor 
$\AAA \in \Fun(\CC,\MM)$ and two natural transformations 
$\mu: \AAA \otimes \AAA \to \AAA$ and $1: \oone \to \AAA$ 
in $\Fun(\CC,\MM)$, subject to the usual associativity 
and unitality axioms. It follows that, for all $c \in \CC$, 
the components $\mu_c: \AAA(c) \otimes \AAA(c) \to \AAA(c)$ 
and $1_c: \oone \to \AAA(c)$ in $\MM$ endow $\AAA(c) \in \MM$ 
with the structure of a monoid $\AAA(c) \in \Mon(\MM)$. 
Furthermore, combining the functoriality of the underlying 
$\AAA \in \Fun(\CC,\MM)$ and the naturality of 
$\mu: \AAA \otimes \AAA \to \AAA$ and $1: \oone \to \AAA$ 
in $\Fun(\CC,\MM)$ promotes the assignment 
$c \in \CC \mapsto \AAA(c) \in \Mon(\MM)$  
to a functor $\AAA: \CC \to \Mon(\MM)$, which is the usual 
concept of a net of algebras.
\sk

A morphism 
$\Phi: \AAA \to \BBB$ in $\Net_{\CC}^{\MM}$ is a natural transformation 
that preserves multiplications and units, i.e.\ whose components 
$\Phi_c: \AAA(c) \to \BBB(c)$ preserve object-wise multiplications 
and units. 
In other words, $\Phi$ is equivalently a natural transformation 
from the functor $\AAA: \CC \to \Mon(\MM)$ to the functor 
$\BBB: \CC \to \Mon(\MM)$, which is the usual concept of a morphism 
of nets of algebras. 
\end{rem}

\begin{ex}\label{ex:Loc}
For applications in the context of Lorentzian quantum field 
theory one often takes the source $\CC = \Loc_m$, $m\geq2$, 
to be the category of oriented and time-oriented $m$-dimensional
globally hyperbolic Lorentzian manifolds 
with morphisms given by orientation and time-orientation preserving 
isometric open embeddings whose image is causally convex. 
This leads to the generally covariant approach 
to algebraic quantum field theory \cite{BrunettiFredenhagenVerch_2003_GenerallyCovariant}. 
Another option is the source category $\CCO(M)$ 
of causally convex open subsets of a fixed $M \in \Loc_m$ 
with subset inclusions as morphisms. 
In this case one obtains nets of algebras 
in the sense of Haag and Kastler \cite{HaagKastler_1964_AlgebraicApproach, Dimock_1980_AlgebrasLocal}. 
For these applications the prime example of target $\MM = \Vec_\bbC$ 
is the symmetric monoidal category of vector spaces over $\bbC$. 
On the other hand, motivated by the Batalin-Vilkovisky formalism,
in homotopy algebraic quantum field theory 
(see Section \ref{sec:intro}) one considers as target $\MM = \Ch_\bbC$ 
the symmetric monoidal category of cochain complexes over $\bbC$. 
\end{ex}

Using the symmetric monoidal functor category $\Fun(\CC,\MM)$, 
it is also possible to give a concise definition of net representations. 

\begin{defi}\label{def:RepA}
Let $\AAA \in \Net_{\CC}^{\MM}$ be an $\MM$-valued net of algebras 
over $\CC$. The category $\RepA$ of {\it $\AAA$-representations} 
is the category $_\AAA\Mod$ of left $\AAA$-modules in the 
symmetric monoidal functor category $\Fun(\CC,\MM)$. 
\end{defi}

\begin{rem}\label{rem:RepA}
Unpacking this compact definition recovers (a generalization of) 
the usual concepts of an $\AAA$-representation and of a morphism 
between $\AAA$-representations, 
see e.g.\ \cite{RuzziVasselli_2012_NewLight, RuzziVasselli_2012_RepresentationsNets}. 
Indeed, $\LLL \in \RepA$ consists of an underlying functor 
$\LLL \in \Fun(\CC,\MM)$ and of a natural transformation 
$\lambda: \AAA \otimes \LLL \to \LLL$ in $\Fun(\CC,\MM)$ 
subject to the usual left $\AAA$-action axioms. 
This structure is equivalently encoded by the following data~(1-2) 
and axioms~(i-ii): 
\begin{enumerate}[label=(\arabic*)]
\item For each $c \in \CC$, a left $\AAA(c)$-module $\LLL_c \in {_{\AAA(c)}}\Mod$. 
\item For each morphism $\gamma: c_1 \to c_2$ in $\CC$, 
a morphism  of left $\AAA(c_2)$-modules 
$\LLL_\gamma: \AAA(c_2) \otimes_{\AAA(c_1)} \LLL_{c_1} \to \LLL_{c_2}$ in ${_{\AAA(c_2)}}\Mod$, 
or equivalently a morphism of left $\AAA(c_1)$-modules 
$\overline{\LLL}_\gamma: \LLL_{c_1} \to \LLL_{c_2}\vert_{\AAA(c_1)}$ in ${_{\AAA(c_1)}}\Mod$. 
\end{enumerate}
\begin{enumerate}[label=(\roman*)]
\item For each $c \in \CC$, 
$\LLL_{\id_c}: \AAA(c) \otimes_{\AAA(c)} \LLL_c \to \LLL_c$ in ${_{\AAA(c)}}\Mod$
coincides with the canonical isomorphism of $\AAA(c)$-modules induced 
by the left $\AAA(c)$-action or, equivalently, 
$\overline{\LLL}_{\id_c} = \id_{\LLL_c}: \LLL_c \to \LLL_c = \LLL_c\vert_{\AAA(c)}$ 
in ${_{\AAA(c)}}\Mod$ coincides with the identity.. 
\item For each pair of composable morphisms 
$\gamma_1: c_1 \to c_2$ and $\gamma_2: c_2 \to c_3$ in $\CC$, 
the diagram 
\begin{equation}
\xymatrix@C=5em{
\AAA(c_3) \underset{\AAA(c_2)}{\otimes} \AAA(c_2) \underset{\AAA(c_1)}{\otimes} \LLL_{c_1} \ar[r]^-{\AAA(c_3) \underset{\AAA(c_2)}{\otimes} \LLL_{\gamma_1}} \ar[d]_-{\cong} & \AAA(c_3) \underset{\AAA(c_2)}{\otimes} \LLL_{c_2} \ar[d]^-{\LLL_{\gamma_2}} \\ 
\AAA(c_3) \underset{\AAA(c_1)}{\otimes} \LLL_{c_1} \ar[r]_-{\LLL_{\gamma_2 \gamma_1}} & \LLL_{c_3}
}
\end{equation}
commutes in the category ${_{\AAA(c_3)}}\Mod$ of left 
$\AAA(c_3)$-modules or, equivalently, the diagram 
\begin{equation}
\xymatrix@C=5em{
\LLL_{c_1} \ar[r]^-{\overline{\LLL}_{\gamma_1}} \ar[d]_-{\overline{\LLL}_{\gamma_2 \gamma_1}} & \LLL_{c_2}\vert_{\AAA(c_1)} \ar[d]^-{\overline{\LLL}_{\gamma_2}\vert_{\AAA(c_1)}} \\ 
\LLL_{c_3}\vert_{\AAA(c_1)} \ar@{=}[r] & \LLL_{c_3}\vert_{\AAA(c_2)}\vert_{\AAA(c_1)}
}
\end{equation}
commutes in the category ${_{\AAA(c_1)}}\Mod$ of left $\AAA(c_1)$-modules. 
\end{enumerate}
This presentation matches the usual description of  
a net representation. 
\sk

A morphism $\FFF: \LLL \to \LLL^\prime$ in $\RepA$ 
is a natural trnasformation that preserves the left $\AAA$-actions. 
This structure is equivalent to a morphism of left $\AAA(c)$-modules 
$\FFF_c: \LLL_c \to \LLL^\prime_c$ in ${_{\AAA(c)}}\Mod$ 
for each $c \in \CC$, subject to 
commutativity of the diagram 
\begin{equation}
\xymatrix@C=5em{
\AAA(c_2) \underset{\AAA(c_1)}{\otimes} \LLL_{c_1} \ar[r]^-{\AAA(c_2) \underset{\AAA(c_1)}{\otimes} \FFF_{c_1}} \ar[d]_-{\LLL_{\gamma}} & \AAA(c_2) \underset{\AAA(c_1)}{\otimes} \LLL^\prime_{c_1} \ar[d]^-{\LLL^\prime_{\gamma}} \\ 
\LLL_{c_2} \ar[r]_-{\FFF_{c_2}} & \LLL^\prime_{c_2}
}
\end{equation}
in the category ${_{\AAA(c_2)}}\Mod$ 
of left $\AAA(c_2)$-modules or, equivalently, subject 
to the commutativity of the diagram 
\begin{equation}
\xymatrix@C=5em{
\LLL_{c_1} \ar[r]^-{\FFF_{c_1}} \ar[d]_-{\overline{\LLL}_{\gamma}} & {\LLL^\prime_{c_1}} \ar[d]^-{\overline{\LLL}^\prime_{\gamma}} \\ 
\LLL_{c_2}\vert_{\AAA(c_1)} \ar[r]_-{\FFF_{c_2}\vert_{\AAA(c_1)}} & \LLL^\prime_{c_2}\vert_{\AAA(c_1)}
}
\end{equation}
in the category ${_{\AAA(c_1)}}\Mod$ of left $\AAA(c_1)$-modules, 
for each morphism $\gamma: c_1 \to c_2$ in $\CC$. 
\end{rem}

\begin{constr}\label{constr:eval-const-adj}
Given a fixed object $\widetilde{c} \in \CC$, there is a useful construction 
of $\AAA$-representations from left $\AAA(\widetilde{c})$-modules 
arising as part of the adjunction 
\begin{equation}\label{eq:eval-const-adj}
\xymatrix{
(-)_{\widetilde{c}}: \RepA \ar@<2pt>[r] & {_{\AAA(\widetilde{c})}\Mod} :(-)^{\widetilde{c}}. \ar@<2pt>[l]
}
\end{equation}
We shall use this construction in Section \ref{sec:Maxwell} 
to construct examples of representations for a net of algebras 
valued in cochain complexes. 
The left adjoint functor $(-)_{\widetilde{c}}$ sends 
an $\AAA$-representation $\LLL \in \RepA$ 
to the left $\AAA(\widetilde{c})$-module 
$\LLL_{\widetilde{c}} \in {_{\AAA(\widetilde{c})}}\Mod$ 
and a morphism of $\AAA$-representations 
$\FFF: \LLL \to \LLL^\prime$ 
to the morphism of left $\AAA(\widetilde{c})$-modules 
$\FFF_{\widetilde{c}}: \LLL_{\widetilde{c}} \to \LLL^\prime_{\widetilde{c}}$ in ${_{\AAA(\widetilde{c})}}\Mod$. 
To define the right adjoint functor 
$(-)^{\widetilde{c}}: {_{\AAA(\widetilde{c})}\Mod} \to \RepA$, 
it is convenient to use the equivalent description of 
$\AAA$-representations from Remark \ref{rem:RepA}. 
$(-)^{\widetilde{c}}$ sends a left $\AAA(\widetilde{c})$-module 
$L \in {_{\AAA(\widetilde{c})}}\Mod$ to the $\AAA$-representation $L^{\widetilde{c}} \in \RepA$ 
consisting of the following data: 
\begin{subequations}
\begin{enumerate}[label=(\arabic*)]
\item For each $c \in \CC$, the left $\AAA(c)$-module 
\begin{equation}
(L^{\widetilde{c}})_c \coloneqq \prod_{\widetilde{\gamma} \in \CC(c,\widetilde{c})} L\vert_{\AAA(c)} \in {_{\AAA(c)}}\Mod.
\end{equation}
\item For each morphism $\gamma: c_1 \to c_2$ in $\CC$, 
the morphism of left $\AAA(c_1)$-modules 
\begin{equation}
\xymatrix{
(\overline{L^{\widetilde{c}}})_\gamma: (L^{\widetilde{c}})_{c_1} = {\displaystyle \prod_{\widetilde{\gamma}_1 \in \CC(c_1,\widetilde{c})} L\vert_{\AAA(c_1)}} \ar[r] & {\displaystyle \prod_{\widetilde{\gamma}_2 \in \CC(c_2,\widetilde{c})} L\vert_{\AAA(c_1)}} = (L^{\widetilde{c}})_{c_2}\vert_{\AAA(c_1)}
}
\end{equation}
in ${_{\AAA(c_1)}}\Mod$ defined by sending to the $\widetilde{\gamma}_2$-component 
of the codomain the $\widetilde{\gamma}_2 \gamma$-component 
of the domain. 
\end{enumerate}
Furthermore, $(-)^{\widetilde{c}}$ sends a morphism 
of left $\AAA(\widetilde{c})$-modules $F: L \to L^\prime$ in ${_{\AAA(\widetilde{c})}}\Mod$
to the morphism of $\AAA$-representations 
$F^{\widetilde{c}}: L^{\widetilde{c}} \to L^{\prime\, \widetilde{c}}$ 
in $\RepA$ consisting of the morphism of left $\AAA(c)$-modules 
\begin{equation}
\xymatrix{
(F^{\widetilde{c}})_c \coloneqq {\displaystyle \prod_{\widetilde{\gamma} \in \CC(c,\widetilde{c})} F\vert_{\AAA(c)}}: (L^{\widetilde{c}})_c = {\displaystyle \prod_{\widetilde{\gamma} \in \CC(c,\widetilde{c})} L\vert_{\AAA(c)}} \ar[r]^-{} & {\displaystyle \prod_{\widetilde{\gamma} \in \CC(c,\widetilde{c})} L^\prime\vert_{\AAA(c)}} = (L^{\prime\, \widetilde{c}})_c
}
\end{equation}
\end{subequations}
in ${_{\AAA(c)}}\Mod$ for each $c \in \CC$. 
It is straightforward to confirm that the above construction 
defines a functor $(-)^{\widetilde{c}}$. To check that the latter is 
right adjoint to $(-)_{\widetilde{c}}$, 
we exhibit the unit $\eta$ and the counit $\varepsilon$ 
of this adjuction. The unit is the natural transformation 
$\eta: \id_{\RepA} \to ((-)_{\widetilde{c}})^{\widetilde{c}}$ 
whose component at the $\AAA$-representation $\LLL \in \RepA$ 
is the morphism of $\AAA$-representations 
$\eta_{\LLL}: \LLL \to (\LLL_{\widetilde{c}})^{\widetilde{c}}$ 
in $\RepA$ consisting of the left $\AAA(c)$-module morphism 
\begin{equation}
\xymatrix{
(\eta_{\LLL})_c \coloneqq (\overline{\LLL}_{\widetilde{\gamma}})_{\widetilde{\gamma} \in \CC(c,\widetilde{c})}: \LLL_c \ar[r]^-{} & {\displaystyle \prod_{\widetilde{\gamma} \in \CC(c,\widetilde{c})} \LLL_{\widetilde{c}}\vert_{\AAA(c)}} = ((\LLL_{\widetilde{c}})^{\widetilde{c}})_c
}
\end{equation}
in ${_{\AAA(c)}}\Mod$ for each $c \in \CC$. The counit is the natural transformation 
$\varepsilon: ((-)^{\widetilde{c}})_{\widetilde{c}} \to \id_{_{\AAA(\widetilde{c})}\Mod}$ 
whose component at the left $\AAA(\widetilde{c})$-module $L \in {_{\AAA(\widetilde{c})}}\Mod$ 
is the morphism of left $\AAA(\widetilde{c})$-modules 
\begin{equation}
\xymatrix{
\varepsilon_L \coloneqq \pr_{\id_{\widetilde{c}}}: (L^{\widetilde{c}})_{\widetilde{c}} = {\displaystyle \prod_{\widetilde{\gamma} \in \CC(\widetilde{c},\widetilde{c})} L\vert_{\AAA(\widetilde{c})}} \ar[r]^-{} & L
}
\end{equation}
in ${_{\AAA(\widetilde{c})}}\Mod$ given by the projection $\pr_{\id_{\widetilde{c}}}$ 
of the $\id_{\widetilde{c}}$-component. 
The triangle identities 
$(\varepsilon_L)^{\widetilde{c}}\, \eta_{L^{\widetilde{c}}} = \id_{L^{\widetilde{c}}}$, for all $L \in {_{\AAA(\widetilde{c})}}\Mod$, 
and $\varepsilon_{\LLL_{\widetilde{c}}}\, (\eta_\LLL)_{\widetilde{c}} = \id_{\LLL_{\widetilde{c}}}$, 
for all $\LLL \in \RepA$, are straightforward, 
which shows that \eqref{eq:eval-const-adj} 
is indeed an adjunction. 
\sk

Recall from Example \ref{ex:Loc} that net of algebras 
in the sense of Haag and Kastler are defined on the 
source category $\CC = \CCO(M)$ of causally convex open 
subsets of a fixed object $M \in \Loc_m$. 
Then, for $\widetilde{c} = M \in \CCO(M)$, 
one recognizes that the right adjoint functor $(-)^M$ in 
\eqref{eq:eval-const-adj} is the well-known construction 
of {\it constant} $\AAA$-representations 
from left $\AAA(M)$-modules over the 
global algebra of observables $\AAA(M) \in \Mon(\MM)$ of 
a net of algebras $\AAA \in \Net_{\CCO(M)}^{\MM}$ over $\CCO(M)$. 
This large class of net representations is the one 
that most frequently occurs in the literature, 
see \cite[Sec.\ 4.2]{RuzziVasselli_2012_NewLight}. 
(Note that what we call constant $\AAA$-representations are frequently 
called Hilbert space representations in the $C^\ast$-setting 
of algebraic quantum field theory. This name is motivated by the fact 
that the whole net of algebras is represented on the same Hilbert space. 
In our framework the latter is replaced by a single left 
$\AAA(M)$-module $L \in {_{\AAA(M)}\Mod}$, 
which defines an $\AAA$-representation $L^M \in \RepA$ 
through the adjunction \eqref{eq:eval-const-adj}.) 
\end{constr}

\subsubsection{Change-of-net adjunction}
In order to compare  categories of net representations 
associated with different nets of algebras we will make extensive use
of the {\it change-of-net} adjunction, i.e.\ the change-of-monoid adjunction\eqref{eq:change-monoid-adj}
applied on the symmetric monoidal category $\Net_\CC^\MM$.
More specifically, given a morphism of nets of algebras $\Phi: \AAA \to \BBB$ 
in $\Net_\CC^\MM$, 
there is an associated 
{\it change-of-net} adjunction 
\begin{equation}\label{eq:change-net-adj}
\xymatrix{
\Ext_\Phi: \RepA \ar@<2pt>[r] & \RepB :\Res_{\Phi}. \ar@<2pt>[l]
}
\end{equation}
In this case,
the {\it extension} $\Ext_\Phi$ and {\it restriction} $\Res_\Phi$ functors
admit also an explicit description in terms of the more elementary 
extension and restriction functors of the change-of-monoid adjunction \eqref{eq:change-monoid-adj}
in the underlying category $\MM$. 
\sk

Specifically, the restriction functor 
$\Res_{\Phi}: \RepB \to \RepA$ assigns to a $\BBB$-representation 
$\MMM \in \RepB$ the $\AAA$-representation 
$\Res_{\Phi} \MMM \in \RepA$ 
consisting of the following data, see Remark \ref{rem:RepA}: 
\begin{subequations}
\begin{enumerate}[label=(\arabic*)]
\item For each $c \in \CC$, the left $\AAA(c)$-module 
\begin{equation}
(\Res_{\Phi} \MMM)_c \coloneqq \MMM_c\vert_{\AAA(c)} \in {_{\AAA(c)}\Mod} 
\end{equation}
obtained restricting the left $\BBB(c)$-module $\MMM_c \in {_{\BBB(c)}\Mod}$ 
along $\Phi_c: \AAA(c) \to \BBB(c)$ in $\Mon(\MM)$. 
\item For each morphism $\gamma: c_1 \to c_2$ in $\CC$, 
the morphism of left $\AAA(c_1)$-modules 
\begin{equation}
\xymatrix{
(\overline{\Res_{\Phi} \MMM})_\gamma \coloneqq \overline{\MMM}_\gamma\vert_{\AAA(c_1)}: (\Res_{\Phi} \MMM)_{c_1} \ar[r]^-{} & \left(\MMM_{c_2}\vert_{\BBB(c_1)}\right)\vert_{\AAA(c_1)} = (\Res_{\Phi} \MMM)_{c_2}\vert_{\AAA(c_1)}
}
\end{equation}
in ${_{\AAA(c_1)}\Mod}$ defined by restricting 
the morphism of left $\BBB(c_1)$-modules 
$\overline{\MMM}_\gamma: \MMM_{c_1} \to \MMM_{c_2}\vert_{\BBB(c_1)}$ 
in ${_{\BBB(c_1)}\Mod}$
along $\Phi_{c_1}: \AAA(c_1) \to \BBB(c_1)$ in $\Mon(\MM)$ 
and then by observing that both iterated restrictions 
on the right hand side coincide with the restriction along the composition 
$\BBB(\gamma)\, \Phi_{c_1} = \Phi_{c_2}\, \AAA(\gamma): \AAA(c_1) \to \BBB(c_2)$ in $\Mon(\MM)$, 
see Remark \ref{rem:change-monoid-composition}. 
\end{enumerate}
Furthermore, $\Res_\Phi$ assigns to a morphism 
$\GGG: \MMM \to \MMM^\prime$ in $\RepB$ the morphism 
$\Res_\Phi \GGG: \Res_\Phi \MMM \to \Res_\Phi \MMM^\prime$ 
in $\RepA$ consisting of the morphisms of left $\AAA(c)$-modules 
\begin{equation}
\xymatrix{
(\Res_\Phi \GGG)_c \coloneqq \GGG_c\vert_{\AAA(c)}: (\Res_\Phi \MMM)_c \ar[r] & (\Res_\Phi \MMM^\prime)_c
}
\end{equation}
\end{subequations}
in ${_{\AAA(c)}\Mod}$ obtained restricting 
the morphism of left $\BBB(c)$-modules 
$\GGG_c: \MMM_c \to \MMM^\prime_c$ 
in ${_{\BBB(c)}\Mod}$ along 
$\Phi_c: \AAA(c) \to \BBB(c)$ in $\Mon(\MM)$, for all $c \in \CC$. 
Using the change-of-monoid adjunction 
\eqref{eq:change-monoid-adj} and its compatibility with 
compositions, see Remark \ref{rem:change-monoid-composition}, 
it is straightforward to check that the data listed above fulfil 
the axioms from Remark \ref{rem:RepA}. 
\sk

Similarly, the extension functor $\Ext_\Phi: \RepA \to \RepB$ 
admits a similar description using the extension functor of 
the change-monoid adjunction\eqref{eq:change-monoid-adj} in $\MM$. 
$\Ext_\Phi$ assigns to an $\AAA$-representation $\LLL \in \RepA$ 
the $\BBB$-representation consisting of the following data: 
\begin{subequations}
\begin{enumerate}[label=(\arabic*)]
\item For each $c \in \CC$, the left $\BBB(c)$-module 
\begin{equation}
(\Ext_\Phi \LLL)_c \coloneqq \BBB(c) \otimes_{\AAA(c)} \LLL_c \in {_{\BBB(c)}\Mod}
\end{equation}
obtained extending the left $\AAA(c)$-module $\LLL_c \in {_{\AAA(c)}\Mod}$ 
along $\Phi_c: \AAA(c) \to \BBB(c)$ 
in $\Mon(\MM)$. 
\item For each morphism $\gamma: c_1 \to c_2$ in $\CC$, 
the morphism of left $\BBB(c_2)$-modules $(\Ext_\Phi \LLL)_\gamma: \BBB(c_2) \otimes_{\BBB(c_1)} (\Ext_\Phi \LLL)_{c_1} \to (\Ext_\Phi \LLL)_{c_2}$ 
in ${_{\BBB(c_2)}\Mod}$ defined by the diagram 
\begin{equation}
\xymatrix@C=0em{
\BBB(c_2) \underset{\BBB(c_1)}{\otimes} \Big( \BBB(c_1) \underset{\AAA(c_1)}{\otimes} \LLL_{c_1} \Big) \ar[rd]_-{\cong} \ar[rr]^-{(\Ext_\Phi \LLL)_\gamma} && \BBB(c_2) \underset{\AAA(c_2)}{\otimes} \LLL_{c_2} 
\\ 
& \BBB(c_2) \underset{\AAA(c_2)}{\otimes} \Big( \AAA(c_2) \underset{\AAA(c_1)}{\otimes} \LLL_{c_1} \Big) \ar[ru]_(.65)*+<1em>{_{\BBB(c_2) \underset{\AAA(c_2)}{\otimes} \LLL_\gamma}}
}
\end{equation}
in ${_{\BBB(c_2)}\Mod}$, where we used 
that both iterated extensions on the left hand side 
are naturally isomorphic to the extension along the composition 
$\BBB(\gamma)\, \Phi_{c_1} = \Phi_{c_2}\, \AAA(\gamma): \AAA(c_1) \to \BBB(c_2)$ 
in $\Mon(\MM)$, 
see Remark \ref{rem:change-monoid-composition}, 
and then we post-composed with the extension of the morphism of left $\AAA(c_2)$-modules 
$\LLL_\gamma: \AAA(c_2) \otimes_{\AAA(c_1)} \LLL_{c_1} \to \LLL_{c_2}$ 
in ${_{\AAA(c_2)}\Mod}$ 
along $\Phi_{c_2}: \AAA(c_2) \to \BBB(c_2)$ in $\Mon(\MM)$. 
\end{enumerate}
Furthermore, $\Ext_\Phi$ assigns to a morphism 
$\FFF: \LLL \to \LLL^\prime$ in $\RepA$ the morphism 
$\Ext_\Phi \FFF: \Ext_\Phi \LLL \to \Ext_\Phi \LLL^\prime$ 
in $\RepB$ consisting of the morphisms of left $\BBB(c)$-modules 
\begin{equation}
\xymatrix{
(\Ext_\Phi \FFF)_c \coloneqq \BBB(c) \otimes_{\AAA(c)} \FFF_c: (\Ext_\Phi \LLL)_c \ar[r] & (\Ext_\Phi \LLL^\prime)_c
}
\end{equation}
\end{subequations}
in ${_{\BBB(c)}\Mod}$
obtained extending the morphism of left $\AAA(c)$-modules 
$\FFF_c: \LLL_c \to \LLL^\prime_c$ in ${_{\AAA(c)}\Mod}$
along $\Phi_c: \AAA(c) \to \BBB(c)$ 
in $\Mon(\MM)$, for all $c \in \CC$. 
Using once again the change-of-monoid adjunction 
\eqref{eq:change-monoid-adj} and its compatibility with 
compositions, see Remark \ref{rem:change-monoid-composition}, 
it is straightforward to check that the above data fulfil 
the axioms from Definition \ref{def:RepA}. 

\begin{rem}\label{rem:change-net-equivalence}
Since the change-of-net adjunction is just a special instance of 
the change-of-monoid adjunction, 
Remark \ref{rem:change-monoid-adjoint-equivalence} applies. 
Therefore, given an isomorphism $\Phi: \AAA \to \BBB$ 
in $\Net_{\CC}^{\MM}$, the change-of-net adjunction 
$\Ext_\Phi \dashv \Res_\Phi$ from \eqref{eq:change-net-adj} 
is an adjoint equivalence. 
\end{rem}

\subsubsection{\label{subsec:Rep-tensoring}\texorpdfstring{$\MM$-tensoring}{Tensoring}, powering and enriched hom on \texorpdfstring{$\RepA$}{the category of net representations}}
This section is devoted to the description of the 
$\MM$-tensoring, powering and enriched hom on the category $\RepA$ 
of representations of a net of algebras $\AAA \in \Net_\CC^\MM$.
These concepts will be handy in endowing the category $\RepA$ with a model structure in Section \ref{subsec:model-structures}.
\sk

Having defined the category of nets of algebras as the category 
of monoids over the closed symmetric monoidal functor category 
$\Fun(\CC,\MM)$, one obtains immediately from Appendix 
\ref{app:tensoring} a $\Fun(\CC,\MM)$-tensoring on the category 
$\RepA$. Then restricting the latter $\Fun(\CC,\MM)$-tensoring 
to constant functors provides an $\MM$-tensoring on $\RepA$. 
Working out the usual adjunctions leads also to the 
$\MM$-powering and enrichement on $\RepA$. 
We present below the relevant constructions 
exploiting the $\MM$-tensoring, powering and enriched hom on the category 
of left modules over a monoid, see Appendix \ref{app:tensoring}. 
\sk

Let us start from the $\MM$-tensoring 
\begin{subequations}\label{eq:Rep-tensoring}
\begin{equation}
\otimes: \RepA \times \MM \longrightarrow \RepA. 
\end{equation}
$\otimes$ assigns to an 
$\AAA$-representation $\LLL \in \RepA$ and an object $V \in \MM$ 
the $\AAA$-representation $\LLL \otimes V \in \RepA$ 
consisting of the following data:

\begin{enumerate}[label=(\arabic*)]
\item For each $c \in \CC$, the left $\AAA(c)$-module 
\begin{equation}
(\LLL \otimes V)_c \coloneqq \LLL_c \otimes V \in {_{\AAA(c)}\Mod}
\end{equation}
obtained evaluating the $\MM$-tensoring 
$\otimes: {_{\AAA(c)}\Mod} \times \MM \to {_{\AAA(c)}\Mod}$ 
on $\LLL_c \in {_{\AAA(c)}\Mod}$ and $V \in \MM$, see \eqref{eq:tensoring}. 
\item For each morphism $\gamma: c_1 \to c_2$ in $\CC$, 
the morphism of left $\AAA(c_2)$-modules 
$(\LLL \otimes V)_\gamma: \AAA(c_2) \otimes_{\AAA(c_1)} (\LLL \otimes V)_{c_1} \to (\LLL \otimes V)_{c_2}$ in ${_{\AAA(c_2)}\Mod}$ defined by the diagram 
\begin{equation}
\xymatrix@C=0em{
\AAA(c_2) \underset{\AAA(c_1)}{\otimes} \Big( \LLL_{c_1} \otimes V \Big) \ar[rd]_-{\cong} \ar[rr]^-{(\LLL \otimes V)_\gamma} && \LLL_{c_2} \otimes V 
\\ 
& {} \Big( \AAA(c_2) \underset{\AAA(c_1)}{\otimes} {\LLL_{c_1}} \Big) \otimes V \ar[ru]_-{\LLL_\gamma \otimes \id_V}
}
\end{equation}
in ${_{\AAA(c_2)}\Mod}$ as the composition of the natural 
isomorphism of left $\AAA(c_2)$-modules from Remark \ref{rem:change-monoid-vs-tensoring-powering} and  
the morphism of $\AAA(c_2)$-modules obtained by evaluating 
the $\MM$-tensoring $\otimes: {_{\AAA(c_2)}\Mod} \times \MM \to {_{\AAA(c_2)}\Mod}$ 
on $\LLL_\gamma: \AAA(c_2) \otimes_{\AAA(c_1)} \LLL_{c_1} \to \LLL_{c_2}$ 
in ${_{\AAA(c_2)}\Mod}$ 
and $\id_V: V \to V$ in $\MM$. 
\end{enumerate}
Futhermore, $\otimes$ assigns to morphisms 
$\FFF: \LLL \to \LLL^\prime$ in $\RepA$ and 
$\xi: V \to V^\prime$ in $\MM$ the morphism 
$\FFF \otimes \xi: \LLL \otimes V \to \LLL^\prime \otimes V^\prime$ 
in $\RepA$ consisting of the morphisms of left $\AAA(c)$-modules 
\begin{equation}
\xymatrix{
(\FFF \otimes \xi)_c \coloneqq \FFF_c \otimes \xi: (\LLL \otimes V)_c \ar[r] & (\LLL^\prime \otimes V^\prime)_c
}
\end{equation}
\end{subequations}
in ${_{\AAA(c)}\Mod}$
obtained evaluating the $\MM$-tensoring 
$\otimes: {_{\AAA(c)}\Mod} \times \MM \to {_{\AAA(c)}\Mod}$ 
on $\FFF_c: \LLL_c \to \LLL^\prime_c$ in ${_{\AAA(c)}\Mod}$ 
and $\xi: V \to V^\prime$ in $\MM$, 
for all $c \in \CC$. Using the $\MM$-tensoring  
for left modules \eqref{eq:tensoring} and its compatibility 
with the change-of-monoid adjuction, 
see Remark \ref{rem:change-monoid-vs-tensoring-powering}, 
it is straightforward to check that the above data fulfil 
the axioms of Remark \ref{rem:RepA} and that 
the above assignment defines a functor $\otimes$. 
\sk

The $\MM$-powering 
\begin{subequations}\label{eq:Rep-powering}
\begin{equation}
(-)^{(-)}: \RepA \times \MM^\op \longrightarrow \RepA 
\end{equation}
on $\RepA$ can be described explicitly as follows. 
$(-)^{(-)}$ assigns to an 
$\AAA$-representation $\LLL \in \RepA$ and an object $V \in \MM$ 
the $\AAA$-representation $\LLL^V \in \RepA$ consisting of the following data: 
\begin{enumerate}[label=(\arabic*)]
\item For each $c \in \CC$, the left $\AAA(c)$-module 
\begin{equation}
(\LLL^V)_c \coloneqq (\LLL_c)^V \in {_{\AAA(c)}\Mod}
\end{equation}
obtained evaluating the powering 
$(-)^{(-)}: {_{\AAA(c)}\Mod} \times \MM^\op \to {_{\AAA(c)}\Mod}$
on $\LLL_c \in {_{\AAA(c)}\Mod}$ and $V \in \MM$, see \eqref{eq:powering}. 
\item For each morphism $\gamma: c_1 \to c_2$ in $\CC$, 
the morphism of left $\AAA(c_1)$-modules 
\begin{equation}
\xymatrix{
(\overline{\LLL^V})_\gamma \coloneqq \overline{\LLL}_\gamma^{\id_V}: (\LLL^V)_{c_1} \ar[r] & (\LLL_{c_2}\vert_{\AAA(c_1)})^V = (\LLL^V)_{c_2}\vert_{\AAA(c_1)}
}
\end{equation}
in ${_{\AAA(c_1)}\Mod}$ defined as the morphism of $\AAA(c_1)$-modules obtained evaluating the powering 
$(-)^{(-)}: {_{\AAA(c_1)}\Mod} \times \MM^\op \to {_{\AAA(c_1)}\Mod}$ 
on $\overline{\LLL}_\gamma: \LLL_{c_1} \to \LLL_{c_2}\vert_{\AAA(c_1)}$ 
in ${_{\AAA(c_1)}\Mod}$ and $\id_V: V \to V$ in $\MM$, 
where we also used the commutative square 
\eqref{eq:restriction-vs-powering} on the right hand side. 
\end{enumerate}
Furthermore, $(-)^{(-)}$ assigns to morphisms 
$\FFF: \LLL \to \LLL^\prime$ in $\RepA$ and 
$\xi: V^\prime \to V$ in $\MM$ the morphism 
$\FFF^\xi: \LLL^V \to {\LLL^\prime}^{V^\prime}$ 
in $\RepA$ consisting of the morphisms of left $\AAA(c)$-modules 
\begin{equation}
\xymatrix{
(\FFF^\xi)_c \coloneqq (\FFF_c)^{\xi}: (\LLL^V)_c \ar[r] & ({\LLL^\prime}^{V^\prime})_c
}
\end{equation}
\end{subequations}
in ${_{\AAA(c)}\Mod}$
obtained evaluating the powering 
$(-)^{(-)}: {_{\AAA(c)}\Mod} \times \MM^\op \to {_{\AAA(c)}\Mod}$ 
on $\FFF_c: \LLL_c \to \LLL^\prime_c$ in ${_{\AAA(c)}\Mod}$ and $\xi: V^\prime \to V$ in $\MM$, for all $c \in C$. 
Using the $\MM$-powering functor for left modules \eqref{eq:powering} 
and its compatibility with the change-of-monoid adjuction, 
see Remark \ref{rem:change-monoid-vs-tensoring-powering}, 
it is straightforward to check that the data listed above fulfil 
the axioms of Remark \ref{rem:RepA} 
and that the above assignment defines a functor $(-)^{(-)}$. 
\sk

For each object $V \in \MM$, partial evaluations 
of the $\MM$-tensoring and of the $\MM$-powering 
on $\RepA$ give rise to the adjunction 
\begin{equation}\label{eq:Rep-tensoring-powering}
\xymatrix{
(-) \otimes V: \RepA \ar@<2pt>[r] & \RepA :(-)^V. \ar@<2pt>[l]
}
\end{equation}

We conclude describing the $\MM$-enriched hom 
\begin{subequations}\label{eq:Rep-enriched-hom}
\begin{equation}
[-,-]_\AAA: \RepA^\op \times \RepA \longrightarrow \MM
\end{equation}
on $\RepA$. $[-,-]_\AAA$ assigns to $\AAA$-representations 
$\LLL, \LLL^\prime \in \RepA$ the equalizer 
\begin{equation}
[\LLL,\LLL^\prime]_\AAA \coloneqq \lim \Big( 
\xymatrix{
{\displaystyle \prod_{c \in \CC} [\LLL_c,\LLL^\prime_c]_{\AAA(c)}} \ar@<2pt>[r]^-{\LLL^\ast} \ar@<-2pt>[r]_-{\LLL^\prime_\ast} & {\displaystyle \prod_{\substack{c_1,c_2 \in \CC\\ \gamma \in \CC(c_1,c_2)}} [\LLL_{c_1},\LLL^\prime_{c_2}\vert_{\AAA(c_1)}]_{\AAA(c_1)}}
}
\Big) \in \MM, 
\end{equation}
where we used the $\MM$-enriched hom 
$[-,-]_A: \AMod^\op \times \AMod \to \MM$ from \eqref{eq:enriched-hom}. 
$\LLL^\ast$ above denotes the morphism in $\MM$ 
defined via the universal property of the product by the diagram 
\begin{equation}
\xymatrix@R=3em@C=3.2em{
{\displaystyle \prod_{c \in \CC} [\LLL_c,\LLL^\prime_c]_{\AAA(c)}} \ar[rr]^-{\LLL^\ast} \ar[d]_-{\pr_{c_2}} && {\displaystyle \prod_{\substack{c_1,c_2 \in \CC\\ \gamma \in \CC(c_1,c_2)}} [\LLL_{c_1},\LLL^\prime_{c_2}\vert_{\AAA(c_1)}]_{\AAA(c_1)}} \ar[d]^-{\pr_{c_1,c_2,\gamma}} \\ 
[\LLL_{c_2},\LLL^\prime_{c_2}]_{\AAA(c_2)} \ar[r]_-{\text{\eqref{eq:restriction-vs-enriched-hom}}} & [\LLL_{c_2}\vert_{\AAA(c_1)},\LLL^\prime_{c_2}\vert_{\AAA(c_1)}]_{\AAA(c_1)} \ar[r]_-{[\overline{\LLL}_\gamma,\id]_{\AAA(c_1)}} & [\LLL_{c_1},\LLL^\prime_{c_2}\vert_{\AAA(c_1)}]_{\AAA(c_1)}
}
\end{equation}
in $\MM$, for each $\gamma: c_1 \to c_2$ in $\CC$. Furthermore, $\LLL^\prime_\ast$ above denotes the morphism in $\MM$ 
defined via the universal property of the product by the diagram 
\begin{equation}
\xymatrix@R=3em@C=4em{
{\displaystyle \prod_{c \in \CC} [\LLL_c,\LLL^\prime_c]_{\AAA(c)}} \ar[r]^-{\LLL^\prime_\ast} \ar[d]_-{\pr_{c_1}} & {\displaystyle \prod_{\substack{c_1,c_2 \in \CC\\ \gamma \in \CC(c_1,c_2)}} [\LLL_{c_1},\LLL^\prime_{c_2}\vert_{\AAA(c_1)}]_{\AAA(c_1)}} \ar[d]^-{\pr_{c_1,c_2,\gamma}} \\ 
[\LLL_{c_1},\LLL^\prime_{c_1}]_{\AAA(c_1)} \ar[r]_-{[\id_{\LLL_{c_1}},\overline{\LLL}^\prime_\gamma]_{\AAA(c_1)}} & [\LLL_{c_1},\LLL^\prime_{c_2}\vert_{\AAA(c_1)}]_{\AAA(c_1)}
}
\end{equation}
\end{subequations}
in $\MM$, for each $\gamma: c_1 \to c_2$ in $\CC$. 
The action of $[-,-]_\AAA$ on morphisms is defined 
combining the universal property of the limit 
in \eqref{eq:Rep-enriched-hom}, the $\MM$-enriched 
hom $[-,-]_A: \AMod^\op \times \AMod \to \MM$ 
in \eqref{eq:enriched-hom} 
and the restriction functor $(-)\vert_A: \BMod \to \AMod$ 
in \eqref{eq:change-monoid-adj}. 

\begin{rem}
Let us provide a more explicit description of the 
$\MM$-enriched hom $[\LLL,\LLL^\prime]_{\AAA} \in \MM$ 
from \eqref{eq:Rep-enriched-hom} when the target 
$\MM = \Vec_\bbK$ is the familiar closed symmetric monoidal 
category of vector spaces over a field $\bbK$. 
In this case the vector space 
$[\LLL,\LLL^\prime]_{\AAA} \in \Vec_\bbK$ 
consists of collections $x = (x_c)_{c \in \CC}$ 
of $\AAA(c)$-linear maps $x_c: \LLL_c \to \LLL^\prime_c$, 
for all $c \in \CC$, such that 
$x_{c_2}\vert_{\AAA(c_1)}\, \overline{\LLL}_\gamma = \overline{\LLL}^\prime_\gamma\, x_{c_1}: \LLL_{c_1} \to \LLL^\prime_{c_2}\vert_{\AAA(c_1)}$ 
coincide as $\AAA(c_1)$-linear maps, 
for all $\gamma: c_1 \to c_2$ in $\CC$. 
A similar description holds true for any concrete 
closed symmetric monoidal category, 
including the category of cochain complexes $\MM = \Ch_\bbK$, 
which will be used in Section \ref{sec:Maxwell}. 
\end{rem}

For each object $\LLL \in \RepA$, partial evaluations 
of the $\MM$-tensoring and of the $\MM$-enriched hom 
give rise to the adjunction 
\begin{equation}\label{eq:Rep-tensoring-enriched-hom}
\xymatrix{
\LLL \otimes (-): \MM \ar@<2pt>[r] & \RepA :[\LLL,-]_{\AAA}. \ar@<2pt>[l]
}
\end{equation}

\begin{rem}\label{rem:change-net-vs-tensoring-powering}
Note that the change-of-net adjunction \eqref{eq:change-net-adj} 
is compatible with the adjunction \eqref{eq:Rep-tensoring-powering} 
in the following sense. 
Given a morphism $\Phi: \AAA \to \BBB$ 
in $\Net_{\CC}^{\MM}$ and an object $V \in \MM$, 
the diagram of right adjoint functors 
\begin{equation}\label{eq:net-restriction-vs-powering}
\xymatrix@C=3em{
\RepB \ar[r]^-{\Res_{\Phi}} \ar[d]_-{(-)^V} & \RepA \ar[d]^-{(-)^V} \\
\RepB \ar[r]_-{\Res_{\Phi}} & \RepA
}
\end{equation}
commutes as a straightforward consequence of their 
definitions. Therefore, the corresponding diagram 
of left adjoint functors commutes 
up to a unique natural isomorphism 
$\Ext_{\Phi} (- \otimes V) \cong \Ext_{\Phi}(-) \otimes V$. 
\end{rem}

\begin{rem}\label{rem:eval-const-vs-tensoring-powering}
Also the adjunction \eqref{eq:eval-const-adj} 
is compatible with the adjunction 
\eqref{eq:Rep-tensoring-powering} in the following sense. 
Given $\widetilde{c} \in \CC$, $\AAA \in \Net_{\CC}^{\MM}$ 
and $V \in \MM$, the diagram of left adjoint functors 
\begin{equation}\label{eq:eval-vs-tensoring}
\xymatrix@C=3em{
\RepA \ar[r]^-{(-)_{\widetilde{c}}} \ar[d]_-{(-) \otimes V} & _{\AAA(\widetilde{c})}\Mod \ar[d]^-{(-) \otimes V} \\
\RepA \ar[r]_-{(-)_{\widetilde{c}}} & _{\AAA(\widetilde{c})}\Mod
}
\end{equation}
commutes as a straightforward consequence of their definitions. 
\end{rem}

\subsection{\label{subsec:model-structures}Model structure for net representations}

As anticipated in Section \ref{sec:intro}, nets of algebras valued in a 
closed symmetric monoidal model category $\MM$ come equipped 
with the projective model structure recalled below.

\begin{defi}\label{def:Net-model-str}
Let $\CC$ be a (small) category and $\MM$ 
a closed symmetric monoidal model category. 
A natural transformation in $\Net_{\CC}^\MM$ is called 
a weak equivalence (fibration) if all its components in $\MM$ 
are weak equivalences (respectively fibrations) and a cofibration 
if it has the left lifting property with respect to all acyclic 
fibrations. 
\end{defi}

It turns out that such weak equivalences between nets of algebras 
do {\it not} induce ordinary categorical equivalences between 
the associated net representation categories, see Example \ref{ex:KG} below. 
The goal of the present section is to repair this shortcoming 
by endowing net representation categories with suitable 
model structures (including a notion of weak equivalence between 
net representations), such that weak equivalences between nets of algebras 
induce Quillen equivalences between the associated model categories of net representations. 
In this way, ordinary categorical equivalence is recovered 
at the level of the homotopy categories of net representations, 
i.e.\ after inverting all weak equivalences between net representations. 

\begin{ex}[Ordinary categorical equivalence fails for weakly equivalent nets of algebras]\label{ex:KG}
To motivate the introduction of model structures on the categories of 
net representations and showcase the relevance of the upcoming 
Proposition \ref{propo:change-net-vs-model-str} 
in the context of algebraic quantum field theory, 
we examine the Klein-Gordon field of mass $m \geq 0$ using two different, 
yet equivalent, descriptions. More precisely, 
we shall construct two $\Ch_\bbC$-valued nets of algebras 
$\AAA$ and $\widetilde{\AAA}$
on $\Loc_m$ describing the Klein-Gordon field and the evident weak equivalence 
$\Phi: \widetilde{\AAA} \to \AAA$ in $\Net_{\Loc_m}^{\Ch_\bbC}$ relating them. 
($\AAA$ will denote the standard Klein-Gordon net 
\cite[Sec.\ 4.2]{FewsterVerch_2015_AlgebraicQuantum}, 
regarded as a $\Ch_\bbC$-valued net of algebras concentrated 
in degree $0$, while $\widetilde{\AAA}$ will denote 
the Klein-Gordon net from the Batalin-Vilkovisky formalism 
\cite{BeniniBruinsmaSchenkel_2020_LinearYang}.) 
From the explicit description of the associated net representation categories 
$\RepA$ and $\Rep(\widetilde{\AAA})$, 
it will be manifest that the restriction functor 
$\Res_\Phi: \Rep(\widetilde{\AAA}) \to \RepA$ 
is not even essentially surjective 
and hence it fails to be an ordinary categorical equivalence. 
This explains why the flexibility of the Batalin-Vilkovisky formalism 
forces one to endow the net representation categories with suitable 
model structures, that come along with the more flexible concept of 
Quillen equivalence. In this way, although being genuinely different 
in the ordinary categorical sense, the net representation categories 
$\RepA$ and $\Rep(\widetilde{\AAA})$, 
associated with the two equivalent descriptions 
$\AAA$ and $\widetilde{\AAA}$ of the Klein-Gordon field 
become the same in the model categorical sense 
due to Proposition \ref{propo:change-net-vs-model-str}. 
\sk

Let us start from the standard Klein-Gordon net of algebras 
$\AAA \in \Net_{\Loc_m}^{\Ch_\bbC}$. To $M \in \Loc_m$ 
it assigns the unital associative differential graded algebra 
\begin{subequations}\label{eq:KG}
\begin{equation}
\AAA(M) \coloneqq T_\bbC \big( \V(M) \big) / J \in \DGA_\bbC \coloneqq \Mon(\Ch_\bbC)
\end{equation}
that is freely generated over $\bbC$ by the cochain complex 
$\V(M) \coloneqq C^\infty_\cc(M)/PC^\infty_\cc(M) \in \Ch_\bbR$ 
concentrated in degree $0$, modulo the two-sided ideal $J \subseteq T_\bbC(\W(M))$ 
generated by the canonical commutation relations 
\begin{equation}
[\varphi_1] \otimes [\varphi_2] - [\varphi_2] \otimes [\varphi_1] - i \int_M \varphi_1\, G \varphi_2\, \vol_M\; \bfone,
\end{equation}
\end{subequations}
for all $[\varphi_1], [\varphi_2] \in \V(M)^0$. 
Here $P \coloneqq \Box - m^2: C^\infty(M) \to C^\infty(M)$ 
denotes the Klein-Gordon operator, 
$G: C^\infty_\cc(M) \to C^\infty(M)$ denotes 
the associated retarded-minus-advanced propagator 
(which vanishes on $PC^\infty_\cc(M)$ and is formally skew-adjoint, 
see \cite[Sec.\ 3.4]{BarGinouxPfaffle_2007_WaveEquations}) 
and $\vol_M$ denotes the volume form on $M$. 
The push-forward of compactly supported smooth functions 
along morphisms in $\Loc_m$ turns the assignment 
$M \in \Loc_m \mapsto \AAA(M) \in \DGA_\bbC $ into a net of algebras 
$\AAA \in \Net_{\Loc_m}^{\Ch_\bbC}$. The Klein-Gordon net 
$\widetilde{\AAA} \in \Net_{\Loc_m}^{\Ch_\bbC}$ 
from the Batalin-Vilkovisky formalism is defined in a similar fashion. 
Explicitly, to $M \in\Loc_m$ it assigns the differential graded algebra 
\begin{subequations}\label{eq:KG-BV}
\begin{equation}
\widetilde{\AAA}(M) \coloneqq T_\bbC \big( \widetilde{\V}(M) \big) / I \in \DGA_\bbC
\end{equation}
that is freely generated by the cochain complex 
\begin{equation}
\widetilde{\V}(M) \coloneqq \big( \xymatrix{C^\infty_\cc(M) \ar[r]^{P} & C^\infty_\cc(M)} \big) \in \Ch_\bbR, 
\end{equation}
resolving the quotient $\V(M)$ and concentrated in degrees $-1$ and $0$, 
modulo the two-sided ideal $I \subseteq \V(M)$ generated by 
the (graded) canonical commutation relations 
\begin{align}
\varphi_1 \otimes \varphi_2 - \varphi_2 \otimes \varphi_1 - i \int_M \varphi_1\, G \varphi_2\, \vol_M\; \bfone, && \varphi \otimes \varphi^\ddagger - \varphi^\ddagger \otimes \varphi, && \varphi^\ddagger_1 \otimes \varphi^\ddagger_2 + \varphi^\ddagger_2 \otimes \varphi^\ddagger_1, 
\end{align}
\end{subequations}
for all $\varphi_1, \varphi_2, \varphi \in \widetilde{\V}(M)^0$ and 
$\varphi^\ddagger, \varphi^\ddagger_1, \varphi^\ddagger_2 \in \widetilde{\V}(M)^{-1}$. 
(The only non-trivial commutation relations involve $0$-cochains.) 
There is an evident morphism 
\begin{equation}\label{eq:KG-compare}
\Phi: \widetilde{\AAA} \longrightarrow \AAA
\end{equation}
in $\Net_{\Loc_m}^{\Ch_\bbC}$, whose $M$-component 
$\Phi_M: \widetilde{\AAA}(M) \to \AAA(M)$ is defined on generators by 
$\Phi_M(\varphi) \coloneqq [\varphi]$ and 
$\Phi_M(\varphi^\ddagger) \coloneqq 0$, for all $M \in \Loc_m$, 
$\varphi \in \widetilde{\V}(M)^0$ 
and $\varphi^\ddagger \in \widetilde{\V}(M)^{-1}$. 
In \cite[Rem.\ 6.20]{BeniniBruinsmaSchenkel_2020_LinearYang} 
it is shown that \eqref{eq:KG-compare} is a weak equivalence. 
\sk

Consider now an $\widetilde{\AAA}$-representation $\widetilde{\LLL} \in \Rep(\widetilde{\AAA})$. 
To each $M \in \Loc_m$, $\widetilde{\LLL}$ assigns a left 
$\widetilde{\AAA}(M)$-module 
$\widetilde{\LLL}(M) \in {_{\widetilde{\AAA}(M)}\Mod}$. 
One can equivalently encode its left $\AAA(M)$-action 
$\widetilde{\lambda}_M: \widetilde{\AAA}(M) \otimes \widetilde{\LLL}(M) \to \widetilde{\LLL}(M)$ in $\Ch_\bbC$ 
as a morphism $\widetilde{\lambda}_M: \widetilde{\AAA}(M) \to [\widetilde{\LLL}(M),\widetilde{\LLL}(M)]$ 
in $\DGA_\bbC$. (Recall that the internal hom 
$[\widetilde{\LLL}(M),\widetilde{\LLL}(M)] \in \Ch_\bbC$ 
carries a canonical monoid structure whose multiplication is the composition 
and whose unit is the identity. Below we will denote its differential 
by $\partial$.) It follows from \eqref{eq:KG-BV}
that $\widetilde{\lambda}_M$ assigns 
(compatibly with the (graded) canonical commutation relations)
to each $\varphi \in \widetilde{\V}(M)^0$ a $0$-cocycle 
$\widetilde{\lambda}_M(\varphi) \in Z^0([\widetilde{\LLL}(M),\widetilde{\LLL}(M)])$ 
and to each $\varphi^\ddagger \in \widetilde{\V}(M)^{-1}$ a $(-1)$-cochain 
$\widetilde{\lambda}_M(\varphi^\ddagger) \in [\widetilde{\LLL}(M),\widetilde{\LLL}(M)]^{-1}$ 
such that 
$\partial(\widetilde{\lambda}_M(\varphi^\ddagger)) = \widetilde{\lambda}_M(P\varphi^\ddagger)$. 
\sk

Compared to an $\widetilde{\AAA}$-representation, 
an $\AAA$-representation $\LLL \in \RepA$ 
consists of fewer data, as explained below. Recalling \eqref{eq:KG}, 
for each $M \in \Loc_m$, the underlying left $\AAA(M)$-action 
$\lambda_M: \AAA(M) \to [\LLL(M),\LLL(M)]$ in $\DGA_\bbC$ 
only assigns (compatibly with the canonical commutation relations) 
to each $[\varphi] \in \V(M)^0$ a $0$-cocycle 
$\lambda_M([\varphi]) \in Z^0([\LLL(M),\LLL(M)])$. 
\sk

Recall now that the restriction functor 
$\Res_\Phi: \RepA \to \Rep(\widetilde{\AAA})$ 
simply restricts the left actions, retaining all other data. 
Given an $\widetilde{\AAA}$-representation 
$\widetilde{\LLL} \in \Rep(\widetilde{\AAA})$ 
such that $\widetilde{\lambda}_M(\varphi^\ddagger) \neq 0 \in [\widetilde{\LLL}(M),\widetilde{\LLL}(M)]^{-1}$ 
for some $M \in \Loc_m$ and $\varphi^\ddagger \in \widetilde{\V}(M)^0$ 
(for instance take $\widetilde{\LLL} = \widetilde{\AAA}$ to be the net 
of algebras $\widetilde{\AAA}$ itself regarded as an 
$\widetilde{\AAA}$-representation), 
one immediately realizes that there cannot exist 
an $\AAA$-representation $\LLL \in \RepA$ and an isomorphism 
$\FFF: \Res_\Phi \LLL \to \widetilde{\AAA}$ in $\Rep(\widetilde{\AAA})$. 
Indeed, by construction of $\Phi$, 
the left actions $\lambda_M \circ \Phi_M$ of $\Res_\Phi \LLL$ 
are such that $\lambda_M(\Phi_M(\varphi^\ddagger)) = 0 \in [\LLL(M),\LLL(M)]^{-1}$ vanishes 
for all $M \in \Loc_m$ and $\varphi^\ddagger \in \widetilde{\V}(M)^{-1}$; 
moreover, any $\widetilde{\AAA}$-representation 
that is isomorphic to $\Res_\Phi \AAA$ shares this feature. 
It follows that $\Res_\Phi$ is not essentially surjective 
and hence $\Ext_\Phi \dashv \Res_\Phi$ cannot be 
an ordinary categorical equivalence.
\end{ex}

The above example indicates that the categories of net representations
associated with weakly equivalent nets of algebras fail to be equivalent 
in the ordinary categorical sense due to the rigidity of the concept of
isomorphism in these categories. This suggests that the concept of 
isomorphism between net representations 
needs to be replaced with an appropriate concept of weak equivalence.
As we will see in the remainder of this section, 
after endowing the net representation categories 
with suitable model structures (including weak equivalences), 
we shall solve the shortcoming 
evidenced in Example \ref{ex:KG} by proving with Proposition 
\ref{propo:change-net-vs-model-str} that $\Ext_\Phi \dashv \Res_\Phi$ 
gives rise to a Quillen equivalence when $\Phi$ is a weak equivalence. 
\sk 

We construct the desired model structures on net representation categories 
$\RepA$, for all $\AAA \in \Net_\CC^\MM$, 
by exploiting (right) transfer techniques 
for cofibrantly generated model categories. 
After recalling some preliminaries about transfer of model structures, 
we will proceed with the above mentioned construction in two steps. 
Even though our results are certainly well-known 
to practitioners, it seems that the question of endowing 
the category of left modules over a monoid in the functor 
category $\Fun(\CC,\MM)$ 
has not received attention so far in the literature 
(at least when finite coproducts fail to exist 
in the source category $\CC$, see Remark \ref{rem:coprod}). 
\sk

In preparation for the transfer theorem 
that we shall recall below, let us introduce some terminology. 
For a category $\EE$ equipped with a terminal object 
$\ast \in \EE$ and with two distinguished classes of morphisms, called weak equivalences and respectively fibrations, 
one says that an object $E \in \EE$ is {\it fibrant} 
when the unique morphism $E \to \ast$ in $\EE$ 
to the terminal object $\ast$ is a fibration. 
A {\it functorial path object} 
is a triplet $(P,w,f)$ consisting of a functor $P: \EE \to \EE$, 
a natural weak equivalence $w: \id_{\EE} \to P$, 
i.e.\ a natural transformation whose components $w_E: E \to P(E)$ 
in $\EE$ are weak equivalences, for all $E \in \EE$, 
and a natural fibration $f: P \to (-)^{\times2}$, 
i.e.\ a natural transformation whose components 
$f_E: P(E) \to E \times E$ in $\EE$ are fibrations, 
for all $E \in \EE$, such that the diagonal 
morphism $(\id_E,\id_E) = f_E\, w_E: E \to E \times E$ 
in $\EE$ factors as the weak equivalence $w_E$
followed by the fibration $f_E$, for all $E \in \EE$. 

\begin{theo}[{\cite[Sec.\ 2.5 and Sec.\ 2.6]{BergerMoerdijk_2003_AxiomaticHomotopy}}]\label{th:transfer}
Let $\EE$ be a complete and cocomplete category, 
$\DD$ a cofibrantly generated model category, 
whose objects are all fibrant, and $F \dashv U: \EE \to \DD$ 
an adjunction. Define a morphism $f: E \to E^\prime$ in $\EE$ 
to be a weak equivalence (fibration) 
if $U(f): U(E) \to U(E^\prime)$ in $\DD$ is a weak equivalence 
(respectively fibration). 
Then this determines a cofibrantly generated 
model structure on $\EE$ under the following hypotheses:
\begin{enumerate}[label=(\roman*)]
\item $F$ preserves small objects, 
\item $\EE$ has a functorial path object $(P,w,f)$.
\end{enumerate}
\end{theo}

\begin{rem}\label{rem:generating-sets}
Even though this is not stated explicitly 
in Theorem \ref{th:transfer}, it is straightforward 
to realize that, given a set $I$ ($J$) of generating 
cofibrations (respectively acyclic cofibrations) for $\DD$, $F(I)$ (respectively 
$F(J)$) is a set of generating cofibrations 
(respectively acyclic cofibrations) for $\EE$. 
Indeed, by the adjunction $F \dashv U$, the right 
lifting property (see e.g.\ \cite[Def.\ 1.1.2]{Hovey_1999_ModelCategories})
of a morphism $f: E \to E^\prime$ in $\EE$ 
against any morphism of $F(I)$ (respectively $F(J)$) 
is equivalent to the right lifting property 
of the morphism $U(f): U(E) \to U(E^\prime)$ in $\DD$ 
against any morphism of $I$ (respectively $J$). 
Since weak equivalences and fibrations in $\EE$ are 
by definition detected by $U$, it follows that $F(I)$ 
(respectively $F(J)$) is a set of generating cofibrations 
(respectively acyclic cofibrations) for $\EE$. 
\end{rem}

\begin{setup}\label{setup}
We assume $\MM$ to be a cofibrantly generated closed 
symmetric monoidal model category with cofibrant unit $\oone$, 
see \cite[Sec.\ 4.2]{Hovey_1999_ModelCategories}, 
that satisfies the monoid axiom, 
see \cite[Def.\ 3.3]{SchwedeShipley_2000_AlgebrasModules}. 
In order to meet the hypotheses of Theorem \ref{th:transfer}, 
we further assume that all objects of $\MM$ are fibrant 
and the existence 
of an {\it interval object} $(I,r,b)$ in $\MM$, 
that consists of a cofibrant object $I \in \MM$ equipped 
with a factorization of the codiagonal morphism 
$\langle \id_{\oone},\id_{\oone} \rangle = r\, b: \oone \sqcup \oone \to \oone$ in $\MM$ 
into a cofibration $b: \oone \sqcup \oone \to I$ in $\MM$ 
followed by a weak equivalence $r: I \to \oone$ in $\MM$. 
\end{setup}

Let us mention that the category $\MM = \Ch_\bbK$ 
of cochain complexes over a field $\bbK$, 
which features in our applications in Section \ref{sec:Maxwell}, 
meets all the requirements of Set-up \ref{setup}. For more 
details see the beginning of Section \ref{sec:Maxwell}. 
\sk

For a given net of algebras $\AAA \in \Net_\CC^\MM$ and 
using both Set-up \ref{setup} and Theorem \ref{th:transfer}, 
we shall now go through a two-step procedure to transfer 
a model structure on the category 
$\EE \coloneqq \RepA$ of $\AAA$-representations from the 
product model category $\DD \coloneqq \prod_{c \in \CC} \MM$. 
The latter is the model category 
whose weak equivalences, fibrations and cofibrations 
are defined component-wise, 
see \cite[Ex.\ 1.1.6]{Hovey_1999_ModelCategories}. 
Since per hypothesis $\MM$ is cofibrantly generated 
and its objects are all fibrant, 
the same holds true for the product model category $\prod_{c \in \CC} \MM$. 

\paragraph{Step 1}
The first step presents the adjunction that will be used 
to transfer the model structure from $\prod_{c \in \CC} \MM$. 
Consider the forgetful functor 
$U: \RepA \to \prod_{c \in \CC} \MM$ 
that sends an $\AAA$-representation $\LLL \in \RepA$ 
to its underlying collection 
$(\LLL_c)_c \in \prod_{c \in \CC} \MM$, where $\LLL_c \in \MM$ denotes here the object of $\MM$ underlying the left $\AAA(c)$-module $\LLL_c \in {_{\AAA(c)}\Mod}$, for all $c \in \CC$. 
We emphasize that $U$ preserves and lifts 
both limits and colimits because 
those can be computed component-wise 
in the product category $\prod_{c \in \CC} \MM$ 
by endowing the resulting collection 
with the induced structure 
of an $\AAA$-representation. 
This shows that $\RepA$ is both complete and cocomplete. 
Furthermore, $U$ is part of the adjunction 
\begin{equation}\label{eq:free-forget-adj}
\xymatrix{
F: \displaystyle\prod\limits_{c \in \CC} \MM \ar@<2pt>[r] & \RepA :U, \ar@<2pt>[l]
}
\end{equation}
whose left adjoint $F$ is defined below. 
$F$ sends a collection $\und{V} \coloneqq (V_c)_c \in \prod_{c \in \CC} M$ 
to the $\AAA$-representation $F(\und{V}) \in \RepA$ 
consisting of the following data, see Definition \ref{def:RepA}: 
\begin{subequations}
\begin{enumerate}[label=(\arabic*)]
\item For each $c \in \CC$, the free left $\AAA(c)$-module 
\begin{equation}
(F(\und{V}))_c \coloneqq \AAA(c) \otimes \coprod_{\substack{\widetilde{c} \in \CC\\ \widetilde{\gamma} \in \CC(\widetilde{c},c)}} V_{\widetilde{c}} \in {_{\AAA(c)}\Mod}. 
\end{equation}
\item For each morphism $\gamma: c_1 \to c_2$ in $\CC$, 
the morphism of left $\AAA(c_2)$-modules 
\begin{equation}
\xymatrix@C=3em{
(F(\und{V}))_{\gamma}: \AAA(c_2) \otimes_{\AAA(c_1)} (F(\und{V}))_{c_1} \cong \AAA(c_2) \otimes \displaystyle\coprod_{\substack{\widetilde{c}_1 \in \CC\\ \widetilde{\gamma} \in \CC(\widetilde{c}_1,c_1)}} V_{\widetilde{c}_1} \ar[r]^-{\id \otimes \gamma_\ast} & (F(\und{V}))_{c_2}
}
\end{equation}
in ${_{\AAA(c_2)}\Mod}$, where 
$\gamma_\ast: \coprod_{\widetilde{c_1},\widetilde{\gamma}_1} V_{\widetilde{c_1}} \to \coprod_{\widetilde{c_2},\widetilde{\gamma}_2} V_{\widetilde{c_2}}$ in $\MM$ 
denotes the morphism that sends the 
$(\widetilde{c}_1,\widetilde{\gamma}_1)$-component of the domain 
to the $(\widetilde{c}_1,\gamma \widetilde{\gamma}_1)$-component of the codomain. 
\end{enumerate}
One easily checks that these data fulfil the axioms 
for an $\AAA$-representation. Furthermore, $F$ sends a morphism 
$\und{\xi} \coloneqq (\xi_c)_c: \und{V} \to \und{V}^\prime$ 
in $\prod_{c \in \CC} M$ to the morphism 
of $\AAA$-representations 
$F(\und{\xi}): F(\und{V}) \to F(\und{V}^\prime)$ in $\RepA$ 
consisting of the morphism of left $\AAA(c)$-modules 
\begin{equation}
\xymatrix{
(F(\und{\xi}))_c \coloneqq \id \otimes \displaystyle\coprod_{\substack{\widetilde{c} \in \CC\\ \widetilde{\gamma} \in \CC(\widetilde{c},c)}} \xi_{\widetilde{c}}: (F(\und{V}))_c \ar[r]^-{} & (F(\und{V}^\prime))_c
}
\end{equation}
\end{subequations}
in ${_{\AAA(c)}\Mod}$, for each $c \in \CC$. 
It is easy to check that these data 
indeed define a morphism of a $\AAA$-representations 
and that $F$ as defined above is a functor.
\sk

In order to show that \eqref{eq:free-forget-adj} 
is an adjunction as claimed, we describe its unit $\eta$ 
and its counit $\varepsilon$. 
The unit is the natural transformation 
\begin{subequations}
\begin{equation}
\xymatrix{\eta: \id_{\prod_{c \in \CC} \MM} \ar[r] & UF,}
\end{equation}
whose component $\eta_{\und{V}}: \und{V} \to UF(\und{V})$ 
in $\prod_{c \in \CC} \MM$ at the collection 
$\und{V} \in \prod_{c \in \CC} \MM$ 
consists of the morphisms in $\MM$ defined by 
\begin{equation}
\xymatrix@C=3em{
(\eta_{\und{V}})_c: V_c \cong \oone \otimes V_c \ar[r]^-{\bfone \otimes \iota_{c,\id_c}} & (UF(\und{V}))_c,
}
\end{equation}
\end{subequations}
for all $c \in \CC$, where 
$\iota_{\widetilde{c},\widetilde{\gamma}}: V_{\widetilde{c}} \to \coprod_{\widetilde{c}^\prime,\widetilde{\gamma}^\prime} V_{\widetilde{c}^\prime}$ 
in $\MM$ denotes the canonical morphism to the coproduct. 
The counit is the natural transformation 
\begin{subequations}
\begin{equation}
\xymatrix{\varepsilon: FU \ar[r] & \id_{\RepA},} 
\end{equation}
whose component $\varepsilon_\LLL: FU(\LLL) \to \LLL$ 
in $\RepA$ at the $\AAA$-representation $\LLL \in \RepA$ 
consists, for each $c \in \CC$, 
of the morphism of left $\AAA(c)$-modules 
$(\varepsilon_\LLL)_c: (FU(\LLL))_c \to \LLL_c$ in ${_{\AAA(c)}}\Mod$ 
defined via the universal property of the coproduct by the diagram 
\begin{equation}
\xymatrix@R=3em{
(FU(\LLL))_c \ar[r]^-{(\varepsilon_\LLL)_c} & \LLL_c \\ 
\AAA(c) \otimes \LLL_{\widetilde{c}} \ar[r] \ar[u]^-{\id \otimes \iota_{\widetilde{c},\widetilde{\gamma}}} & \AAA(c) \underset{\AAA(\widetilde{c})}{\otimes} \LLL_{\widetilde{c}} \ar[u]_-{\LLL_{\widetilde{\gamma}}}
}
\end{equation}
\end{subequations}
in ${_{\AAA(c)}\Mod}$, for all $\widetilde{c} \in \CC$ 
and $\widetilde{\gamma} \in \CC(\widetilde{c},c)$. 
The bottom horizontal arrow denotes the canonical morphism 
to the coequalizer 
$\AAA(c) \otimes_{\AAA(\widetilde{c})} \LLL_{\widetilde{c}} \in {_{\AAA(c)}\Mod}$, 
see \eqref{eq:relative-tensor}. 
The verification of the triangle identities 
$U(\varepsilon)\, \eta_U = \id_U$ and 
$\varepsilon_F\, F(\eta) = \id_F$ is straightforward, 
thus proving that \eqref{eq:free-forget-adj} 
is an adjunction, as claimed. 
\sk

\begin{rem}\label{rem:transfer-vs-powering}
We observe incidentally that the right adjoint functor $U$ 
preserves the $\MM$-powerings on $\RepA$ and 
on $\prod_{c \in \CC} \MM$ respectively. The latter is obtained by acting 
component-wise with the internal hom of $\MM$. 
It follows then by definition of the $\MM$-powering functor on $\RepA$, 
see \eqref{eq:Rep-powering}, that 
for each $V \in \MM$, the diagram of functors 
\begin{equation}
\xymatrix@C=4em{
\RepA \ar[r]^-{U} \ar[d]_-{(-)^V} & {\displaystyle \prod_{c \in \CC} \MM} \ar[d]^-{(-)^V} \\ 
\RepA \ar[r]_-{U} & {\displaystyle \prod_{c \in \CC} \MM}
}
\end{equation}
commutes. As a consequence, the diagram of left adjoint functors 
commutes up to a unique natural isomorphism, 
i.e.\ there exists a unique natural isomorphism 
$F(-) \otimes V \cong F((-) \otimes V)$ between functors 
from $\prod_{c \in \CC} \MM$ to $\RepA$, 
where the left hand side displays the $\MM$-tensoring 
on $\RepA$ from \eqref{eq:Rep-tensoring} 
and the right hand side displays the $\MM$-tensoring 
on $\prod_{c \in \CC} \MM$ 
defined component-wise in $\CC$ by the tensor product of $\MM$. 
\end{rem}

\begin{defi}\label{def:Rep-we-fib}
A morphism $\FFF: \LLL \to \LLL^\prime$ in $\RepA$ 
is a {\it weak equivalence} ({\it fibration}) if 
$U(\FFF): U(\LLL) \to U(\LLL^\prime)$ in $\prod_{c \in \CC} \MM$ 
is a weak equivalence (respectively fibration), 
i.e.\ if $\FFF_c: \LLL_c \to \LLL^\prime_c$ 
is a weak equivalence (respectively fibration) 
in the underlying model category $\MM$, for all $c \in \CC$. 
\end{defi}

Note that all objects of $\RepA$ are fibrant because, 
as previously explained, all objects of 
$\prod_{c \in \CC} \MM$ are fibrant. 

\paragraph{Step 2}
The second step checks that the hypotheses 
of Theorem \ref{th:transfer} are met. 
Hypothesis~(i), i.e.\ that $F$ preserves small objects, 
follows by recalling that, as explained in Step 1, 
$U$ preserves colimits and $F$ is left adjoint to $U$. 
We show that also hypothesis~(ii) 
of Theorem \ref{th:transfer} is met, 
i.e.\ we construct a functorial path object $(P,w,f)$ in $\RepA$. 
This is achieved by a standard construction 
that uses the interval object $(I,r,b)$ in $\MM$, 
see Set-up \ref{setup}, 
the closed symmetric monoidal model structure on $\MM$, 
the hypothesis that all objects of $\MM$ are fibrant 
and the $\MM$-powering on $\RepA$ from \eqref{eq:Rep-powering}. 
Explicitly, we consider the functor 
\begin{subequations}\label{eq:P-w-f}
\begin{equation}
\xymatrix{P \coloneqq (-)^I: \RepA \ar[r] & \RepA}
\end{equation}
and the natural transformations 
\begin{align}
\xymatrix{w: \id_\RepA \cong (-)^{\oone} \ar[r]^-{(-)^r} & P,} && \xymatrix{f: P \ar[r]^-{(-)^b} & (-)^{\oone \sqcup \oone} \cong (-)^{\times2}.}
\end{align}
\end{subequations}
By definition of the $\MM$-powering \eqref{eq:Rep-powering}, 
for each $\LLL \in \RepA$ and up to the evident isomorphisms, 
$U(w_\LLL): U(\LLL) \to U(\LLL^I)$ consists of the components 
$[r,\LLL_c]: [\oone,\LLL_c] \to [I,\LLL_c]$ in $\MM$, for all $c \in \CC$, and 
$U(f_\LLL): U(\LLL^I) \to U(\LLL \times \LLL)$ consists of the components 
$[b,\LLL_c]: [I,\LLL_c] \to [\oone \sqcup \oone,\LLL_c]$ 
in $\MM$, for all $c \in \CC$. 
(Here $[-,-]: \MM^\op \times \MM \to \MM$ denotes 
the internal hom of the closed symmetric monoidal 
model category $\MM$.) Since by Set-up \ref{setup} 
all objects $V \in \MM$ are fibrant, 
$[-,V]: \MM^\op \to \MM$ sends weak equivalences 
between cofibrant objects in $\MM$ to weak equivalences 
in $\MM$ and cofibrations in $\MM$ to fibrations in $\MM$, 
see \cite[Sec.\ 4.2]{Hovey_1999_ModelCategories}. 
Since $r: I \to \oone$ in $\MM$ is a weak equivalence 
between cofibrant objects and 
$b: \oone \sqcup \oone \to I$ in $\MM$ is a cofibration, 
recalling Definition \ref{def:Rep-we-fib} 
we conclude that $w_\LLL: \LLL \to \LLL^I$ in $\RepA$ 
is a weak equivalence and 
$f_\LLL: \LLL^I \to \LLL \times \LLL$ in $\RepA$ 
is a fibration. Furthermore, 
since by Set-up \ref{setup} the codiagonal morphism factors as 
$\langle \id_\oone,\id_\oone \rangle = r\, b : \oone \sqcup \oone \to \oone$ 
in $\MM$, it follows that the diagonal morphism factors as 
$(\id_\LLL,\id_\LLL) = f_\LLL\, w_\LLL: \LLL \to \LLL \times \LLL$ 
in $\RepA$, for all $\LLL \in \RepA$. This shows that 
the functor $P$ and the natural transformations $w$ and $f$ 
introduced  in \eqref{eq:P-w-f} define 
a functorial path object $(P,w,f)$ in $\RepA$, hence 
also hypothesis~(ii) of Theorem \ref{th:transfer} is met. 
The next corollary summarizes the conclusions so far, 
taking into account also Remark \ref{rem:generating-sets}. 

\begin{cor}\label{cor:Rep-model-str}
Under the assumptions stated in Set-up \ref{setup}, 
the notions of weak equivalences and fibrations 
from Definition \ref{def:Rep-we-fib} determine a
cofibrantly generated model structure 
on the category $\RepA$ of $\AAA$-representations. 
Furthermore, given a set of generating (acyclic) cofibrations $I$ 
for $\MM$, $F(\prod_{c \in \CC} I)$ 
is a set of generating (respectively acyclic) cofibrations for $\RepA$. 
\end{cor}

\begin{rem}\label{rem:coprod}
When the source category $\CC$ admits finite coproducts, 
the same model structures on $\Net_{\CC}^{\MM}$ and $\RepA$ 
can be more concisely obtained by applying 
Appendix \ref{subsec:model-str} to the functor category $\Fun(\CC,\MM)$. 
Indeed, as explained below, when $\CC$ admits finite coproducts, 
$\Fun(\CC,\MM)$ meets the requirements of Appendix \ref{subsec:model-str}. 
\sk 

Assuming, as in Set-up \ref{setup:model-str}, that $\MM$ is 
a cofibrantly generated closed symmetric monoidal model category 
that satisfies the monoid axiom and whose objects are all small, 
one gets the standard cofibrantly generated projective model structure 
on $\Fun(\CC,\MM)$. Since all objects of $\MM$ are small, 
all objects of $\Fun(\CC,\MM)$ are small too. Furthermore, 
if the object-wise symmetric monoidal structure 
from Section \ref{subsec:AQFT} and the projective model structure 
are suitably compatible, $\Fun(\CC,\MM)$ becomes a closed symmetric 
monoidal model category that inherits the monoid axiom from $\MM$. 
It only remains to check that 
the object-wise symmetric monoidal structure 
and the projective model structure interact nicely, 
i.e.\ that the pushout-product axiom holds. 
The latter follows from the assumption 
that the source category $\CC$ admits finite coproducts,
see (the arXiv version of) the proof of 
\cite[Prop.\ 7.9]{PavlovScholbach_2018_AdmissibilityRectification}.\footnote{For 
completeness, we recall here the core of this argument. 
Since the projective model structure on $\Fun(\CC,\MM)$ 
is cofibrantly generated, it suffices to check the pushout-product axiom 
on the generating (acyclic) cofibrations $\coprod_{\gamma \in \CC(c,-)} i$, 
where $c \in \CC$ is any object and $i$ is any generating (acyclic) 
cofibration of $\MM$. Since colimits in $\MM$ commute with each other and with the tensor product, 
the existence of finite coproducts in $\CC$ entails
that the pushout-product of $\coprod_{\gamma_1 \in \CC(c_1,-)} i_1$ and 
$\coprod_{\gamma_2 \in \CC(c_2,-)} i_2$ in $\Fun(\CC,\MM)$ is given by 
$\coprod_{\gamma \in \CC(c_1 \coprod c_2,-)} i_1 \square i_2$, where $\square$ 
denotes the pushout-product in $\MM$. (Finite coproducts in $\CC$ 
are responsible for the natural isomorphisms 
$\CC(c_1,-) \times \CC(c_2,-) \cong \CC(c_1 \coprod c_2,-)$.) 
This shows that the pushout-product axiom of $\MM$ entails 
that of $\Fun(\CC,\MM)$.} 
To the best of our knowledge this is the only proof of 
the pushout-product axiom in this setting. 
\sk

Unfortunately, finite coproducts do not exist in all the source 
categories of interest in the context of algebraic quantum field theory, 
see Example \ref{ex:Loc}. 
For instance, recalling the category $\Loc_m$ of globally hyperbolic 
Lorentzian manifolds, 
one easily realizes that binary coproducts often fail to exist. 
On the other hand, finite coproducts exist in the source category 
$\CC = \CCO(M)$ of causally convex open subsets 
of a fixed $M \in \Loc_m$, 
where they are given by the {\it causally convex hull} 
$U_1 \coprod U_2 := I^+_M(U_1 \cup U_2) \cap I^-_M(U_1 \cup U_2)$ 
of $U_1, U_2 \in \CCO(M)$. 
(Here $I^\pm_M$ denotes the chronological future/past 
of a subset in $M$.) 
\end{rem}

\begin{rem}\label{rem:eval-const-adj-vs-model-str}
Combining the above model structure on $\RepA$ 
with the model structure on categories of left modules 
from Proposition \ref{propo:model-str}, one easily recognizes that, 
for each $\widetilde{c} \in \CC$, the adjunction 
$(-)_{\widetilde{c}} \dashv(-)^{\widetilde{c}}: {_{\AAA(\widetilde{c})}}\Mod \to \RepA$ 
from \eqref{eq:eval-const-adj} is a Quillen adjunction, 
see \cite[Sec.\ 1.3.1]{Hovey_1999_ModelCategories}. 
This follows from the fact that the right adjoint functor 
$(-)^{\widetilde{c}}$ sends (acyclic) fibrations 
in ${_{\AAA(\widetilde{c})}}\Mod$, 
which are detected in $\MM$, 
to (acyclic) fibrations in $\RepA$, 
which are detected  in $\MM$ component-wise in $\CC$. 
Note that the Quillen adjunction \eqref{eq:eval-const-adj} 
is compatible with the $\MM$-tensoring, powering and enriched hom 
due to Remark \ref{rem:eval-const-vs-tensoring-powering}. 
(In the language of \cite[Def.\ 4.2.18]{Hovey_1999_ModelCategories} 
the left adjoint $(-)_{\widetilde{c}}$ is an $\MM$-Quillen functor.)
\end{rem}

With Corollary \ref{cor:Rep-model-str}, 
we have established a model structure on $\RepA$. 
We now investigate its compatibility with 
the $\MM$-tensoring, powering and enriched hom on $\RepA$ from 
Section \ref{subsec:Rep-tensoring}, 
i.e.\ we check whether $\RepA$ is an $\MM$-model category 
in the sense of \cite[Def.\ 4.2.18]{Hovey_1999_ModelCategories}. 
This amounts to showing that the $\MM$-tensoring 
\eqref{eq:Rep-tensoring} is a Quillen bifunctor, 
see \cite[Def.\ 4.2.1]{Hovey_1999_ModelCategories}. 
On account of \cite[Cor.\ 4.2.5]{Hovey_1999_ModelCategories} 
and Corollary \ref{cor:Rep-model-str}, this is equivalent 
to proving that the pushout-product morphism 
$F(\und{\eta})\, \square\, \xi$ is a cofibration
(acyclic cofibration) when 
$F(\und{\eta}) \in F(\prod_{c \in \CC} I)$ 
and $\xi \in I$ are both generating cofibrations 
(respectively $F(\und{\eta}) \in F(\prod_{c \in \CC} J)$ 
is a generating acyclic cofibration and $\xi \in I$ 
is a generating cofibration or 
$F(\und{\eta}) \in F(\prod_{c \in \CC} I)$ 
is a generating cofibration and $\xi \in J$ 
is a generating acyclic cofibration). 
Recalling that $F$ preserves the $\MM$-tensorings 
on $\prod_{c \in \CC} \MM$ and on $\RepA$, 
see Remark \ref{rem:transfer-vs-powering}, 
and mimicking the construction of the pushout-product morphism 
in $\AMod$ from \eqref{eq:pushout-product}, 
one finds that the pushout-product morphism 
$F(\und{\eta})\, \square\, \xi$ constructed in $\RepA$ coincides 
(up to the evident isomorphisms) with the image 
$F(\und{\eta}\, \square\, \xi)$ under $F$ of the 
pushout-product morphism $\und{\eta}\, \square\, \xi$ 
constructed in $\prod_{c \in \CC} \MM$. 
(Here we used also that $F$ preserves colimits 
because it is a left-adjoint functor.) Since $\MM$ is a closed 
symmetric monoidal model category 
and the pushout-product in $\prod_{c \in \CC} \MM$ 
is computed component-wise, 
$\und{\eta}\, \square\, \xi$ in $\prod_{c \in \CC} \MM$ 
is a cofibration (respectively acyclic cofibration). 
Since weak equivalences and fibrations in $\RepA$ 
are by definition detected by $U$, whose left adjoint is $F$, 
it follows that $F(\und{\eta}\, \square\, \xi)$ in $\RepA$ 
has the left lifting property against all acyclic fibrations 
(respectively fibrations) in $\RepA$. 
This means that $F(\und{\eta}\, \square\, \xi)$, 
and hence also $F(\und{\eta})\, \square\, \xi$, in $\RepA$ 
is a cofibration (respectively acyclic cofibration). 
The conclusions of this paragraph are summarized below. 

\begin{propo}[$\RepA$ as an $\MM$-model category]
Under the assumptions stated in Set-up \ref{setup}, 
the model structure from Corollary \ref{cor:Rep-model-str} 
and the $\MM$-tensoring from \eqref{eq:Rep-tensoring} 
endow the category $\RepA$ of 
$\AAA$-representations with an $\MM$-model structure. 
(For the concept of $\MM$-model category, refer e.g.\ 
to \cite[Def.\ 4.2.18]{Hovey_1999_ModelCategories}.) 
\end{propo}

We complete the analysis of the model structures 
on the categories of net representations 
from Corollary \ref{cor:Rep-model-str} 
discussing how they interact with the change-of-net 
adjunction 
$\Ext_\Phi \dashv \Res_\Phi: \RepB \to \RepA$ 
from \eqref{eq:change-net-adj} associated with 
a morphism of nets $\Phi: \AAA \to \BBB$ 
in $\Net_\CC^\MM$. Since weak equivalences and 
fibrations both in $\RepA$ and in $\RepB$ 
are by definition detected at the level of 
the underlying collections in $\prod_{c \in \CC} \MM$, 
it follows immediately that $\Res_\Phi$ preserves 
both weak equivalences and fibrations, which entails 
that the change-of-net adjunction 
\eqref{eq:change-net-adj} is a Quillen adjunction. 
\sk

Let us now also suppose that $\Phi: \AAA \to \BBB$ 
in $\Net_\CC^\MM$ is a weak equivalence, see Definition \ref{def:Net-model-str}. 
Our aim is to show that in this case the change-of-net adjunction 
$\Ext_\Phi \dashv \Res_\Phi: \RepB \to \RepA$ 
from \eqref{eq:change-net-adj} is a Quillen equivalence, 
see \cite[Sec.\ 1.3.3]{Hovey_1999_ModelCategories}. 
By definition of the model structures 
on $\RepA$ and $\RepB$, $\Res_\Phi$ detects 
weak equivalences and, as a consequence of Set-up 
\ref{setup}, all net representations are fibrant. 
Therefore, by \cite[Cor.\ 1.3.16]{Hovey_1999_ModelCategories}, 
in order to conclude that the change-of-net 
adjunction $\Ext_\Phi \dashv \Res_\Phi$ is a Quillen equivalence, 
it suffices to show that the components of its unit 
$\LLL \to \Res_\Phi\, \Ext_\Phi\, \LLL$ in $\RepA$ 
are weak equivalences for all cofibrant 
$\AAA$-representations $\LLL \in \RepA$. 
For this purpose we need to show that, for each $c \in \CC$, 
the morphism $\LLL_c \to (\Res_\Phi\, \Ext_\Phi\, \LLL)_c$ 
in ${_{\AAA(c)}}\Mod$ is a weak equivalence. 
(Refer to Section \ref{subsec:model-str} 
for the model structure on ${_{\AAA(c)}}\Mod$.) 
Recalling the construction of the change-of-net adjunction 
\eqref{eq:change-net-adj}, the latter morphism is just the component 
$\LLL_c \to (\BBB(c) \otimes_{\AAA(c)} \LLL_c)\vert_{\AAA(c)}$ 
in ${_{\AAA(c)}}\Mod$ of the unit of the change-of-monoid adjunction 
\eqref{eq:change-monoid-adj} associated with 
$\Phi_c: \AAA(c) \to \BBB(c)$ in $\Mon(\MM)$. Since $\LLL \in \RepA$ 
is cofibrant by hypothesis and 
$(-)_{c} \dashv (-)^{c}: {_{\AAA(c)}}\Mod \to \RepA$ 
is a Quillen adjunction, 
see Remark \ref{rem:eval-const-adj-vs-model-str}, 
$\LLL_c \in {_{\AAA(c)}}\Mod$ is cofibrant too. 
Therefore, we would be able to conclude 
if the change-of-monoid adjunction 
associated with $\Phi_c$ 
were a Quillen equivalence, 
for all $c \in \CC$. This result is achieved by 
Proposition \ref{propo:change-monoid-vs-model-str}, 
under the additional assumption that, 
for each monoid $A \in \Mon(\MM)$ and each cofibrant left 
$A$-module $L$, the functor $(-) \otimes_A L: \Mod_A \to \MM$ 
sends weak equivalences in $\Mod_A$ to weak equivalences in $\MM$. 
(Here $\otimes_A: \Mod_A \times \AMod \to \MM$ 
denotes the familiar relative tensor product over $A$ 
between right and left $A$-modules. Note that 
the model structure on $\Mod_A$ 
is completely analogous to the one on $\AMod$ 
from Proposition \ref{propo:model-str}.) 
We emphasize that this additional assumption 
holds true in many examples, 
see \cite[Secs.\ 4 and 5]{SchwedeShipley_2000_AlgebrasModules},  
including the closed symmetric monoidal model 
category of cochain complexes $\MM = \Ch_\bbK$ 
over a field $\bbK$, which is briefly recalled 
at the beginning of Section \ref{sec:Maxwell}. 
We summarize below the conclusions 
of the last two paragraphs. 

\begin{propo}[Change-of-net as a Quillen adjunction]\label{propo:change-net-vs-model-str}
Under the assumptions stated in Set-up \ref{setup}, 
let $\Phi: \AAA \to \BBB$ in $\Net_\CC^\MM$ 
be a morphism of nets. 
Then the change-of-net adjunction 
$\Ext_\Phi \dashv \Res_\Phi: \RepB \to \RepA$ 
from \eqref{eq:change-net-adj} is a Quillen adjunction, 
which is compatible with the $\MM$-tensoring, powering and enriched hom, 
see Remark \ref{rem:change-net-vs-tensoring-powering}. 
(In the language of \cite[Def.\ 4.2.18]{Hovey_1999_ModelCategories} 
the extension functor $\Ext_{\Phi}$ is an $\MM$-Quillen functor.) 
Furthermore, $\Ext_\Phi \dashv \Res_\Phi$ is also a Quillen equivalence 
if $\Phi$ is a weak equivalence and the following additional hypothesis n holds: 
for each monoid $A \in \Mon(\MM)$ 
and each cofibrant left $A$-module $L \in \AMod$, 
the relative tensor product $(-) \otimes_A L: \Mod_A \to \MM$ 
sends weak equivalences in $\Mod_A$ to weak equivalences in $\MM$. 
\end{propo}

\begin{ex}\label{ex:KG-fixed}
As an immediate consequence of Proposition \ref{propo:change-net-vs-model-str}, 
the weak equivalence $\Phi: \widetilde{\AAA} \to \AAA$ 
in $\Net_{\Loc_m}^{\Ch_\bbC}$ from Example \ref{ex:KG} 
between the two nets of algebras describing the Klein-Gordon field 
gives rise to a Quillen equivalence 
$\Ext_\Phi \dashv \Res_\Phi: \RepA \to \Rep(\widetilde{\AAA})$ 
between the associated model categories of net representations. 
In other words, we solved the shortcoming evidenced in Example \ref{ex:KG} 
by endowing the categories of net representations with suitable 
model structures and by formalizing the notion of ``being the same'' 
by the concept of Quillen equivalence. Let us stress once more that 
ordinary categorical equivalence is recovered by passing 
to the associated homotopy categories, i.e.\ by inverting 
all weak equivalences between net representations. 
\end{ex}


\section{\label{sec:Maxwell}Net representations for Maxwell \texorpdfstring{$p$}{p}-forms}
Section \ref{sec:Rep} developed the homotopy theory of net representations 
with values in a closed symmetric monoidal model category 
$\MM$. We now move on to the problem of 
constructing explicit examples of such net representations. To address this question 
we take $\MM = \Ch_\bbC$ to be the closed symmetric monoidal 
model category of cochain complexes, which is the relevant 
one for applications in the context of the BV formalism 
and of homotopy algebraic quantum field theory, 
see Section \ref{sec:intro}. 
As an instructive example we shall consider the net 
of algebras associated with Maxwell $p$-forms 
\cite{HenneauxTeitelboim_1986_FormElectrodynamics, HenneauxTeitelboim_1992_QuantizationGauge}, 
which we construct and study in detail 
through Sections \ref{subsec:Complexes}, 
\ref{subsec:ret-adv-homotopies} and \ref{subsec:Poisson}. 
The actual construction of an explicit net representation 
for Maxwell $p$-forms is presented in Section 
\ref{subsec:rep-construction} and goes through 
the construction of a two-point function $\omega_2$, 
which for our purposes consists of a cochain map: 
it is the preservation of differentials 
that encodes the compatibility both with the equation of motion 
and with the action of gauge transformations. 
Note that Maxwell $1$-forms recover linear Yang-Mills theory, 
i.e.\ the electromagnetic vector potential, 
and in this case our $\omega_2$ extends (in a sense 
made precise by Remark \ref{rem:omega_2-microlocal}) 
a Hadamard two-point function 
constructed in \cite{FewsterPfenning_2003_QuantumWeak} 
for the gauge invariant on-shell linear observables 
of the electromagnetic vector potential. 
\sk

Before delving into the main subject of this section, 
let us recall some facts about the closed symmetric monoidal 
model category $\Ch_\bbK$ of cochain complexes over a 
field $\bbK$ (either $\bbK=\bbR$ or $\bbK=\bbC$ for our purposes). 
A cochain complex $V \in \Ch_\bbK$ consists of vector spaces 
$V^n$, for all $n \in \bbZ$, and of a differential $\dd_V$, 
i.e.\ a collection of linear maps $\dd_V^n: V^n \to V^{n+1}$, 
for all $n \in \bbZ$, such that $\dd_V^{n+1}\, \dd_V^n = 0$. 
A cochain map $f: V_1 \to V_2$ in $\Ch_\bbC$ 
consists of linear maps $f^n: V_1^n \to V_2^n$, 
for all $n \in \bbZ$, that preserve the differentials 
$f^{n+1}\, \dd_{V_1}^n = \dd_{V_2}^n\, f^n$. 
Given $V \in \Ch_\bbK$, $V[k] \in \Ch_\bbK$ 
denotes the $k$-shifted cochain complex defined 
by $V[k]^n \coloneqq V^{n+k}$, for all $n \in \bbZ$, 
with differential $\dd_{V[k]} \coloneqq (-1)^k\, \dd_{V}$. 
The weak equivalences are the quasi-isomorphisms, i.e.\ 
the cochain maps inducing isomorphisms in cohomology, 
the fibrations are the degree-wise surjective cochain maps 
and the cofibrations are the degree-wise injective cochain maps. 
This endows $\Ch_\bbK$ with a cofibrantly generated 
model structure 
such that all objects are both fibrant and cofibrant, 
see \cite[Sec.\ 2.3]{Hovey_1999_ModelCategories} 
and also \cite[Rem.\ 1.8]{BarthelMayRiehl_2014_SixModel}. 
The monoidal structure on $\Ch_\bbK$ is given 
by the tensor product $V \otimes W \in \Ch_\bbK$ 
of the cochain complexes $V, W \in \Ch_\bbK$, 
which is defined degree-wise for all $n \in \bbZ$ by 
\begin{equation}
(V \otimes W)^n \coloneqq \bigoplus_{k \in \bbZ} V^k \otimes W^{n-k},
\end{equation}
with differential given by the graded Leibniz rule 
$\dd_\otimes := \dd_V \otimes \id_W + \id_V \otimes \dd_W$. 
The monoidal unit $\bbK \in \Ch_\bbK$ is the ground 
field regarded as a cochain complex 
concentrated in degree $0$ and the symmetric braiding 
$V \otimes W \cong W \otimes V$ 
on $\Ch_\bbK$ is defined by the Koszul sign rule $v \otimes w \mapsto (-1)^{|v| |w|} w \otimes v$. 
The internal hom $[V,W] \in \Ch_\bbK$ is defined 
degree-wise for all $n \in \bbZ$ by 
\begin{equation}
[V,W]^n \coloneqq \prod_{k \in \bbZ} \operatorname{Lin}_\bbK(V^k,W^{n+k}),
\end{equation}
where $\operatorname{Lin}_\bbK(-,-)$ 
denotes the vector space of linear maps, with differential
$\partial \coloneqq [\dd,-]$ given 
by the graded commutator with the original differentials. 
This endows $\Ch_\bbK$ with a closed symmetric monoidal 
model structure 
\cite[Prop.\ 4.2.13]{Hovey_1999_ModelCategories}, 
which fulfils the assumptions listed in Set-up \ref{setup}, 
see \cite[Rem.\ 3.4 and Sec.\ 5]{SchwedeShipley_2000_AlgebrasModules}. 
For instance, as interval object one can take 
the cochain complex $I \in \Ch_\bbK$ 
that consists of $\bbK$ in degree $-1$ 
and $\bbK \oplus \bbK$ in degree $0$, 
with differential defined by $\dd(1) = 1 \oplus (-1)$. 
Furthermore, the additional hypothesis of Proposition 
\ref{propo:change-net-vs-model-str} is met, see 
\cite[Secs.\ 4 and 5]{SchwedeShipley_2000_AlgebrasModules}.

\subsection{\label{subsec:Complexes}Cochain complexes of solutions and linear observables}
Let $M$ be an oriented and time-oriented globally hyperbolic 
Lorentzian manifold of dimension $m \geq 2$.
We consider the field complex $\FFF(M) \in \Ch_\bbR$ 
of $p$-forms on $M$, for $p < m$, defined by 
\begin{equation}
\FFF(M) \coloneqq \Bigg( 
\xymatrix{
{\overset{(-p)}{\Omega^0(M)}} \ar[r]^-{\dd} & \cdots \ar[r]^-{\dd} & {\overset{(n)}{\Omega^{p+n}(M)}} \ar[r]^-{\dd} & \cdots \ar[r]^-{\dd} & {\overset{(0)}{\Omega^p(M)}}
} 
\Bigg). 
\end{equation}
(Our convention is that the non-displayed degrees, 
in this case $n < -p$ and $n > 0$, and differentials 
vanish.) In degree $0$ sit the gauge fields 
$A \in \FFF(M)^0 = \Omega^p(M)$, in degree $-1$ sit 
the ghosts $g^{-1} \in \FFF(M)^{-1} = \Omega^{p-1}(M)$ 
(gauge transformations) and, more generally, in lower 
degrees $n < -1$ sit higher ghosts 
$g^{n} \in \FFF(M)^{n} = \Omega^{p+n}(M)$. 
For $p=1$ $\FFF(M)$ is the well-known complex of 
linear Yang-Mills theory, with connection $1$-forms 
sitting in degree $0$ and gauge transformations 
sitting in degree $-1$, whose action on connections 
is determined by the de Rham differential $\dd$. 
The dynamics are encoded by the linear differential operator 
\begin{equation}
\delta \dd: \FFF(M)^0 \rightarrow \FFF(M)^0.
\end{equation}
One can interpret this linear differential operator 
as the equation of motion arising from the variation 
of the (formal) action functional 
\begin{equation}
S(A) \coloneqq \frac{1}{2} \int_M A \wedge \ast\, \delta\dd A,
\end{equation}
where $\ast$ denotes the Hodge star operator defined 
by the metric and the orientation of $M$ 
and $\delta \coloneqq (-1)^k \ast^{-1} \dd\, \ast$ 
denotes the codifferential on $\Omega^k(M)$. 
The action functional $S$ is manifestly gauge-invariant 
since $\delta\dd\, \dd = 0$ and $\delta\dd$ is formally self-adjoint 
with respect to the integral pairing 
$\int (-) \wedge \ast (-)$ displayed above. 
Varying $S$ leads to a section 
$\delta^{\mathrm{v}} S : \FFF(M) \to T^*\FFF(M)$ in $\Ch_\bbR$ 
of the ``cotangent bundle'' $T^*\FFF(M) \in \Ch_\bbR$ 
over $\FFF(M)$, defined as the product 
\begin{equation}
T^*\FFF(M) \coloneqq \FFF(M) \times \FFF_\cc(M)^\ast
\end{equation}
of the cochain complex $\FFF(M) \in \Ch_\bbR$, 
interpreted as the base, and the cochain complex 
\begin{equation}
\xymatrix{
\FFF_\cc(M)^\ast \coloneqq \Bigg(
{\overset{(0)}{\Omega^p(M)}} \ar[r]^-{-\delta} & \cdots \ar[r]^-{-\delta} & {\overset{(n)}{\Omega^{p-n}(M)}} \ar[r]^-{-\delta} & \cdots \ar[r]^-{-\delta} & {\overset{(p)}{\Omega^0(M)}}
}
\Bigg) \in \Ch_\bbR, 
\end{equation}
interpreted as the fiber. 
(The notation $\FFF_\cc(M)^\ast$ is motivated 
by the existence of an evaluation pairing 
$\ev: \FFF_\cc(M)^\ast \otimes \FFF_\cc(M) \to \bbR$ 
in $\Ch_\bbR$ against the cochain complex 
$\FFF_\cc(M) \in \Ch_\bbR$ of compactly 
supported fields, which is defined by 
$\ev(\Phi \otimes \varphi) \coloneqq (-1)^{\lfloor n/2 \rfloor} \int_M \Phi \wedge \ast \varphi$,
for all $n = 0, \ldots, p$, $\Phi \in (\FFF_\cc(M)^\ast)^n$ 
and $\varphi \in \FFF_\cc(M)^{-n}$, 
where $\lfloor - \rfloor$ 
denotes the floor function.) 
The derived critical locus of $S$, 
defined as the homotopy pull-back of 
$\delta^{\mathrm{v}} S: \FFF(M) \to T^\ast \FFF(M)$ 
in $\Ch_\bbR$ along the zero-section of 
$T^*\FFF(M) \in \Ch_\bbR$, determines the 
{\it solution complex} $\Sol(M) \in \Ch_\bbR$. 
Computing the latter explicitly 
as in \cite[Sec.\ 3.4]{BeniniSchenkel_2019_HigherStructures} 
leads to the cochain complex 
\begin{equation}\label{eq:Sol}
\Sol(M) \coloneqq \Bigg(
\xymatrix{
{\overset{(-p)}{\Omega^0(M)}} \ar[r]^-{\dd} & \cdots \ar[r]^-{\dd} & {\overset{(0)}{\Omega^p(M)}} \ar[r]^-{\delta \dd} & {\overset{(1)}{\Omega^p(M)}} \ar[r]^-{\delta} & \cdots \ar[r]^-{\delta} & {\overset{(p+1)}{\Omega^0(M)}}
}
\Bigg) \in \Ch_\bbR.
\end{equation}
The {\it observable complex} $\LLL(M) \in \Ch_\bbR$ 
is defined as the $1$-shifted compactly supported analog 
of $\Sol(M)$. Explicitly, this reads as 
\begin{equation}\label{eq:L}
\LLL(M) \coloneqq \Bigg(
\xymatrix{
{\overset{(-p-1)}{\Omega_\cc^0(M)}} \ar[r]^-{-\dd} & \cdots \ar[r]^-{-\dd} & {\overset{(-1)}{\Omega_\cc^p(M)}} \ar[r]^-{-\delta \dd} & {\overset{(0)}{\Omega_\cc^p(M)}} \ar[r]^-{-\delta} & \cdots \ar[r]^-{-\delta} & {\overset{(p)}{\Omega_\cc^0(M)}}
}
\Bigg) \in \Ch_\bbR.
\end{equation}
The cochains in $\LLL(M)$ are interpreted as linear 
functionals on the solution complex $\Sol(M)$ 
via the evaluation pairing 
\begin{subequations}\label{eq:ev}
\begin{equation}
\ev: \Sol(M) \otimes \LLL(M) \longrightarrow \bbR
\end{equation}
in $\Ch_\bbR$, whose only non vanishing degree 
\begin{equation}
\ev^0: \bigoplus_{k \in \bbZ} \big( \Sol(M)^k \otimes \LLL(M)^{-k} \big) \longrightarrow \bbR
\end{equation}
is defined component-wise for all $k \in \bbZ$ by 
\begin{equation}
\ev^0_k(\Phi \otimes \varphi) \coloneqq (-1)^{\lfloor k/2 \rfloor} \int_M \Phi \wedge \ast \varphi,
\end{equation}
\end{subequations}
for all $\Phi \in \Sol(M)^k$ 
and $\varphi \in \LLL(M)^{-k}$.

\subsection{\label{subsec:ret-adv-homotopies}Retarded and advanced trivializations and initial data complex}
We equip the observable complex with a 
Poisson structure following the approach of 
\cite{BeniniBruinsmaSchenkel_2020_LinearYang}, 
see also \cite{BeniniMusanteSchenkel_2022_GreenHyperbolic} 
for a related, yet more conceptual, approach. 
In order to achieve this goal, 
let us consider the analog of the solution 
complex with an appropriate support restriction. Explicitly, 
we write $\Omega^k_{\sc}(M)$ for the vector space of $k$-forms 
whose support is spacelike compact, 
i.e.\ contained in the causal shadow 
$J_M(K) \coloneqq J^+_M(K) \cup J^-_M(K) \subseteq M$ 
of some compact subset $K \subseteq M$. 
Implementing the same support restriction on the solution complex 
$\Sol(M)$, we define  the solution complex with spacelike compact 
support $\Sol_{\sc}(M) \in \Ch_\bbR$. 
\sk

Since by definition $K \subseteq J_M(K)$, 
there are obvious inclusions 
$\Omega^k_\cc(M) \subseteq \Omega^k_{\sc}(M)$, 
for all $k = 0, \ldots, p$. 
These inclusions assemble to form the cochain map 
$j: \LLL(M) \to \Sol_{\sc}(M)[1]$ in $\Ch_\bbK$ 
to the $1$-shifted solution complex with spacelike compact support. 
Equivalently, this is a $1$-cocycle 
$j \in [\LLL(M),\Sol_{\sc}(M)]^1$ 
in the internal hom. 
As it is shown below, it turns out that $j$ is 
homotopic to $0$ in two inequivalent ways. 
Explicitly, this means that there exist $0$-cochains 
$\Lambda_\pm \in [\LLL(M),\Sol_{\sc}(M)]^0$ 
in the internal hom, 
called {\it retarded and advanced trivializations}, 
whose differentials 
$\partial \Lambda_\pm = \dd_{\Sol} \, \Lambda_\pm - \Lambda_\pm \, \dd_{\LLL} = j$ 
trivialize the $1$-cocycle $j$. 
Let us construct these retarded and advanced trivializations 
$\Lambda_\pm$ for Maxwell $p$-forms. 
To achieve this goal we consider the retarded 
and advanced Green's operators $G_\pm^{(k)}$ 
for the d'Alembert operator 
$\square \coloneqq \delta \dd + \dd \delta: \Omega^k(M) \to \Omega^k(M)$ 
on $k$-forms, $k=0,\ldots,p$. Recall from 
\cite{BarGinouxPfaffle_2007_WaveEquations,Bar_2015_GreenHyperbolicOperators} that a {\it retarded/advanced 
Green's operator} $G^{(k)}_\pm: \Omega_\cc^k(M) \to \Omega^k(M)$ 
for $\square$ is a linear map such that, 
for all $\omega \in \Omega_\cc^k(M)$, it holds that 
(i)~$\square\, G^{(k)}_\pm\, \omega = \omega$, 
(ii)~$G^{(k)}_\pm\, \square\, \omega = \omega$ and 
(iii)~$\supp(G^{(k)}_\pm\, \omega) \subseteq J_M^\pm(\supp (\omega))$. 
It is well-known that retarded and advanced Green's 
operators for $\square$ exist, are unique 
and commute both with the de Rham differential 
$\dd\, G^{k}_\pm = G^{k+1}_\pm\, \dd$ and 
codifferential $\delta\, G^{k}_\pm = G^{k-1}_\pm\, \delta$. 
With these preparations, we define 
the {\it retarded and advanced trivializations} 
$\Lambda_\pm \in [\LLL(M),\Sol_{\sc}(M)]^0$ degree-wise by 
\begin{align}\label{eq:Lambdapm}
\Lambda_\pm^n \coloneqq 
\begin{cases}
G_\pm^{(p+n)}\, \delta, & n = -p, \ldots, -1, \\ 
G_\pm^{(p)}, & n = 0, \\ 
\dd\, G_\pm^{(p-n)}, & n = 1, \ldots, p. 
\end{cases}
\end{align}
(All other components $\Lambda_\pm^n$, $n \leq -p-1$ 
and $n \geq p+1$, necessarily vanish.) 
Note that property~(iii) ensures that the output of $\Lambda_\pm$ 
has the required support. Furthermore, 
direct inspection using properties~(i) and~(ii) 
and the commutation rules of $G_\pm^{(k)}$ with $\dd$ and $\delta$  
shows that $\partial \Lambda_\pm = j$. 
\sk

Taking the difference 
of the retarded and the advanced trivializations 
from \eqref{eq:Lambdapm} defines the cochain map  
\begin{equation}\label{eq:Lambda}
\Lambda \coloneqq \Lambda_+ - \Lambda_-: \LLL(M) \longrightarrow \Sol_\sc(M) 
\end{equation}
in $\Ch_\bbR$. 
($\Lambda$ is indeed a cochain map because 
$\partial \Lambda = j - j = 0 \in [\LLL(M),\Sol_\sc(M)]^1$ 
vanishes when regarded as a $1$-cochain 
in the internal hom.) 

\begin{rem}\label{rem:Lambda-qiso}
The cochain map $\Lambda$ in \eqref{eq:Lambda} is 
a quasi-isomorphism since it induces the isomorphism 
$H(\Lambda): H(\LLL(M)) \to H(\Sol_\sc(M))$ in cohomology, 
which in terms of the de Rham cohomologies of $M$ 
with compact support $H_{\dR,\cc}(M)$ 
and with spacelike compact support $H_{\dR,\sc}(M)$ 
reads as 
\begin{equation}
\begin{cases}
H_{\dR,\cc}^{p+1+n}(M) \cong H_{\dR,\sc}^{p+n}(M), & n=-p-1,\ldots,-1, \\
\Omega_{\cc\,\delta}^p(M)/\delta\dd\Omega_\cc^p(M) \cong \Omega^p_{\sc\,\delta\dd}(M)/\dd\Omega^{p-1}_\sc(M), & n=0, \\
H_{\dR,\cc}^{m-p+n}(M) \cong H_{\dR,\sc}^{m-p-1+n}(M), & n=1,\ldots,p+1. 
\end{cases}
\end{equation}
(Note that the Hodge star operator $\ast$ has been 
implicitly used to identify the cohomology of the 
codifferential $\delta$ with the more familiar 
de Rham cohomology.) 
We refer to \cite{Khavkine_2016_CohomologyCausally} 
for the definition of the de Rham cohomologies 
with spacelike compact support $H^k_{\dR,\sc}(M)$ 
of an $m$-dimensional oriented and time-oriented 
globally hyperbolic Lorentzian manifold $M$ 
and for the isomorphisms in degrees $n\neq0$,
while we refer to \cite{Benini_2016_OptimalSpace} 
for the isomorphism in degree $n=0$ between 
``gauge invariant linear observables modulo equations of motion'' 
$\Omega_{\cc\,\delta}^p(M)/\delta\dd\Omega_\cc^p(M)$ 
and ``spacelike compact on-shell fields modulo gauge transformations'' 
$\Omega^p_{\sc\,\delta\dd}(M)/\dd\Omega^{p-1}_\sc(M)$. 
See \cite{BeniniMusanteSchenkel_2022_GreenHyperbolic} 
for a more conceptual proof of the fact that $\Lambda$ 
is a quasi-isomorphism. 
\end{rem}

For any choice of a spacelike Cauchy surface $\Sigma \subseteq M$, 
let us also consider the compactly supported 
{\it initial data complex} 
\begin{equation}\label{eq:Data}
\Data_\cc(\Sigma) \coloneqq \Bigg(
\xymatrix{
{\overset{(-p)}{\Omega_\cc^0(\Sigma)}} \ar[r]^-{\dd_\Sigma} & \cdots \ar[r]^-{\dd_\Sigma} & {\overset{(-1)}{\Omega_\cc^{p-1}(\Sigma)}} \ar[r]^-{(\dd_\Sigma,0)} & {\overset{(0)}{\Omega_\cc^p(\Sigma)^2}} \ar[r]^-{\delta_\Sigma \pr_2} & {\overset{(1)}{\Omega_\cc^{p-1}(\Sigma)}} \ar[r]^-{\delta_\Sigma} & \cdots \ar[r]^-{\delta_\Sigma} & {\overset{(p)}{\Omega_\cc^0(\Sigma)}}
}
\Bigg).
\end{equation}
(Here the notation $\dd_\Sigma$ and 
$\delta_\Sigma \coloneqq (-1)^k \ast_\Sigma^{-1} \dd_\Sigma\, \ast_\Sigma$ 
is used to emphasize that these differential operators 
are defined with respect to the geometry of $\Sigma$.) 
The compactly supported initial data complex 
is related to the spacelike compactly supported 
solution complex via the initial data map 
$\data: \Sol_\sc(M) \to \Data_\cc(\Sigma)$ in $\Ch_\bbR$ 
defined degree-wise by 
\begin{equation}\label{eq:data}
\data^n \coloneqq 
\begin{cases}
\iota^\ast, & n=-p,\ldots,-1, \\
(\iota^\ast,\ast_\Sigma^{-1}\, \iota^\ast \ast \dd), & n=0, \\
(-1)^n \ast_\Sigma^{-1}\, \iota^\ast\, \ast, & n=1,\ldots,p. 
\end{cases}
\end{equation}
where $\iota: \Sigma \to M$ denotes the embedding 
of the chosen spacelike Cauchy surface. 
(All other components $\data^n$, $n \leq -p-1$ and $n \geq p+1$, necessarily vanish.) 

\begin{rem}\label{rem:data-qiso}
The cochain map $\data$ in \eqref{eq:data} is 
a quasi-isomorphism since it induces the cohomology isomorphism 
$H(\data): H(\Sol_\sc(M)) \overset{\cong}{\to} H(\Data_\cc(M))$, 
which in terms of the de Rham cohomology of $M$ 
with spacelike compact support and the de Rham 
cohomology of $\Sigma$ with compact support reads as 
\begin{equation}
\begin{cases}
H_\sc^{p+n}(M) \cong H_\cc^{p+n}(\Sigma), & n=-p,\ldots,-1, \\
\Omega^p_{\sc\,\delta\dd}(M)/\dd\Omega^{p-1}_\sc(M) \cong \big( \Omega_\cc^p(\Sigma)/\dd_\Sigma\Omega_\cc^{p-1}(\Sigma) \big) \times \Omega_{\cc\,\delta_\Sigma}^p(\Sigma), & n=0, \\
H_\sc^{m-p-1+n}(M) \cong H_\cc^{m-p-1+n}(\Sigma), & n=1,\ldots,p+1. 
\end{cases}
\end{equation}
We refer to \cite{Benini_2016_OptimalSpace, Khavkine_2016_CohomologyCausally} 
for the isomorphisms in degrees $n\neq0$ 
and to \cite{SandersDappiaggiHack_2014_ElectromagnetismLocal} for the isomorphism in degree $n=0$, 
where it is stated in the form of the well-posed 
initial value problem for gauge classes of on-shell Maxwell $p$-forms. 
This perspective suggests the interpretation of 
the quasi-isomorphism $\data$ as a refinement 
of this well-posed initial value problem. 
\sk

Combining the quasi-isomorphisms $\data$ and $\Lambda$ 
from Remarks \ref{rem:Lambda-qiso} and \ref{rem:data-qiso}, 
we obtain a quasi-isomorphism 
$\data\, \Lambda: \LLL(M) \to \Data_\cc(\Sigma)$, 
which provides an explicit computation of the 
cohomology of the observable complex $\LLL(M)$ for Maxwell $p$-forms 
in terms of the compactly supported de Rham cohomology 
of a spacelike Cauchy surface $\Sigma \subseteq M$ 
and of the initial data for gauge classes of on-shell Maxwell $p$-forms. 
\end{rem}

\subsection{\label{subsec:Poisson}Poisson structure and quantization}
This section is devoted to the construction 
of a $\Ch_\bbC$-valued net of algebras, 
associated with Maxwell $p$-forms, 
over the category $\Loc_m$ of oriented and time-oriented $m$-dimensional 
globally hyperbolic Lorentzian manifolds, 
see Example \ref{ex:Loc}. 
The first step constructs a Poisson structure 
$\tau_M: \LLL(M)^{\wedge2} \to \bbR$ in $\Ch_\bbR$ 
on the observable complex 
$\LLL(M) \in \Ch_\bbR$, for each $M \in \Loc_m$. 
This defines a Poisson complex $(\LLL(M),\tau_M)$. 
The second step quantizes this Poisson complex 
using canonical commutation relations (CCR). 
The third step exhibits the net structure. 
Incidentally, we will observe that the resulting net 
of algebras is actually a homotopy algebraic quantum field theory, 
i.e.\ it fulfils both the Einstein's causality axiom 
and the homotopy time-slice axiom, 
see \cite{BeniniBruinsmaSchenkel_2020_LinearYang,BeniniSchenkelWoike_2019_HomotopyTheory}. 
Since we will work with the
closed symmetric monoidal model category 
of cochain complexes $\MM = \Ch_\bbC$, 
we shall replace the term ``monoid'' with the more 
familiar ``differential graded algebra''. 
Accordingly, we shall adopt the familiar notation 
$\DGA_\bbC \coloneqq \Mon(\Ch_\bbC)$. 
\sk

Combining \eqref{eq:ev} and \eqref{eq:Lambda} 
allows us to equip the observable complex $\LLL(M)$ 
with the {\it Poisson structure} 
\begin{equation}\label{eq:PoissonStr}
\tau_M \coloneqq -\ev (\Lambda \otimes \id): \LLL(M) ^{\wedge2} \longrightarrow \bbR 
\end{equation}
in $\Ch_\bbR$. (The obvious inclusion 
$\Sol_\sc(M) \to \Sol(M)$ in $\Ch_\bbR$ 
is implicit in the definition above.) 
Note that $\tau_M$ is graded anti-symmetric, 
as implicitly claimed in \eqref{eq:PoissonStr}. 
This follows because, with respect to the pairing 
$\int_M (-) \wedge \ast (-)$, the codifferential $\delta$ is the formal adjoint of the de Rham 
differential $\dd$ , $\square$ is formally 
self-adjoint and, as a consequence, also the retarded 
and advanced Green's operators are each 
the formal adjoint of the other. 
\sk

We quantize the Poisson complex $(\LLL(M),\tau_M)$,
consisting of the observable complex 
$\LLL(M) \in \Ch_\bbR$ from \eqref{eq:L} 
and the Poisson structure 
$\tau_M: \LLL(M)^{\wedge2} \to \bbR$ in $\Ch_\bbR$ 
from \eqref{eq:PoissonStr}, 
applying the CCR quantization functor of 
\cite{BeniniBruinsmaSchenkel_2020_LinearYang}. 
Explicitly, we consider the free differential graded algebra 
\begin{subequations}\label{eq:CCR}
\begin{equation}
T_\bbC \big( \LLL(M) \big) \coloneqq \bigoplus_{m \geq 0} \LLL(M)_\bbC^{\otimes m} \in \DGA_\bbC
\end{equation}
generated by the complexification $\LLL(M)_\bbC \coloneqq \LLL(M) \otimes_\bbR \bbC \in \Ch_\bbC$ 
of the observable complex $\LLL(M) \in \Ch_\bbR$. 
The multiplication $\mu$ is defined by juxtaposition 
$\varphi_1 \otimes \cdots \otimes \varphi_k \otimes \psi_1 \otimes \cdots \otimes \psi_m$ 
of words $\varphi_1 \otimes \cdots \otimes \varphi_k, 
\psi_1 \otimes \cdots \otimes \psi_m$ in $T_\bbC(\LLL(M))$
and the unit $\bfone$ corresponds to the 
length $0$ word $1 \in \bbC \subseteq T_\bbC(\LLL(M))$. 
Taking the quotient by the two-sided 
ideal $I_\tau \subseteq T_\bbC(\LLL(M))$ generated by the elements 
\begin{equation}
\varphi_1 \otimes \varphi_2 - (-1)^{|\varphi_1| |\varphi_2|}\, \varphi_2 \otimes \varphi_1 - i \tau_M(\varphi_1,\varphi_2) \bfone,
\end{equation}
for all homogeneous cochains $\varphi_1, \varphi_2 \in \LLL(M)$ 
defines the differential graded algebra 
\begin{equation}
\AAA(M) \coloneqq T_\bbC \big( \LLL(M) \big) / I_\tau \in \DGA_\bbC.
\end{equation}
\end{subequations}

\begin{rem}\label{rem:involution}
It is straightforward to endow the differential graded 
algebra $\AAA(M) \in \DGA_\bbC$ with a 
multiplication reversing $\ast$-involution 
arising from complex conjugation $\overline{(-)}$ 
on $\bbC$. (We refer to  
\cite{Jacobs_2012_InvolutiveCategories,BeniniSchenkelWoike_2019_InvolutiveCategories} 
for a more detailed discussion on reversing $\ast$-monoids 
in an involutive symmetric monoidal category.) 
Indeed, one defines $\ast$ on the free 
differential graded algebra $T_\bbC(\LLL(M)) \in \DGA_\bbC$ 
as the unique multiplication reversing graded 
$\bbC$-antilinear map extending the complex conjugation 
$\varphi \mapsto \overline{\varphi}$ on length $1$ words 
$\varphi$ in $\LLL(M)_\bbC \subseteq T_\bbC(\LLL(M))$. 
The graded anti-symmetry of $\tau$ entails that 
$I_\tau \subseteq T_\bbC(\LLL(M))$ is a two-sided  $\ast$-ideal, 
therefore $\ast$ descends to the quotient  $\AAA(M) \in \DGA_\bbC$. 
In other words, $\AAA(M)$ is a differential graded 
unital and associative $\ast$-algebra. 
\end{rem}

We focus now on the construction of the net structure. 
Recall that differential forms with compact support 
can be extended by zero along open embeddings 
$f: M \to N$. Let us denote the corresponding
extension-by-zero map by 
$f_\ast: \Omega_\cc^k(M) \to \Omega_\cc^k(N)$. 
Since all morphisms $f: M \to N$ in $\Loc_m$ are 
in particular open embeddings, one defines the cochain maps 
\begin{equation}\label{eq:LLL-morphism}
\LLL(f): \LLL(M) \longrightarrow \LLL(N) 
\end{equation} 
in $\Ch_\bbR$ degree-wise by 
$\LLL(f)^n \coloneqq f_\ast$ for $n = -p-1, \ldots, p$
and $\LLL(f)^n \coloneqq 0$ else. 
Together with the assignment of the observable complex 
$M \in \Loc_m \mapsto \LLL(M) \in \Ch_\bbR$, 
this defines the functor $\LLL: \Loc_m \to \Ch_\bbR$. 
The naturality of the de Rham differential $\dd$ 
and of the Hodge star operator $\ast$ with respect to morphisms 
in $\Loc_m$ entails the naturality of the linear differential 
operator $\square: \Omega^k(M) \to \Omega^k(M)$. 
From the uniqueness of the associated retarded 
and advanced Green's operators 
$G^{(k)}_\pm: \Omega_\cc^k(M) \to \Omega^k(M)$, 
it follows that 
$f^\ast\, G^{(k)}_\pm\, f_\ast = G^{(k)}_\pm$, 
for all $f: M \to N$ in $\Loc_m$, 
where $f^\ast: \Omega^k(N) \to \Omega^k(M)$ 
denotes the pull-back of $k$-forms. 
This fact entails that also the Poisson 
structure is natural, i.e.\ 
$\tau_N\, (\LLL(f) \otimes \LLL(f)) = \tau_M$, 
for all $f: M \to N$ in $\Loc_m$. 
In other words, one obtains a functor $(\LLL,\tau)$ 
that assigns to each $M \in \Loc_m$ the Poisson complex $(\LLL(M),\tau_M)$ 
and to each morphism $f: M \to N$ in $\Loc_m$ 
the Poisson structure preserving cochain map 
$\LLL(f): (\LLL(M),\tau_M) \to (\LLL(N),\tau_N)$. 
The CCR quantization recalled in \eqref{eq:CCR} 
is manifestly functorial, see 
\cite{BeniniBruinsmaSchenkel_2020_LinearYang}. 
As a result, composing the functor $(\LLL,\tau)$  
and the CCR quantization functor we obtain a functor 
$\AAA: \Loc_m \to \DGA_\bbC$, i.e.\ the net of algebras 
\begin{equation}\label{eq:Maxwell-net}
\AAA \in \Net_{\Loc_m}^{\Ch_\bbC},
\end{equation} 
see Remark \ref{rem:net}, 
associated with Maxwell $p$-forms. 

\begin{rem}
We observe incidentally that the net 
$\AAA \in \Net_{\Loc_m}^{\Ch_\bbC}$ constructed above, 
endowed with the $\ast$-involution 
from Remark \ref{rem:involution}, 
is actually a homotopy algebraic quantum field theory, 
see \cite{BeniniBruinsmaSchenkel_2020_LinearYang,BeniniSchenkelWoike_2019_HomotopyTheory,BeniniSchenkel_2019_HigherStructures}. 
In other words, this net fulfils both the Einstein's causality 
axiom and the homotopy time-slice axiom. 
The Einstein's causality axiom, 
i.e.\ the fact that the graded commutator 
\begin{equation}
[\AAA(f_1),\AAA(f_2)] = 0: \AAA(M_1) \otimes \AAA(M_2) \longrightarrow \AAA(N)
\end{equation}
in $\Ch_\bbC$ vanishes for all pairs of morphisms 
$f_1: M_1 \to N \leftarrow M_2: f_2$ in $\Loc_m$ 
with causally disjoint images, 
is a straightforward consequence 
of the support properties of the retarded and advanced 
Green's operators that enter the definition of the Poisson 
structure through the cochain map $\Lambda$, 
see \eqref{eq:Lambdapm}, \eqref{eq:Lambda} and \eqref{eq:PoissonStr}. 
\sk

It is slightly more involved to check the homotopy time-slice axiom, 
i.e.\ that the 
morphism $\AAA(f): \AAA(M) \to \AAA(N)$ in $\DGA_\bbC$ 
is a weak equivalence whenever $f: M \to N$ 
in $\Loc_m$ is a Cauchy morphism. 
Recall first that weak equivalences in $\DGA_\bbC$ 
are just quasi-isomorphisms between the underlying 
cochain complexes. The fact that the cochain map 
underlying $\AAA(f)$ is indeed a quasi-isomorphism 
can be best understood taking a closer look 
at the cochain map $\LLL(f): \LLL(M) \to \LLL(N)$ 
in $\Ch_\bbR$ from \eqref{eq:LLL-morphism}. Recalling 
that per hypothesis $f$ is a Cauchy morphism, 
i.e.\ the image 
$f(M) \subseteq N$ contains a spacelike Cauchy surface 
$\Sigma \subseteq f(M)$ of the codomain $N$, 
one obtains the commutative diagram 
\begin{equation}
\xymatrix{
\LLL(M) \ar[rr]^-{\LLL(f)} \ar[dr]_-{\data\, \Lambda} && \LLL(N) \ar[dl]^-{\data\, \Lambda} \\
& \Data_\cc(\Sigma)
}
\end{equation}
in $\Ch_\bbR$ involving the passage to the initial data 
complex $\Data_\cc(\Sigma)$, 
see \eqref{eq:Data} and \eqref{eq:data}. 
(Note that we identified the spacelike Cauchy surface 
$\Sigma \subseteq N$ with its preimage in $M$ via $f$, 
which is automatically a spacelike Cauchy surface of $M$.) 
As observed in Remarks \ref{rem:Lambda-qiso} 
and \ref{rem:data-qiso}, both $\Lambda$ and $\data$ 
are quasi-isomorphisms. The fact that 
quasi-isomorphisms are closed under composition 
and fulfil the two-out-of-three 
property entails that $\LLL(f)$ 
is a quasi-isomorphism too. Summing up, 
$\LLL(f): (\LLL(M),\tau_M) \to (\LLL(N),\tau_N)$ 
is a Poisson structure preserving quasi-isomorphism. 
By \cite[Prop.\ 5.3]{BeniniBruinsmaSchenkel_2020_LinearYang}, 
see also \cite{BruinsmaSchenkel_2019_AlgebraicField}, 
the CCR quantization functor, 
which enters the construction of the Maxwell $p$-forms net $\AAA$ 
from \eqref{eq:Maxwell-net}, maps Poisson structure preserving 
quasi-isomorphisms to weak equivalences. 
Therefore, it follows that $\AAA(f): \AAA(M) \to \AAA(N)$ in $\DGA_\bbC$ 
is a weak equivalence for all Cauchy morphisms 
$f: M \to N$ in $\Loc_m$. This proves that the Maxwell $p$-forms net 
$\AAA$ fulfils also the homotopy time-slice axiom holds 
and hence it defines a homotopy algebraic quantum field theory 
in the sense of  \cite{BeniniBruinsmaSchenkel_2020_LinearYang,BeniniSchenkelWoike_2019_HomotopyTheory,BeniniSchenkel_2019_HigherStructures}. 
\end{rem}

\subsection{\label{subsec:rep-construction}Construction of a net representation}
The goal of this section is to construct a 
representation of the $\Ch_\bbC$-valued (Haag-Kastler) net 
\begin{equation}\label{eq:HKnet}
\AAA_M \coloneqq \AAA\, \iota_M \in \Net_{\CCO(M)}^{\Ch_\bbC}
\end{equation}
on $\CCO(M)$, see Example \ref{ex:Loc}, 
that describes quantized Maxwell $p$-form fields 
on a fixed oriented and time-oriented 
globally hyperbolic Lorentzian manifold $M \in \Loc_m$. 
$\AAA_M$ is obtained by restricting the (generally covariant) 
net $\AAA \in \Net_{\Loc_m}^{\Ch_{\bbC}}$ 
from \eqref{eq:Maxwell-net} along the functor 
$\iota_M: \CCO(M) \to \Loc_m$ that sends 
a causally convex open subsets $U \in \CCO(M)$ 
to the oriented and time-oriented globally 
hyperbolic Lorentzian manifolds $U \in \Loc_m$ 
defined by endowing $U$ with the restriction 
of the geometry of $M$. 
We shall obtain a representation of the net $\AAA_M$ 
from a two-point function 
$\omega_2: \LLL(M) \otimes \LLL(M) \to \bbC$ 
in $\Ch_\bbR$
on the observable complex $\LLL(M)$ from \eqref{eq:L}, i.e.\ a cochain 
map whose anti-symmetric part agrees with the Poisson structure 
$\tau_M$ from \eqref{eq:PoissonStr}. 
Mimicking the usual construction of a quasi-free 
state from a two-point function, see 
e.g.\ \cite[Sec.\ 5.2.4]{KhavkineMoretti_2015_AlgebraicQFT},
we shall use $\omega_2$ to define a linear functional 
$\omega: \AAA(M) \to \bbC$ in $\Ch_\bbC$. 
The first step of the Gelfand-Naimark-Segal 
construction applied to $\omega$ shall then provide 
a representation of the global algebra of observables 
$\AAA(M) = \AAA_M(M)$. The latter defines 
a constant net representation of $\AAA_M$ 
via Construction \ref{constr:eval-const-adj}. 

\begin{rem}\label{rem:two-point function}
Usually two-point functions are not only required to have 
anti-symmetric part matching the Poisson structure of interest, 
but are also required to be bisolutions of the equation of motion 
that governs the field theoretic model of interest. 
In our framework the latter requirement is replaced and 
generalized by the condition that $\omega_2$ is a cochain map. 
Notably, this condition takes also care 
of the compatibility with the action of gauge transformations. 
\end{rem}

In order to simplify the presentation and to better 
highlight the main features of our construction, 
we shall assume that $M \in \Loc_m$ is of the form 
\begin{subequations}
\begin{equation}
M = \bbR \times \Sigma,
\end{equation}
with $\Sigma$ a {\it compact} spacelike Cauchy surface, 
and that the metric $g$ is ultra-static, i.e.\ 
\begin{equation}
g = - dt^2 + h,
\end{equation}
\end{subequations}
with $h$ a Riemannian metric on $\Sigma$ 
(constant with respect to $t \in \bbR$). 
\sk

The explicit construction of the two-point function 
$\omega_2: \LLL(M)^{\otimes 2} \to \bbC$ in $\Ch_\bbR$ 
partly follows the lines of \cite{BeniniCapoferriDappiaggi_2017_HadamardStates}. 
First, since $M = \bbR \times \Sigma$ is a product, 
$k$-forms on $M$ decompose into sections 
of the pullbacks along the projection $\pr_2: M \to \Sigma$ 
of the bundles $\Lambda^k \Sigma$ of $k$-forms 
and $\Lambda^{k-1} \Sigma$ of $(k-1)$-forms over $\Sigma$. 
By abuse of notation, we denote the pullback bundles over $M$ 
again by $\Lambda^k \Sigma$ and $\Lambda^{k-1} \Sigma$, 
so that the above mentioned decomposition takes the form 
\begin{align}\label{eq:forms-decomposition}
\Omega^k(M) = \Gamma(M,\Lambda^k \Sigma) \oplus dt \wedge \Gamma(M,\Lambda^{k-1} \Sigma), && \alpha = \alpha_S + dt \wedge \alpha_T. 
\end{align}
The first summand corresponds to $k$-forms on $M$ having only ``spatial legs'', 
while the second summand corresponds to $k$-forms 
on $M$ having precisely one ``time leg''. We shall use 
the subscripts $_S$ and $_T$ to refer to the space 
$\alpha_S \in \Gamma(M,\Lambda^k \Sigma)$ 
and respectively time 
$\alpha_T \in \Gamma(M,\Lambda^{k-1} \Sigma)$ 
parts of a $k$-form $\alpha \in \Omega^k(M)$. 
Since the ultra-static metric $g = - dt^2 + h$ decomposes 
into time and space parts, the de Rham differential $\dd$, 
the codifferential $\delta$ and the d'Alembert operator 
$\square$ acting on $k$-forms on $M$ admit 
a decomposition compatible with \eqref{eq:forms-decomposition}. 
Explicitly, denoting with $\dd_S$ the de Rham differential, 
with $\delta_S$ the codifferential and with 
$\triangle \coloneqq \delta_S \dd_S + \dd_S \delta_S$
the Laplace operator, all acting on differential forms on $\Sigma$, 
and with $\partial_t$ the time-derivative, one finds 
\begin{subequations}\label{eq:operators-decomposition}
\begin{align}
\dd \alpha &= \dd_S \alpha_S + dt \wedge (\partial_t \alpha_S - \dd_S \alpha_T), \\
\delta \alpha &= (\partial_t \alpha_T + \delta_S \alpha_S) - dt \wedge \delta_S \alpha_T, \\
\delta \dd \alpha &= (\partial_t^2 \alpha_S - \dd_S \partial_t \alpha_T + \delta_S \dd_S \alpha_S) + dt \wedge (\delta_S \dd_S \alpha_T - \delta_S \partial_t \alpha_S), \\
\square \alpha &= (\partial_t^2 \alpha_S + \triangle \alpha_S) + dt \wedge (\partial_t^2 \alpha_T + \triangle \alpha_T), 
\end{align}
\end{subequations}
for all $\alpha \in \Omega^k(M)$. 
Combining \eqref{eq:forms-decomposition} 
and \eqref{eq:operators-decomposition} 
with the Hodge decomposition for differential $k$-forms on $\Sigma$ 
\begin{align}\label{eq:harmonic-decomposition}
\Omega^k(\Sigma) = \calH^k(\Sigma) \oplus \calH_\perp^k(\Sigma), && \sigma = \sigma_{\calH} + \sigma_\perp,
\end{align}
into (spatially) harmonic $\sigma_{\calH} \in \calH^k(\Sigma)$ ($\triangle \sigma_{\calH} = 0$)
and orthogonal $\sigma_\perp \in \calH_\perp^k(\Sigma)$ 
($\sigma_\perp = \dd_S \sigma_1 + \delta_S \sigma_2$) parts, 
one obtains an explicit formula 
for the retarded-minus-advanced propagator 
\begin{subequations}\label{eq:G-explicit}
\begin{equation}
G^{(k)} \coloneqq G^{(k)}_+ - G^{(k)}_-: \Omega^k_\cc(M) \longrightarrow \Omega^k(M)
\end{equation}
associated with $\Box: \Omega^k(M) \to \Omega^k(M)$, 
see Section \ref{subsec:ret-adv-homotopies}, given by 
\begin{equation}
G^{(k)}\alpha = G \alpha_S + dt \wedge G \alpha_T, 
\end{equation}
for all $\alpha \in \Omega^k_\cc(M)$, where 
\begin{align}
(G \beta) (t,\cdot) \coloneqq \int_\bbR dt^\prime\, (t-t^\prime) \beta_\calH(t^\prime,\cdot) + \int_\bbR dt^\prime\, \Big( \triangle^{-\frac{1}{2}} \sin \big( \triangle^{\frac{1}{2}} (t-t^\prime) \big) \beta_\perp \Big) (t^\prime,\cdot),
\end{align}
\end{subequations}
with $\beta \in \Gamma_\cc(M,\Lambda^j \Sigma)$, $j=k,k-1$. 
(Note that the Laplacian $\triangle$ on the orthogonal part 
$\calH_{\perp}^j(\Sigma)$ has strictly positive spectrum, 
hence one can consider $\triangle^{\frac{1}{2}}$, 
as well as its inverse $\triangle^{-\frac{1}{2}}$. 
Incidentally, we observe that $G$, 
is the retarded-minus-advanced propagator for the 
normally hyperbolic linear differential operator 
$\partial_T^2 + \triangle: \Gamma(M,\Lambda^j \Sigma) \to \Gamma(M,\Lambda^j \Sigma)$.) 
\eqref{eq:G-explicit} can be used to compute $\Lambda$ 
from \eqref{eq:Lambda} more explicitly, 
which leads to an explicit formula for the Poisson 
structure $\tau_M$ from \eqref{eq:PoissonStr} 
that makes the so-called ``positive and negative 
frequency contributions'' manifest. 
\sk

With these preparations, we construct a two-point 
function $\omega_2$ by mimicking the usual prescription, 
which amounts to selecting the ``positive frequency 
contribution'' in \eqref{eq:G-explicit}. 
Explicitly, we introduce 
\begin{subequations}\label{eq:omega_2}
\begin{equation}
\omega_2: \LLL(M)^{\otimes 2} \longrightarrow \bbC
\end{equation}
in $\Ch_\bbR$ defining the only non-vanishing component 
\begin{equation}
\omega_2^0: \bigoplus_{m=-p}^p \big( \LLL(M)^m \otimes \LLL(M)^{-m} \big) \longrightarrow \bbC
\end{equation}
component-wise for all $m=-p,\ldots,p$ by 
\begin{equation}
(\omega_2^0)_m \coloneqq 
\begin{cases}
(-1)^{\lfloor m/2 \rfloor}\, W^{(p+m)}\, (\delta \otimes \id), & m = -p, \ldots, -1, \\
W^{(p)}, & m = 0, \\
(-1)^{\lfloor m/2 \rfloor}\, W^{(p-m)}\, (\id \otimes \delta), & m = 1, \cdots, p,
\end{cases}
\end{equation}
\end{subequations}
where the linear map 
\begin{subequations}\label{eq:W^(k)}
\begin{equation}
W^{(k)}: \Omega^k_\cc(M)^{\otimes 2} \longrightarrow \bbC, 
\end{equation}
for $k = 0, \ldots, p$, is defined 
for all $\alpha, \alpha^\prime \in \Omega^k_\cc(M)$ by 
\begin{equation}
W^{(k)}(\alpha \otimes \alpha^\prime) \coloneqq W_\calH(\alpha_{S} \otimes \alpha^\prime_{S}) + W_\perp(\alpha_{S} \otimes \alpha^\prime_{S}) - W_\calH(\alpha_{T} \otimes \alpha^\prime_{T}) - W_\perp(\alpha_{T} \otimes \alpha^\prime_{T}),
\end{equation}
with
\begin{flalign}
W_\calH(\beta \otimes \beta^\prime) &\coloneqq \frac{1}{2} \int_\bbR dt\, \int_\bbR dt^\prime\, \Big\langle \beta_{\calH}(t,\cdot), i (t-t^\prime) \beta^\prime_{\calH}(t^\prime,\cdot) \Big\rangle, \\
W_\perp(\beta \otimes \beta^\prime) &\coloneqq \frac{1}{2} \int_\bbR dt\, \int_\bbR dt^\prime\, \Big\langle \beta_{\perp}(t,\cdot), \big( \triangle^{-\frac{1}{2}} \exp \big( i \triangle^{\frac{1}{2}} (t-t^\prime) \big) \beta^\prime_{\perp} \big) (t^\prime,\cdot) \Big\rangle,
\end{flalign}
\end{subequations}
for all $\beta,\beta^\prime \in \Gamma_\cc(M,\Lambda^j \Sigma)$. 
Here the pairing $\langle\cdot,\cdot\rangle$ denotes 
the usual scalar product between $\bbC$-valued 
$j$-forms on $\Sigma$ for $j=k,k-1$.
\sk

Now it remains to verify that $\omega_2$ is a cochain map whose anti-symmetric part 
matches the Poisson structure of interest, as specified in Remark \ref{rem:two-point function}.
\sk

First, observe that $W^{(k)}$ is a $\square$-bisolution 
for all $k=0,\ldots,p$, i.e.\ 
\begin{equation}\label{eq:W^(k)-bisol}
W^{(k)}(\square \otimes \id) = 0 = W^{(k)}(\id \otimes \square),
\end{equation}
which follows from the fact that both $W_{\calH}$ 
and $W_{\perp}$ are by construction 
$(\partial_t^2 + \triangle)$-bisolutions. Furthermore, 
combining \eqref{eq:operators-decomposition} 
and \eqref{eq:W^(k)} one checks that 
\begin{align}\label{eq:W^(k)-d-delta}
W^{(k)}(\dd \otimes \id) = W^{(k-1)}(\id \otimes \delta), && W^{(k)}(\id \otimes \dd) = W^{(k-1)}(\delta \otimes \id), 
\end{align}
for all $k=1,\ldots,p$.
\sk

Let us now confirm that $\omega_2$ is indeed a cochain map. 
Explicitly, this is equivalent to the conditions 
\begin{equation}\label{eq:omega_2-cochain-map}
\begin{cases}
W^{(0)}(\delta\dd \otimes \id) = 0, & m = -p, \\
W^{(p+m)}(\delta\dd \otimes \id) + W^{(p+m-1)}(\delta \otimes \delta) = 0, & m = -p+1, \ldots, -1, \\
W^{(p)}(\delta\dd \otimes \id) + W^{(p-1)}(\delta \otimes \delta) = 0, & m = 0, \\
W^{(p-1)}(\delta \otimes \delta) + W^{(p)}(\id \otimes \delta\dd) = 0, & m = 1, \\
W^{(p-m)}(\delta \otimes \delta) + W^{(p-m+1)}(\id \otimes \delta \dd) = 0, & m = 2, \ldots, p, \\
W^{(p)}(\id \otimes \delta\dd) = 0, & m = p+1.
\end{cases}
\end{equation}
These conditions are a direct consequence of \eqref{eq:W^(k)-bisol} 
and \eqref{eq:W^(k)-d-delta},
hence $\omega_2$ is a cochain map, as claimed.
\sk 

Moreover, one can verify that 
the anti-symmetric part of $\omega_2$ 
agrees with $\tau_M$, namely 
\begin{equation}\label{eq:omega_2-vs-tau}
\omega_2(\varphi_1 \otimes \varphi_2) - (-1)^{|\varphi_1| |\varphi_2|}\, \omega_2(\varphi_2 \otimes \varphi_1) = i\, \tau_M(\varphi_1 \otimes \varphi_2),
\end{equation}
for all homogeneous cochains $\varphi_1,\varphi_2 \in \LLL(M)$. This follows directly from 
\eqref{eq:omega_2} and \eqref{eq:W^(k)} recalling
\eqref{eq:Lambda}, \eqref{eq:PoissonStr} 
and \eqref{eq:G-explicit}.
This property will ensure that the cochain map 
$\omega: \AAA(M) \to \bbC$ in $\Ch_\bbC$ 
defined in \eqref{eq:omega} is compatible with 
the canonical commutation relations \eqref{eq:CCR} 
and hence well-defined. 

\begin{rem}
Note that in \eqref{eq:W^(k)} one could add any 
(time-constant) symmetric operator $A_j$ acting 
on harmonic $j$-forms on $\Sigma$ by replacing 
$i (t-t^\prime)$ with $i (t-t^\prime) + A_j$. 
Since $\Sigma$ is assumed to be compact, 
harmonic $j$-forms on $\Sigma$ form 
a finite dimensional Hilbert space. 
Therefore $A_j$ is just any symmetric matrix. 
\end{rem}

\begin{rem}\label{rem:omega_2-microlocal}
We emphasize that the integral kernel of $W_\perp$ 
from \eqref{eq:W^(k)} is a bidistribution 
fulfilling the microlocal spectrum condition, 
see e.g.\ \cite[Sec.\ 5.3.4]{KhavkineMoretti_2015_AlgebraicQFT}. 
In particular, this entails that 
the induced two-point function 
$H^0(\omega_2): H^0(\LLL(M)) \otimes H^0(\LLL(M)) \to \bbC$ 
on the degree $0$ cohomology of the observable complex $\LLL(M)$ 
fulfils the microlocal spectrum condition. 
Note that in degree $0$ cohomology and for $p=1$ our construction 
reproduces gauge-invariant on-shell linear 
observables for the electromagnetic vector potential. 
Indeed, the two-point function 
$H^0(\omega_2): H^0(\LLL(M)) \otimes H^0(\LLL(M)) \to \bbC$ 
agrees with the Hadamard two-point function of
\cite[Sec.\ IV.C]{FewsterPfenning_2003_QuantumWeak}. 
\end{rem}

The next step uses $\omega_2$ from \eqref{eq:omega} 
to define a linear functional 
\begin{subequations}\label{eq:omega}
\begin{equation}
\omega: \AAA(M) \longrightarrow \bbC
\end{equation}
in $\Ch_\bbC$ on (the cochain complex underlying) 
the quantized differential graded algebra 
$\AAA(M)$ from \eqref{eq:CCR}. The latter is specified 
on words of arbitrary length $m \geq 0$ 
($m=0$ corresponds to the unit $\bfone \in \AAA(M)$) by 
\begin{equation}
\omega(\varphi_1 \otimes \cdots \otimes \varphi_m) \coloneqq 
\begin{cases}
1, & m=0, \\ 
0, & m=2k-1, k \geq 1, \\
\displaystyle \sum_{\sigma \in P} \operatorname{sign}(\sigma; |\varphi_1|, \ldots, |\varphi_{2k}|) \prod_{i=1}^k \omega(\varphi_{\sigma_{i1}} \otimes \varphi_{\sigma_{i2}}), & m=2k, k \geq 1, 
\end{cases}
\end{equation}
\end{subequations}
for all homogeneous cochains 
$\varphi_1, \ldots, \varphi_m \in \LLL(M)$, 
see \cite[Sec.\ 5.2.4]{KhavkineMoretti_2015_AlgebraicQFT}. 
In the previous formula $P$ denotes the set of all 
partitions $\sigma = \{\sigma_1,\ldots,\sigma_k\}$ 
of the ordered set $\{1 < \ldots < 2k\}$ into $k$ 
ordered pairs $\sigma_i = \{\sigma_{i1} < \sigma_{i2}\}$, 
$i=1,\ldots,k$. Furthermore, 
$\operatorname{sign}(\sigma; |\varphi_1|, \ldots, |\varphi_{2k}|)$ 
denotes the Koszul sign obtained by permuting 
the letters $\varphi_1, \ldots, \varphi_{2k}$ 
according to the permutation associated with $\sigma$. 
Let us confirm that $\omega$ is a well-defined cochain 
map. First, combining \eqref{eq:omega_2-vs-tau} 
and the explicit formula in \eqref{eq:omega}, 
it follows that $\omega$ vanishes on the ideal 
generated by the canonical commutation relations 
\eqref{eq:CCR} and hence it descends to 
the quotient $\AAA(M)$. 
Furthermore, the fact that $\omega_2$ is a cochain map 
entails that $\omega$ is a cochain map too. 
(To this end Koszul signs play a crucial role.) 

\begin{rem}\label{rem:omega-inv}
Since $\omega$ is a cochain map, passing 
to degree $0$ cohomology one obtains a linear functional 
$H^0(\omega): H^0(\AAA(M)) \to \bbC$. 
It is straightforward to check that the algebra 
$H^0(\AAA(M))$ contains the CCR algebra 
$\AAA^{\mathrm{inv}}(M) \subseteq H^0(\AAA(M))$ associated with the 
degree $0$ cohomology $H^0(\LLL(M))$ of the observable 
complex, consisting of gauge-invariant on-shell linear 
observables, endowed with the induced Poisson structure 
$H^0(\tau_M): H^0(\LLL(M))^{\wedge2} \to \bbR$. 
Recalling Remark \ref{rem:omega_2-microlocal}, 
one observes that the restriction 
$\omega^{\mathrm{inv}}: \AAA^{\mathrm{inv}}(M) \to \bbC$ 
of $H^0(\omega)$ is a Hadamard state. 
In particular, for $p=1$, this produces a Hadamard state 
on the CCR algebra $\AAA^{\mathrm{inv}}(M)$ 
of gauge-invariant on-shell linear observables 
for the electromagnetic vector potential. 
As we shall explain in Remark \ref{rem:H0-vs-inv}, 
the algebra $H^0(\AAA(M))$ is often strictly larger than 
the algebra $\AAA^{\mathrm{inv}}(M)$. 
As a consequence $H^0(\omega)$ is often richer 
than its restriction $\omega^{\mathrm{inv}}$. 
\end{rem}

The second step of our construction 
of a net representation mimics the first part of the 
Gelfand-Naimark-Segal construction. This yields 
a representation of the differential graded algebra 
$\AAA_M(M) = \AAA(M) \in \DGA_\bbC$ induced by $\omega$ 
from $\eqref{eq:omega}$. Introducing the subcomplex 
$R_\omega \subseteq \AAA_M(M) \in \Ch_\bbC$ defined degree-wise 
for all $n \in \bbZ$ by 
\begin{subequations}
\begin{equation}
R_\omega^n \coloneqq \big\{ a \in \AAA_M(M)^n:\: \omega(ba) = 0,\; \forall b \in \AAA_M(M)^{-n} \big\},
\end{equation}
one immediately observes that $R_\omega$ is a left
$\AAA_M(M)$-ideal. Therefore, the quotient 
of $\AAA_M(M)$ by $R_\omega$ defines the left 
$\AAA_M(M)$-module
\begin{equation}
V_\omega \coloneqq \AAA_M(M) / R_\omega \in {_{\AAA_M(M)}\Mod}.
\end{equation}
\end{subequations}
In other words, $V_\omega$ defines a representation of the global algebra of observables $\AAA_M(M)$.
\sk

Finally, in order to obtain a net representation of $\AAA_M$ we exploit Construction \ref{constr:eval-const-adj}.
Evaluating the right adjoint functor 
$(-)^M: {_{\AAA_M(M)}\Mod} \to \Rep(\AAA_M)$ 
on $V_\omega$ defines the constant net representation 
\begin{equation}
\VVV_\omega \coloneqq V_\omega^M \in \Rep(\AAA_M) 
\end{equation}
of the net $\AAA_M \in \Net_{\CCO(M)}^\Ch$ 
from \eqref{eq:HKnet}. 
Explicitly, for each causally convex open subset 
$U \in \CCO(M)$, $\VVV_\omega(U) = V_\omega\vert_{\AAA_M(U)} \in {_{\AAA_M(U)}\Mod}$ 
is just the restriction of the left $\AAA_M(M)$-module 
$V_\omega$ along the morphism 
$\AAA_M(\iota_U^M): \AAA_M(U) \to \AAA_M(M)$ 
in $\DGA_\bbC$ associated with the 
inclusion $\iota_U^M: U \to M$ in $\CCO(M)$.

\subsection{2-dimensional example}
In this section we shall compute explicitly the global algebra 
of observables $\AAA_M(M)$ of the net 
$\AAA_M \in \Net_{\CCO(M)}^{\Ch_\bbC}$ from \eqref{eq:HKnet} and describe its constant net representations. 
We shall do so in the simplest scenario, namely setting $m=2$, $p=1$ 
and choosing as oriented and time-oriented globally 
hyperbolic Lorentzian manifold the $2$-dimensional 
flat Lorentz cylinder $M \in \Loc_2$ consisting of 
the manifold $\bbR \times S^1$ endowed with 
the constant metric $g = - dt^2 + d\theta^2$, 
with the time-orientation determined by the vector field 
$\partial_t$ and with the counter-clockwise orientation on the unit length circle $S^1$. 
Concretely, we shall construct a differential graded 
algebra $A \in \DGA_\bbC$ that is weakly 
equivalent to the original one $\AAA_M(M)$, but has 
the advantages of having finitely many generators 
and trivial differential. 
These features make its category of left modules 
$_{A}\Mod$ easier to describe than the category 
$_{\AAA_M(M)}\Mod$, which is Quillen equivalent to it 
on account of Proposition
\ref{propo:change-monoid-vs-model-str}, 
but practically less accessible. 
This fact will allow us to obtain a very explicit description 
of all the constant net representations of $\AAA_M$ 
up to weak equivalence. 
\sk

To construct the differential graded algebra 
$A \in \DGA_\bbC$, weakly equivalent 
to $\AAA_M(M)$, we proceed in three steps. 
First, we shall construct a cochain complex 
$L \in \Ch_\bbC$ with trivial differential 
and a quasi-isomorphism $c: L \to \LLL(M)$ in $\Ch_\bbR$ 
to the observable complex $\LLL(M)$ from \eqref{eq:L}. 
Second, we shall endow $L$ with the Poisson 
structure $\widetilde{\tau} \coloneqq \tau_M (c \otimes c)$ 
induced by $\tau_M$ from \eqref{eq:PoissonStr}. 
This defines a new Poisson complex 
$(L,\widetilde{\tau})$ and a quasi-isomorphism 
$c: (L,\widetilde{\tau}) \to (\LLL(M),\tau_M)$ 
that by construction preserves the Poisson structures. 
In the third and last step we shall quantize the Poisson 
complex $(L,\widetilde{\tau})$ by means of the canonical 
commutation relations, as in \eqref{eq:CCR}. Since by  
\cite[Prop.\ 5.3]{BeniniBruinsmaSchenkel_2020_LinearYang} 
the CCR quantization functor preserves weak 
equivalences, from 
$c: (L,\widetilde{\tau}) \to (\LLL(M),\tau_M)$ 
we obtain a differential graded algebra 
$A \in \DGA_\bbC$ and a weak equivalence $q: A \to \AAA_M(M)$ in $\DGA_\bbC$, 
to the original differential graded algebra 
$\AAA_M(M) \in \DGA_\bbC$. 
\sk

We consider the cochain complex with vanishing differential 
\begin{equation}\label{eq:tildeL}
L \coloneqq \Bigg(
\xymatrix{
{\overset{(-1)}{\underset{e^\ddagger}{\calH^0(S^1)}}} \ar[r]^-{0} & {\overset{(0)}{\underset{a_1}{\calH^1(S^1)} \oplus \underset{a_2}{\calH^1(S^1)}}} \ar[r]^-{0} & {\overset{(1)}{\underset{e}{\calH^0(S^1)}}}
}
\Bigg) \in \Ch_\bbR,
\end{equation}
which is generated as a graded vector space by the harmonic 
forms $e^\ddagger \coloneqq 1, a_1 \coloneqq d\theta , a_2 \coloneqq d\theta, e \coloneqq 1$. 
Furthermore, choosing any compactly 
supported function $f \in C^\infty_\cc(\bbR)$ 
such that $\int_\bbR dt\, f = 1$, 
we define the cochain map 
\begin{subequations}\label{eq:c}
\begin{equation}
c: L \longrightarrow \LLL(M)
\end{equation}
in $\Ch_\bbR$ degree-wise by 
\begin{align}
c^{-1}(e^\ddagger) & \coloneqq f(t)\, dt \in \LLL(M)^{-1}, \\ 
c^{0}(a_1) & \coloneqq f(t)\, d\theta + \Big( \int_\bbR ds\, s\, f(s) \Big)\, f^\prime(t)\, d\theta, & c^{0}(a_2) \in \LLL(M)^0 & \coloneqq - f^\prime(t)\, d\theta \in \LLL(M)^0, \\ 
c^{1}(e) & \coloneqq f(t) \in \LLL(M)^1.
\end{align}
\end{subequations}
Recalling \eqref{eq:L} one easily checks that those 
on the right hand sides are cocycles in $\LLL(M)$ 
and therefore $c$ is a well-defined cochain map, 
as claimed. Even more, $c$ is a quasi-isomorphism. 
To check this fact, we consider also the cochain map 
\begin{subequations}\label{eq:tildec}
\begin{flalign}
\widetilde{c}: \LLL(M) \longrightarrow L
\end{flalign}
in $\Ch_\bbR$ defined degree-wise by 
\begin{align}
\widetilde{c}^{-2}(\varepsilon^\ddagger) & \coloneqq 0, \\ 
\widetilde{c}^{-1}(\alpha^\ddagger) & \coloneqq \Big( \int_\bbR dt\, \alpha^\ddagger_{T \calH} \Big)\, e^\ddagger, \\
\widetilde{c}^{0}(\alpha) & \coloneqq \int_\bbR dt\, \alpha_{S \calH} \oplus \int_\bbR dt\, t\, \alpha_{S \calH}, \\
\widetilde{c}^{1}(\varepsilon) & \coloneqq \Big( \int_\bbR dt\, \varepsilon_{\calH} \Big)\, e,
\end{align}
\end{subequations}
for all $\varepsilon^\ddagger \in \LLL(M)^{-2} = \Omega^0_\cc(M)$, 
$\alpha^\ddagger \in \LLL(M)^{-1} = \Omega^1_\cc(M)$, 
$\alpha \in \LLL(M)^0 = \Omega^1_\cc(M)$ and 
$\varepsilon \in \LLL^1(M) = \Omega^0_\cc(M)$. 
Note that the above definition of $\widetilde{c}$ 
uses also the decompositions 
in \eqref{eq:forms-decomposition} and 
\eqref{eq:harmonic-decomposition}. 
Using \eqref{eq:operators-decomposition} to compute 
the relevant differentials, one checks that 
$\widetilde{c}$ as defined above is indeed a cochain 
map. Furthermore, one immediately realizes that  
$\widetilde{c}\, c = \id$, 
hence $H(\widetilde{c})\, H(c) = \id$ in cohomology. 
On the other hand direct inspection shows also that 
$H(c)\, H(\widetilde{c}) = \id$. 
Therefore $H(c): H(L) \to H(\LLL(M))$ 
is an isomorphism, i.e.\ $c$ is a quasi-isomorphism. 

\begin{rem}
The previous argument can be refined 
by constructing a homotopy $\eta$ that exhibits 
$\widetilde{c}$ as a quasi-inverse of $c$. 
More in detail, $\eta \in [\LLL(M),\LLL(M)]^{-1}$
is a $(-1)$-cochain in the internal hom, 
whose differential $\partial \eta = c\, \widetilde{c} - \id$ 
controls to what extent $c\, \widetilde{c}$ differs from $\id$. 
The homotopy $\eta$ and the equations $\widetilde{c}\, c = \id$ 
and $c\, \widetilde{c} = \id + \partial \eta$ 
witness that $\widetilde{c}$ is 
a quasi-inverse of $c$. 
We do not provide an explicit choice of $\eta$ 
as it is not needed in the sequel. Let us just mention 
that it can be constructed for instance using 
the decompositions \eqref{eq:forms-decomposition} 
and \eqref{eq:harmonic-decomposition} in a way similar 
to the construction of $\widetilde{c}$ 
in \eqref{eq:tildec}. 
\end{rem}

We now endow the cochain complex $L$ 
with the transferred Poisson structure 
\begin{subequations}
\begin{equation}
\widetilde{\tau} \coloneqq \tau_M\, (c \otimes c): L^{\wedge 2} \longrightarrow \bbR, 
\end{equation}
which reads explicitly as 
\begin{align}
\widetilde{\tau}(e,e^\dagger) = 1, && 
\widetilde{\tau}(a_2,a_1) = 1, && \widetilde{\tau}(a_i,e^{(\ddagger)}) = 0.
\end{align}
\end{subequations}
This result follows from the 
definition of the Poisson structure $\tau_M$ 
in \eqref{eq:PoissonStr} and the explicit formula for the retarded-minus-advanced 
propagator $G^{(k)}$ from \eqref{eq:G-explicit}. 
(One also uses that $S^1$ has unit length
and that $f$ is such that $\int_\bbR dt\, f = 1$.) 
\sk

Summarizing, 
$c: (L,\widetilde{\tau}) \to (\LLL(M),\tau_M)$ 
is a quasi-isomorphism that preserves 
the Poisson structures. Recalling 
\cite[Prop.\ 5.3]{BeniniBruinsmaSchenkel_2020_LinearYang}, 
one finds that the CCR quantization functor sends 
$c$ to the weak equivalence of differential graded 
algebras $q: A \to \AAA_M(M)$ in $\DGA_\bbC$, where 
\begin{subequations}\label{eq:CCR-explicit}
\begin{equation}
A \coloneqq T_\bbC(L) / I_{\widetilde{\tau}} \in \DGA_\bbC
\end{equation}
is the differential graded algebra 
generated by $L$ and subject 
to the canonical commutation relations encoded 
by the two-sided ideal $I_{\widetilde{\tau}}$ generated by 
\begin{align}
e \otimes e^\dagger + e^\ddagger \otimes e - i, && a_2 \otimes a_1 - a_1 \otimes a_2 - i, && a_i\, e^{(\ddagger)} - e^{(\ddagger)}\, a_i, 
\end{align}
\end{subequations}
and $q$ extends $c$ from generators. (Compare the 
above construction of $A$ to \eqref{eq:CCR} 
and note that $q$ is indeed well-defined because 
$c$ preserves the Poisson structures.) 

\begin{rem}\label{rem:H0-vs-inv}
Recalling Remark \ref{rem:omega-inv}, one realizes that 
the CCR algebra $\AAA^{\mathrm{inv}}(M)$ 
generated by gauge-invariant on-shell linear observables 
is isomorphic to the subalgebra of $A$ 
generated by $a_1, a_2$ only. 
In contrast, the cohomology $H(\AAA_M(M))$ 
of the global algebra $\AAA_M(M)$ is isomorphic to 
$A$ (seen as a graded algebra, i.e.\ forgetting 
its differential). In particular, the degree $0$ cohomology algebra 
$H^0(\AAA_M(M)) \cong A^0$ is strictly larger 
than $\AAA^{\mathrm{inv}}(M)$. 
Indeed, along with the generators $a_1$ and $a_2$, 
which are also contained in $\AAA^{\mathrm{inv}}(M)$, 
$A^0$ has an additional generator $e^\ddagger\, e$ 
that commutes with both $a_1$ and $a_2$. 
The latter is a composite observable 
formed by an antifield observable $e^\dagger$ 
and a ghost observable $e$. Those may be regarded as  
``gauge-invariant'' observables in the broader sense 
of being cohomologically non-trivial. 
\end{rem}

While being Quillen equivalent to $_{\AAA_M(M)}\Mod$ due to 
Proposition \ref{propo:change-monoid-vs-model-str}, 
the category of left $A$-modules $_A\Mod$ is 
considerably easier to describe. 
As it is always the case, a left $A$-module $V = (V,\nu) \in \AMod$ 
consists of a cochain complex $V \in \Ch_\bbC$ 
and of a left $A$-action $\nu: A \otimes V \to V$ 
in $\Ch_\bbC$, which is the same datum as a morphism 
$\nu: A \to [V,V]$ in $\DGA_\bbC$ to the 
internal endomorphism algebra\footnote{Given 
any cochain complex $V \in \Ch_\bbC$, 
the internal hom $[V,V] \in \Ch_\bbC$ carries 
a natural differential graded algebra structure 
consisting of the usual internal hom differential 
complemented with the multiplication given by composition 
and the unit given by $\id_V$. When endowed 
with this structure we refer to $[V,V] \in \DGA_\bbC$ 
as the internal endomorphism algebra of $V \in \Ch_\bbC$.} 
$[V,V] \in \DGA_\bbC$. The specific form of $A$, 
however, allows us to describe $\nu$ in terms of 
very few concrete data. 
Since $A = T_\bbC(L) / I_{\widetilde{\tau}}$ is defined 
as the differential graded algebra that is generated 
by the $(-1)$-cocycle $e^\ddagger \in L^{-1}$, 
the $0$-cocycles $a_1, a_2 \in L^0$ 
and the $1$-cocycle $e \in L^1$ 
and that is subject to the canonical commutation 
relations from \eqref{eq:CCR-explicit}, 
it follows that the left $A$-action $\nu$ is pinned down by 
the $(-1)$-cocycle $\nu(e^\ddagger) \in [V,V]^{-1}$, 
the $0$-cocycles $\nu(a_1), \nu(a_2) \in [V,V]^0$ 
and the $1$-cocycle $\nu(e) \in [V,V]^1$ 
in the internal endomorphism algebra $[V,V] \in \DGA_\bbC$. 
Furthermore, these cocycles must fulfil the relations 
$\nu(e)\, \nu(e^\ddagger) + \nu(e^\ddagger)\, \nu(e) = i\, \id_V$, 
$\nu(a_2)\, \nu(a_1) - \nu(a_1)\, \nu(a_2) = i\, \id_V$ 
and $\nu(a_i)\, \nu(e^{(\ddagger)}) = \nu(e^{(\ddagger)})\, \nu(a_i)$. 
Similarly, a left $A$-module morphism $F: V \to V^\prime$ in $\AMod$ 
consists of a cochain map $F: V \to V^\prime$ in $\Ch_\bbC$ 
that preserves the cocycles above, 
i.e.\ $F\, \nu(e^\ddagger) = \nu^\prime(e^\ddagger)\, F$, 
$F\, \nu(a_1) = \nu^\prime(a_1)\, F$, 
$F\, \nu(a_2) = \nu^\prime(a_2)\, F$ 
and $F\, \nu(e) = \nu^\prime(e)\, F$. 
\sk

The above very explicit description of left 
$A$-modules gives us a handle to construct 
all constant representations of the net $\AAA_M$ 
up to weak equivalence. 
The latter form the essential image of the total 
right derived functor of the right Quillen functor 
$(-)^M: {_{\AAA_M(M)}\Mod} \to \Rep(\AAA_M)$ 
from Construction \ref{constr:eval-const-adj} and 
Remark \ref{rem:eval-const-adj-vs-model-str}. 
Since $q: A \to \AAA_M(M)$ in $\DGA_\bbC$ 
is a weak equivalence, it follows by Proposition 
\ref{propo:change-monoid-vs-model-str} that 
$\Ext_q \dashv \Res_q: {_{\AAA_M(M)}\Mod} \to {_A\Mod}$ 
is a Quillen equivalence and hence induces 
an equivalence of homotopy categories. 
Combining these facts, the essential image of the 
total right derived functor $(-)^M$ 
is equivalent to the homotopy category of $\AMod$. 
In other words, left $A$-modules provide all constant 
representations of the net $\AAA_M$ 
up to weak equivalence. (For the notions of homotopy 
category associated with a model category and of total 
left (right) derived functor associated with a left 
(respectively right) Quillen functor we refer the 
reader to \cite[Ch.\ 1]{Hovey_1999_ModelCategories}.)


\section*{Acknowledgments}
We are grateful to the referee for his valuable comments 
and in particular for observing that nets of algebras 
are the same as monoids in a functor category, 
and hence that net representations are the same as left modules therein, 
which helped us to condense many of the concepts and constructions 
appearing in Section \ref{subsec:AQFT}. 
A.A.\ is supported by a PhD scholarship of the University of Genoa (IT).

\section*{Data availability statement}
Data sharing not applicable to this article as no datasets were generated or analysed during the current study.


\appendix

\section{\label{app:monoids-modules}Monoids and modules in a symmetric monoidal category}
We collect here the basics of the theory of monoids 
and of modules over a monoid in a complete and cocomplete 
closed symmetric monoidal category $\MM = (\MM,\otimes,\oone)$. 
Standard textbook references for these topics are 
\cite{MacLane_1978_CategoriesWorking, Borceux_1994_HandbookCategorical, EtingofGelakiNikshychEtAl_2015_TensorCategories}. 
The prime example of a complete and cocomplete 
closed symmetric monoidal category is the category $\Vec_\bbK$ 
of vector spaces over a field $\bbK$. 
Another important example is the category $\Ch_\bbK$ 
of cochain complexes, which is recalled at the beginning of 
Section \ref{sec:Maxwell}, where it will be extensively used. 

\begin{defi}\label{defi:monoid}
A {\it monoid} $A = (A,\mu,\bfone)$ in $\MM$ 
consists of an object $A \in \MM$ endowed with two morphisms 
$\mu: A \otimes A \to A$ and $\bfone: \oone \to A$ in $M$, 
called {\it multiplication} and {\it unit} respectively, 
subject to the usual associativity 
$\mu (\mu \otimes \id_A) = \mu (\id_A \otimes \mu)$ 
and unitality $\mu (\bfone \otimes \id_A) = \id_A = \mu (\id_A \otimes \bfone)$ axioms. 
Furthermore, one defines a {\it morphism of monoids} 
$\varphi: A \to B$ as a morphism $\varphi: A \to B$ in $\MM$ 
that preserves multiplications 
($\mu_B\, (\varphi \otimes \varphi) = \varphi\, \mu_A$) 
and units ($\bfone_B = \varphi\, \bfone_A$). 
$\Mon(\MM)$ denotes the category of monoids in $\MM$. 
\end{defi}

\begin{defi}\label{defi:module}
Given a monoid $A \in \Mon(\MM)$, a {\it left $A$-module} 
$L = (L,\lambda)$ in $\MM$ consists of an object $L \in \MM$ 
endowed with a morphism $\lambda: A \otimes L \to L$ in $\MM$, 
called {\it left $A$-action}, subject to the usual axioms 
$\lambda (\id_A \otimes \lambda) = \lambda (\mu \otimes \id_L)$ 
and $\lambda (\bfone \otimes \id_L) = \id_L$. Furthermore, 
one defines a {\it morphism of left $A$-modules} 
$F: L \to L^\prime$ as a morphism $F: L \to L^\prime$ in $\MM$ 
that is compatible with the left $A$-actions 
$\lambda^\prime (\id_A \otimes F) = F \lambda$. 
This defines the category $\AMod$ of left $A$-modules in $\MM$. 
\end{defi}
One defines the category $\Mod_A$ of {\it right} $A$-modules 
in a similar fashion. 

\subsection{\label{app:change-monoid-adj}Change-of-monoid adjunction}
For the constructions recalled below, 
we refer the reader again to the textbooks
\cite{MacLane_1978_CategoriesWorking, Borceux_1994_HandbookCategorical, EtingofGelakiNikshychEtAl_2015_TensorCategories}.
Associated with a morphism $\varphi: A \to B$ in $\Mon(\MM)$, 
one has the so-called {\it change-of-monoid} adjunction 
\begin{equation}\label{eq:change-monoid-adj}
\xymatrix{
\Ext_{\varphi} \coloneqq B \underset{A}{\otimes} (-): \AMod \ar@<2pt>[r] & \BMod :(-)\vert_{A} \eqqcolon \Res_{\varphi}, \ar@<2pt>[l]
}
\end{equation}
whose right adjoint $\Res_{\varphi}$ and left adjoint $\Ext_{\varphi}$ 
restrict and respectively extend the module structure along $\varphi$. 
Explicitly, the {\it restriction} functor 
\begin{equation}
\xymatrix{\Res_{\varphi} \coloneqq (-)\vert_{A}: \BMod \ar[r] & \AMod}
\end{equation}
assigns to a left $B$-module $M=(M,\lambda) \in \BMod$ 
the left $A$-module 
$M\vert_A \coloneqq (M,\lambda (\varphi \otimes \id_M)) \in \AMod$ 
defined restricting the left $B$-module action 
$\lambda: B \otimes M \to M$ along $\varphi$. 
Furthermore, $\Res_{\varphi}$ assigns to a morphism 
of left $B$-modules $G: M \to M^\prime$ in $\BMod$ 
the morphism of left $A$-modules 
$G\vert_A: M\vert_A \to M^\prime\vert_A$ in $\AMod$ 
whose underlying morphism in $\MM$ coincides 
with the one underlying $G$. 
The left adjoint to $\Res_{\varphi}$ 
is the {\it extension} functor 
\begin{subequations}\label{eq:relative-tensor}
\begin{equation}
\xymatrix{\Ext_{\varphi} := B \otimes_A (-): \AMod \ar[r] & \BMod}
\end{equation}
defined below. 
$\Ext_{\varphi}$ assigns to a left $A$-module $L = (L,\lambda) \in \AMod$ the relative 
tensor product $B \otimes_A L \in \BMod$, 
defined as the coequalizer 
\begin{equation}
B \underset{A}{\otimes} L \coloneqq \colim \Big(
\xymatrix@C=6.5em{
B \otimes A \otimes L \ar@<2pt>[r]^-{\mu_B(\id_B \otimes \varphi) \otimes \id_L} \ar@<-2pt>[r]_-{\id_B \otimes \lambda} & B \otimes L
}
\Big) \in \BMod. 
\end{equation} 
Furthermore, $\Ext_{\varphi}$ assigns to a 
morphism of left $A$-modules $F: L \to L^\prime$ in $\AMod$ 
the morphism of left $B$-modules 
$B \otimes_A F: B \otimes_A L \to B \otimes_A L^\prime$ 
in $\BMod$ defined via the universal property 
of the coequalizer by the diagram 
\begin{equation}
\xymatrix@C=6.5em{
B \otimes A \otimes L \ar@<2pt>[r]^-{\mu_B(\id_B \otimes \varphi) \otimes \id_L} \ar@<-2pt>[r]_-{\id_B \otimes \lambda} \ar[d]_-{\id_B \otimes \id_A \otimes F} & B \otimes L \ar[d]^-{\id_B \otimes F} \\ 
B \otimes A \otimes L^\prime \ar@<2pt>[r]^-{\mu_B(\id_B \otimes \varphi) \otimes \id_{L^\prime}} \ar@<-2pt>[r]_-{\id_B \otimes \lambda^\prime} & B \otimes L^\prime
}
\end{equation}
\end{subequations}
in $\BMod$. From the definitions above one easily obtains a bijection 
\begin{equation}
\BMod(\Ext_{\varphi} L, M) \cong \AMod(L, \Res_{\varphi} M),
\end{equation} 
which is natural with respect to both $L \in \AMod$ 
and $M \in \BMod$, hence proving the adjunction \eqref{eq:change-monoid-adj}. 

\begin{rem}\label{rem:change-monoid-composition}
Given composable morphisms $\varphi: A \to B$ 
and $\psi: B \to C$ in $\Mon(\MM)$, 
one immediately realizes that the restriction 
$\Res_{\psi}: {_C\Mod} \to \BMod$ followed 
by the restriction $\Res_{\varphi}: \BMod \to \AMod$ 
coincides with the restriction 
$\Res_{\psi\, \varphi}: {_C\Mod} \to \AMod$ 
along the composition $\psi\, \varphi$. 
From the change-of-monoid adjunction \ref{eq:change-monoid-adj} 
it then follows that the extension $\Ext_{\varphi}: \AMod \to \BMod$ 
followed by the extension $\Ext_{\psi}: \BMod \to {_C\Mod}$ 
is naturally isomorphic to the extension 
$\Ext_{\psi\, \varphi}: \AMod \to {_C\Mod}$ 
along the composition $\psi\, \varphi$.
\end{rem}

\begin{rem}\label{rem:change-monoid-adjoint-equivalence}
Given an isomorphism $\varphi: A \to B$ in $\Mon(\MM)$, 
the corresponding change-of-monoid adjunction 
\eqref{eq:change-monoid-adj} is actually an adjoint 
equivalence. This follows from the straightforward 
observation that in this case both the unit and the counit of 
the change-of-monoid adjunction are natural isomorphisms. 
\end{rem}

\subsection{\label{app:tensoring}\texorpdfstring{$\MM$-tensoring}{Tensoring}, powering and enriched hom on \texorpdfstring{$\AMod$}{the category of modules}}
For a given monoid $A \in \Mon(\MM)$, the category 
of left $A$-modules $\AMod$ 
admits the canonical $\MM$-tensoring 
\begin{equation}\label{eq:tensoring}
\otimes: \AMod \times \MM \longrightarrow \AMod, 
\end{equation}
that assigns to a left $A$-module $L = (L,\lambda) \in \AMod$ 
and an object $V \in \MM$ the left $A$-module 
$L \otimes V \in \AMod$ consisting of the object 
$L \otimes V \in \MM$ equipped with the left $A$-module action 
$\lambda \otimes \id_V: A \otimes L \otimes V \to L \otimes V$, 
and to a morphism $F: L \to L^\prime$ in $\AMod$ and a morphism 
$\xi: V \to V^\prime$ in $\MM$ the morphism 
$F \otimes \xi: L \otimes V \to L^\prime \otimes V^\prime$ 
in $\AMod$ whose underlying morphism in $\MM$ is the tensor 
product of the underlying morphism $F$ in $\MM$ with $\xi$. 
\sk

Given $V \in \MM$, the $\MM$-tensoring 
$(-) \otimes V: \AMod \to \AMod$ is part of the adjunction 
\begin{equation}\label{eq:tensoring-powering}
\xymatrix{
(-) \otimes V: \AMod \ar@<2pt>[r] & \AMod :(-)^V, \ar@<2pt>[l]
}
\end{equation}
whose right adjoint $(-)^V$ is the partial evaluation of the $\MM$-powering
\begin{subequations}\label{eq:powering}
\begin{equation}
(-)^{(-)}: \AMod \times \MM^\op \longrightarrow \AMod. 
\end{equation}
The latter is defined through the internal hom 
$[-,-]: \MM^\op \otimes \MM \to \MM$ 
of the closed symmetric monoidal category $\MM$. 
Explicitly, $(-)^{(-)}$ assigns to a left $A$-module $L=(L,\lambda) \in \AMod$ and an object $V \in \MM$ 
the left $A$-module $L^V \in \AMod$ consisting of 
the internal hom $[V,L] \in \MM$ 
and the left $A$-module action 
$A \otimes [V,L] \to [V,L]$ in $\MM$ defined 
as the adjunct of the morphism 
\begin{equation}
\xymatrix{A \otimes [V,L] \otimes V \ar[r]^-{\id \otimes \ev} & A \otimes L \ar[r]^-{\lambda} & L}
\end{equation}
\end{subequations}
in $\MM$ with respect to the adjunction 
$(-) \otimes V \dashv [V,-]: \MM \to \MM$. 
Furthermore, $(-)^{(-)}$ assigns 
to a morphism $F: L \to L^\prime$ in $\AMod$ and 
a morphism $\xi: V^\prime \to V$ in $\MM$ 
the morphism $F^\xi: L^V \to {L^\prime}^{V^\prime}$ in $\AMod$ 
whose underlying morphism is 
$[\xi,F]: [V,L] \to [V^\prime,L^\prime]$ in $\MM$. 
\sk

Similarly, given $L \in \AMod$, the other partial 
evaluation of the $\MM$-tensoring 
$L \otimes (-): \MM \to \AMod$ 
is part of the adjunction 
\begin{equation}\label{eq:tensoring-enriched-hom}
\xymatrix{
L \otimes (-): \MM \ar@<2pt>[r] & \AMod: [L,-]_A, \ar@<2pt>[l]
}
\end{equation}
whose right adjoint $[L,-]_A$ is 
the partial evaluation of the $\MM$-enriched hom 
\begin{subequations}\label{eq:enriched-hom}
\begin{equation}
[-,-]_A: \AMod^\op \times \AMod \longrightarrow \MM. 
\end{equation}
The latter assigns to left $A$-modules $L = (L,\lambda), L^\prime = (L^\prime,\lambda^\prime) \in \AMod$ the equalizer 
\begin{equation}\label{eq:enriched-hom-equalizer}
[L,L^\prime]_A \coloneqq \lim \Big(
\xymatrix@C=2.5em{
[L,L^\prime] \ar@<2pt>[r]^-{[\lambda,\id]} \ar@<-2pt>[r]_-{\lambda^\prime_\ast} & [A \otimes L,L^\prime]
}
\Big) \in \MM, 
\end{equation}
where $\lambda^\prime_\ast$ denotes the adjunct with respect to the adjunction 
$(-) \otimes A \otimes L \dashv [A \otimes L,-]: \MM \to \MM$ of the morphism 
\begin{equation}
\xymatrix@C=3em{
[L,L^\prime] \otimes A \otimes L \ar[r]^-{\cong} & A \otimes [L,L^\prime] \otimes L \ar[r]^-{\id \otimes \ev} & A \otimes L^\prime \ar[r]^-{\lambda^\prime} & L^\prime
}
\end{equation}
\end{subequations}
in $\MM$. 
The action of $[-,-]_A$ on morphisms is defined 
combining the universal property of the equalizer 
in \eqref{eq:enriched-hom-equalizer} 
and the functoriality both of the tensor product 
$\otimes$ and of the internal hom $[-,-]$ of $\MM$. 

\begin{rem}\label{rem:change-monoid-vs-tensoring-powering}
Note that the adjunction \eqref{eq:tensoring-powering} 
is compatible with the change-of-monoid adjunction 
\eqref{eq:change-monoid-adj} in the following sense. 
Given a morphism of monoids $\varphi: A \to B$ 
in $\Mon(\MM)$ and an object $V \in \MM$, 
the diagram of right adjoint functors 
\begin{equation}\label{eq:restriction-vs-powering}
\xymatrix@C=3em{
\BMod \ar[r]^-{\Res_{\varphi}} \ar[d]_-{(-)^V} & \AMod \ar[d]^-{(-)^V} \\
\BMod \ar[r]_-{\Res_{\varphi}} & \AMod
}
\end{equation}
commutes as a straightforward consequence of their 
definitions. Therefore, the corresponding diagram 
of left adjoint functors commutes 
up to a unique natural isomorphism 
$\Ext_{\varphi} (- \otimes V) \cong (\Ext_{\varphi}(-)) \otimes V$. 
\end{rem}

\begin{rem}
One easily checks that the functors 
\eqref{eq:tensoring}, \eqref{eq:powering} 
and \eqref{eq:enriched-hom} form a two-variable adjunction 
with isomorphisms 
\begin{equation}
\AMod(L_1 \otimes V, L_2) \cong \AMod(L_1, L_2^V) \cong \MM\left(V, [L_1, L_2]_A\right),
\end{equation}
natural in $L_1, L_2 \in \AMod$ and $V \in \MM$, 
given by the adjunctions 
\eqref{eq:tensoring-powering}, \eqref{eq:tensoring-enriched-hom} and 
\begin{equation}\label{eq:enriched-hom-powering}
\xymatrix{
[-,L]_A: \AMod \ar@<2pt>[r] & \MM^\op :L^{(-)},  \ar@<2pt>[l]
}
\end{equation}
for $L \in \AMod$. 
Given a morphism of monoids $\varphi: A \to B$ in $\Mon(\MM)$, 
the latter adjunction allows us to describe 
the interplay between the $\MM$-enriched hom
$[-,-]_A: \AMod^\op \times \AMod \to \MM$ on $\AMod$, 
the $\MM$-enriched hom 
$[-,-]_B: \BMod^\op \times \BMod \to \MM$ on $\BMod$ 
and the restriction $\Res_{\varphi}: \BMod \to \AMod$ 
along $\varphi$ in terms of the natural transformation 
\begin{equation}\label{eq:restriction-vs-enriched-hom}
[-,-]_B \longrightarrow [-,-]_A \circ (\Res_{\varphi} \times \Res_{\varphi})
\end{equation}
between functors from $\BMod^\op \times \BMod$ to $\MM$. 
Explicitly, its component for $M,M^\prime \in \BMod$ 
is the morphism 
$[M,M^\prime]_B \to [M\vert_A,M^\prime\vert_A]_A$ in $\MM$ 
defined by the following three-step construction: (1)~Consider the $M$-component 
$M \to M^{\prime\,[M,M^\prime]_B}$ 
in $\BMod$ of the unit of the adjunction 
$[-,M^\prime]_B \dashv {M^\prime}^{(-)}: \MM^{\op} \to \BMod$. 
(2)~Construct from the latter the morphism 
$M\vert_A \to {M^{\prime}\vert_A}^{[M,M^\prime]_B}$ 
in $\AMod$ by applying the restriction 
$\Res_{\varphi}: \BMod \to \AMod$ along $\varphi$ 
and by recalling the commutative square 
in \eqref{eq:restriction-vs-powering}. 
(3)~Recalling also the adjunction 
$[-,M^\prime\vert_A]_A \dashv {M^\prime\vert_A}^{(-)}: \MM^{\op} \to \AMod$, 
define the morphism 
$[M\vert_A,M^\prime\vert_A]_A \to [M,M^\prime]_B$ 
in the opposite category $\MM^{\op}$, 
which is equivalent to the desired morphism 
$[M,M^\prime]_B \to [M\vert_A,M^\prime\vert_A]_A$ 
in $\MM$. (Note that we used here two different 
instances of the adjunction displayed in 
\eqref{eq:enriched-hom-powering}.) 
\end{rem}

\subsection{\label{subsec:model-str}Model structure for modules over monoids}
We recall from \cite{SchwedeShipley_2000_AlgebrasModules} 
some fundamental facts about the model structures 
on the category $\Mon(\MM)$ of monoids and on the category $\AMod$ 
of left $A$-modules over a monoid $A \in \Mon(\MM)$. 
For this purpose we assume that the symmetric monoidal category $\MM$ 
is endowed with a suitable model structure as explained below. 

\begin{setup}\label{setup:model-str}
Suppose $\MM$ is a cofibrantly generated closed symmetric 
monoidal model category, 
see \cite[Sec.\ 4.2]{Hovey_1999_ModelCategories}, 
that satisfies the monoid axiom and whose objects are small, 
see \cite[Secs.\ 2 and 3]{SchwedeShipley_2000_AlgebrasModules}. 
\end{setup}

\begin{ex}
The assumptions listed above are met, 
e.g., by the closed symmetric monoidal category $\MM=\Ch_\bbK$ 
of (unbounded) cochain complexes over a field $\bbK$ 
equipped with the projective model structure. 
In particular, the monoid axiom follows from 
the fact that all cochain complexes 
$V \in \Ch_\bbK$ are cofibrant (and fibrant), 
see \cite[Rem.\ 3.4]{SchwedeShipley_2000_AlgebrasModules} 
and also the beginning of Section \ref{sec:Maxwell}. 
\end{ex}

\begin{defi}\label{def:model-str}
A morphism $\varphi: A \to B$ in the category $\Mon(\MM)$ 
of monoids in $\MM$ is called a {\it weak equivalence} ({\it fibration}) 
if the underlying morphism in the model category $\MM$ 
is a weak equivalence (respectively fibration) 
or a {\it cofibration} if it has the left lifting 
property (see e.g.\ \cite[Def.\ 1.1.2]{Hovey_1999_ModelCategories}) 
with respect to all {\it acyclic fibrations} in $\Mon(\MM)$, 
i.e.\ morphisms that are simultaneously 
fibrations and weak equivalences. 
\sk

Furthermore, let $A \in \Mon(\MM)$ be a monoid in $\MM$ and consider 
the category $\AMod$ of left $A$-modules in $\MM$. 
A morphism $F:L \to L^\prime$ in $\AMod$ 
is called a {\it weak equivalence} ({\it fibration}) 
if the underlying morphism in the model category $\MM$ 
is a weak equivalence (respectively fibration) 
or a {\it cofibration} if it has the left lifting 
property with respect to all {\it acyclic fibrations} in $\AMod$. 
\end{defi}

It follows from \cite[Th.\ 4.1]{SchwedeShipley_2000_AlgebrasModules} 
that the previous definition equips both $\Mon(\MM)$ and $\AMod$ 
with cofibrantly generated model structures. 
We record this fact in the next statement. 

\begin{propo}[$\Mon(\MM)$ and $\AMod$ as model categories]\label{propo:model-str}
Let $\MM$ be as in Set-up \ref{setup:model-str} 
and consider a monoid $A \in \Mon(\MM)$. 
With the previous Definition \ref{def:model-str} 
of weak equivalences, fibrations and cofibrations, 
the category $\Mon(M)$ of monoids in $\MM$ 
and the category $\AMod$ of left $A$-modules 
become cofibrantly generated model categories. 
\end{propo}

In a similar fashion one endows the category $\Mod_A$ 
of right $A$-modules with a cofibrantly generated model 
category structure. 
\sk

Having established a model structure on $\AMod$, 
we investigate its compatibility with the $\MM$-tensoring, 
powering and enriched hom on $\AMod$. Namely we 
ask whether $\AMod$ is an $\MM$-model category, 
see \cite[Def.\ 4.2.18]{Hovey_1999_ModelCategories}. 
This amounts to showing that 
the $\MM$-tensoring \eqref{eq:tensoring} is a Quillen bifunctor, 
see \cite[Def.\ 4.2.1]{Hovey_1999_ModelCategories}. 
For $F:L \to L^\prime$ in $\AMod$ and $\xi:V \to V^\prime$ in $\MM$, 
the {\it pushout-product} morphism 
\begin{subequations}\label{eq:pushout-product}
\begin{equation}
F\, \square\, \xi: P(F,\xi) \to L^\prime \otimes V^\prime
\end{equation}
in $\AMod$ is determined by the commutative diagram 
\begin{equation}
\xymatrix@C=3em{
L \otimes V \ar[r]^-{F \otimes \id_V} \ar[d]_-{\id_L \otimes \xi} & L^\prime \otimes V \ar[d]^-{\id_{L^\prime} \otimes \xi} \\
L \otimes V^\prime \ar[r]_-{F \otimes \id_{V^\prime}} & L^\prime \otimes V^\prime 
}
\end{equation}
\end{subequations}
in $\AMod$ and by the universal property of the pushout 
$P(F,\xi) \coloneqq \colim (L \otimes V^\prime \leftarrow L \otimes V \rightarrow L^\prime \otimes V) \in \AMod$. 
Let $I$ ($J$) be a set of generating cofibrations 
(respectively acyclic cofibrations) for $\MM$. 
By \cite[Th.\ 4.1 and Lem.\ 2.3]{SchwedeShipley_2000_AlgebrasModules}, 
applying the free left $A$-module 
functor $A \otimes (-): \MM \to \AMod$ to $I$ ($J$) determines 
a set of generating cofibrations 
$I_A \coloneqq A \otimes I$ (respectively acyclic 
cofibrations $J_A \coloneqq A \otimes J$) 
for $\AMod$. In order to show that the $\MM$-tensoring is 
a Quillen bifunctor, it suffices to check that 
the pushout-product $F\, \square\, \xi$ is a cofibration 
(acyclic cofibration) in $\AMod$ when $F \in I_A$ and $\xi \in I$ 
(respectively $F \in J_A$ and $\xi \in I$ or $F \in I_A$ and 
$\xi \in J$), 
see \cite[Cor.\ 4.2.5]{Hovey_1999_ModelCategories}. 
Since $F \in I_A$ ($F \in J_A$) is of the form 
$F = \id_A \otimes \eta$, with $\eta \in I$ 
(respectively $\eta \in J$), one obtains the isomorphism 
$P(F,\xi) \cong A \otimes P(\eta,\xi)$ in $\AMod$ 
and hence the pushout-product morphism 
$F\, \square\, \xi = A \otimes (\eta\, \square\, \xi)$ 
can be computed from the pushout-product morphism 
$\eta\, \square\, \xi$ in $\MM$. Since $\MM$ is 
per hypothesis a closed symmetric monoidal model category, 
$\eta\, \square\, \xi$ is a cofibration 
(respectively acyclic cofibration) in $\MM$. 
Furthermore, the free left $A$-module functor 
$A \otimes (-): \MM \to \AMod$ is a left Quillen functor
since it is left adjoint to the forgetful functor $\AMod \to \MM$, 
which detects fibrations and weak equivalences 
by definition of the model structure on $\AMod$. 
It follows that $F\, \square\, \xi$ is a cofibration (respectively acyclic 
cofibration) in $\AMod$, as desired. 
For later reference it is convenient to summarize 
the conclusion of this paragraph in a proposition.

\begin{propo}[$\AMod$ as an $\MM$-model category]\label{propo:tensoring-vs-model-str}
Let $\MM$ be as in Set-up \ref{setup:model-str} 
and $A \in \Mon(\MM)$ be a monoid in $\MM$. 
Then $\AMod$ becomes an $\MM$-model category, 
see \cite[Def.\ 4.2.18]{Hovey_1999_ModelCategories}, 
when equipped with the $\MM$-tensoring 
\eqref{eq:tensoring} 
and the model structure from Proposition \ref{propo:model-str}. 
\end{propo}

We conclude our preliminaries recalling that the model structures 
on monoids and modules interact well with the change-of-monoid 
adjunction \eqref{eq:change-monoid-adj} associated with 
a morphism of monoids $\varphi: A \to B$ in $\Mon(\MM)$. 
By definition, weak equivalences and fibrations 
in both $\AMod$ and $\BMod$ are detected 
in the underlying model category $\MM$. 
Since the right adjoint functor $\Res_{\varphi}$ followed by the 
forgetful functor $\AMod \to \MM$ coincides with the forgetful 
functor $\BMod \to \MM$, 
$\Res_{\varphi}$ preserves both fibrations and weak equivalences. 
In particular, the change-of-monoid adjunction 
\eqref{eq:change-monoid-adj} is a Quillen adjunction. 
(For the concept of Quillen adjunction 
refer e.g.\ to \cite[Sec.\ 1.3.1]{Hovey_1999_ModelCategories}.) 
Furthermore, by \cite[Th.\ 4.3]{SchwedeShipley_2000_AlgebrasModules}, 
the latter is also a Quillen equivalence 
(see \cite[Sec.\ 1.3.3]{Hovey_1999_ModelCategories}) 
when $\varphi$ is a weak equivalence in $\Mon(\MM)$ 
and an additional technical assumption is met. 
These facts are recorded below. 

\begin{propo}[Change-of-monoid as a Quillen adjunction]\label{propo:change-monoid-vs-model-str}
Let $\MM$ be as in Set-up \ref{setup:model-str} and 
consider a morphism of monoids $\varphi: A \to B$ in $\Mon(\MM)$. 
Then the change-of-monoid adjunction 
$\Ext_{\varphi} \dashv \Res_{\varphi}: \BMod \to \AMod$ from 
\eqref{eq:change-monoid-adj} is a Quillen adjunction, which is 
compatible with $\MM$-tensoring, powering and enriched hom, 
see Remark \ref{rem:change-monoid-vs-tensoring-powering}. 
(In the language of \cite[Def.\ 4.2.18]{Hovey_1999_ModelCategories} 
the extension functor $\Ext_{\varphi}$ is an $\MM$-Quillen functor.) 
Additionally, if $\varphi$ is a weak equivalence and, 
for each cofibrant left $A$-module $L \in \AMod$, 
the relative tensor product $(-) \otimes_A L: \Mod_A \to \MM$ 
sends weak equivalences in $\Mod_A$ to weak equivalences in $\MM$, 
then the change-of-monoid adjunction 
$\Ext_{\varphi} \dashv \Res_{\varphi}$ is a Quillen equivalence. 
\end{propo}


\printbibliography[category=cited]

\end{document}